\RequirePackage{ifpdf}

\documentclass[aps,preprintnumbers,groupedaddress,nofootinbib,notitlepage]{revtex4-1}
\pdfoutput=1
\usepackage{ae,aecompl}
\usepackage{bm,bbm}
\usepackage{latexsym}
\usepackage{dcolumn}
\usepackage{amsfonts,amssymb}
\usepackage{graphicx,epsfig}
\usepackage{psfrag}
\usepackage{amsmath,amssymb,multirow,array,mathtools}
\usepackage{bbold}
\usepackage{tabularx}
\usepackage{float}
\usepackage{wrapfig}
\usepackage{caption}
\usepackage{subcaption}
\usepackage{subcaption}
\renewcommand{\arraystretch}{1.3}
\usepackage{enumerate}

\numberwithin{equation}{section}

\usepackage{youngtab} 
\usepackage{young} 
\usepackage{ytableau}

\usepackage{hyperref} 
\hypersetup{
	colorlinks=true,	
	linkcolor=blue,		
	citecolor=red,		
	filecolor=blue,		
	urlcolor=blue		
}

\usepackage{cleveref}
\usepackage{soul, todonotes}

\newcommand{\Tr}{\text{Tr}}
\newcommand{\ex}{\text{e}}
\newcommand{\vect}[1]{\vec{#1}}
\newcommand{\av}[1]{\<{#1}\>}
\newcommand{\seff}[1]{S_{\text{eff}}({#1})}

\newcommand{\nn}{\nonumber}

\newcommand{\ben}{\begin{eqnarray}\displaystyle}
\newcommand{\een}{\end{eqnarray}}

\newcommand{\be}{\begin{equation}}
\newcommand{\ee}{\end{equation}}

\newcommand{\bc}{\begin{center}}
\newcommand{\ec}{\end{center}}

\newcommand{\Rmnum}[1]{\expandafter\@slowromancap\romannumeral #1@}

\renewcommand{\b}{\beta}		
\renewcommand{\d}{\delta}	

\newcommand{\e}{\epsilon}

\newcommand{\g}{\gamma}

\renewcommand{\k}{\kappa}	
\renewcommand{\l}{\lambda}	
\renewcommand{\o}{\omega}	

\newcommand{\s}{\sigma}

\newcommand{\bab}{\beta_{ab}}

\newcommand{\I}{\infty}
\renewcommand{\L}{\Lambda}	

\newcommand{\cC}{\mathcal{C}}
\newcommand{\cD}{\mathcal{D}}

\newcommand{\cN}{\mathcal{N}}
\newcommand{\cO}{\mathcal{O}}

\newcommand{\ra}{\rightarrow}
\newcommand{\Ra}{\Rightarrow}

\newcommand{\lB}{\left [}
\newcommand{\rB}{\right ]}
\newcommand{\lb}{\left (}
\newcommand{\rb}{\right )}

\newcommand{\<}{\left\langle}
\renewcommand{\>}{\right\rangle}		

\newcommand{\Rp}{R'(0)}

\newcommand{\type}[1]{$\mathbb{Type \ #1}$}
\newcommand{\tp}[1]{$\mathbb{T#1}$}

\newcommand{\class}[1]{$\mathbb{Class \ #1}$}
\newcommand{\cl}[1]{$\mathbb{C#1}$}

\newcommand{\ds}{Dyson-Schwinger}

\newcommand{\hp}{h_{+}(\theta)}
\newcommand{\hmm}{h_{-}(\theta)}

\newcommand{\hmt}{$h_{-}(\theta)$}

\newcommand{\ads}{{\it AdS}}

\def\Xint#1{\mathchoice
{\XXint\displaystyle\textstyle{#1}}%
{\XXint\textstyle\scriptstyle{#1}}%
{\XXint\scriptstyle\scriptscriptstyle{#1}}%
{\XXint\scriptscriptstyle\scriptscriptstyle{#1}}%
\!\int}
\def\XXint#1#2#3{{\setbox0=\hbox{$#1{#2#3}{\int}$ }
\vcenter{\hbox{$#2#3$ }}\kern-.6\wd0}}

\newcommand{\pa}{\partial}

\begin{document}

\title{Phase Space Distribution for Two-Gap Solution in Unitary Matrix
  Model} \author{Parikshit Dutta} \email[]{pd@iiserb.ac.in}
\author{Suvankar Dutta} \email[]{suvankar@iiserb.ac.in}
\affiliation{Department of Physics, Indian Institute of Science
  Education and Research Bhopal\\
Bhopal Bypass, Bhopal 462066}


\begin{abstract}
  {\bf Abstract :} We analyze the dynamics of weakly coupled finite
  temperature $U(N)$ gauge theories on $S^3$ by studying a class of
  effective unitary matrix model. Solving Dyson-Schwinger equation at
  large $N$, we find that different phases of gauge theories are
  characterized by gaps in eigenvalue distribution over a unit
  circle. In particular, we obtain no-gap, one-gap and two-gap
  solutions at large $N$ for a class of matrix model we are
  considering. The same effective matrix model can equivalently be
  written as a sum over representations (or Young diagrams) of unitary
  group.  We show that at large $N$, Young diagrams corresponding to
  different phases can be classified in terms of discontinuities in
  number of boxes in two consecutive rows. More precisely, the
  representation, where there is no discontinuity, corresponds to
  no-gap and one-gap solution, where as, a diagram with one
  discontinuity corresponds to two-gap phase, mentioned above. This
  observation allows us to write a one to one relation between
  eigenvalue distribution function and Young tableaux distribution
  function for each saddle point, in particular for two-gap
  solution. We find that all the saddle points can be described in
  terms of free fermions with a phase space distribution for no-gap,
  one-gap and two-gap phases.
\end{abstract}


\maketitle

\tableofcontents


\section{Introduction and summary}\label{intro}

The large $N$ limit of $SU(N)$ gauge theories has many qualitative
features similar to QCD, primarily confinement-deconfinement
transitions. Moreover, from the {\it AdS/CFT} perspective, large $N$
gauge theories are conjectured to be dual to certain string theories
on an $AdS$ background. Different phases of strongly coupled large $N$
theories correspond to different phases of asymptotically
\ads\ black holes. For example, confinement/deconfinement transitions
in large $N$ gauge theories on sphere are identified with
gravitational phase transitions between \ads \ black hole and global
\ads \ (known as {\it Hawking-Page} phase transition). This conjecture
has opened up a new avenue to study different thermodynamic properties
and phase diagrams of large $N$ Yang Mills theories on a compact
manifold.

There are different ways to study non-abelian gauge theories on a
compact manifold. In \cite{shiraz}, authors argued that the thermal
partition function of a gauge theory with gauge group $SU(N)$ on a
compact manifold can be written in terms of a unitary matrix model.
\cite{Sundborg:1999ue} was the first to obtain the unitary matrix
model of the finite temperature partition function counting the number
of states in free theory. To be precise, let us consider an Euclidean
gauge theory on $S^3$ with an Euclidean time direction $S^1$. $S^1$
has a circumference $\beta$ which is inversely proportional to the
temperature of the theory. In zero 'tHooft coupling limit ($\lambda
=g_{YM}^2 N \ra 0$) all the modes (expanding the fields in spherical
harmonics) become massive, with mass scale proportional to inverse of
the radius of $S^3$, except the time component of the gauge field,
\begin{eqnarray}
\alpha=\frac{1}{V_{S^{3}}}\int_{S^{3}}A_{0}.
\end{eqnarray}
$\alpha$ is strongly interacting even in zero coupling limit. One can
integrate out all the massive modes and the theory can be described by
an effective action written purely in terms of $\alpha$. In fact the
actual variable turns out to be a $N\times N$ unitary matrix
\begin{eqnarray}\label{eq:holonomy}
U=e^{i \beta \alpha}.
\end{eqnarray}
Finally, the thermal partition function in terms of holonomy matrix $U$ is
given by,
\be
Z(\beta) = \int [dU] \exp \lB \sum_{n=1}^{\infty} \frac{a_n(\beta)}{n}
\Tr U^n \Tr U^{\dagger n}\rB
\ee
where $[dU]$ is the Haar measure and the coefficients $a_n(\beta)$ are
given by
\be\label{pf0}
a_n(\beta) = z_B(x^n) + (-1)^{n+1}z_F(x^n), \qquad x= e^{-\beta}.
\ee
Here $z_B(x)$ and $z_F(x)$ are single particle partition functions of
the bosonic and fermionic modes respectively. They completely capture
the field content of the gauge theory.

Moving to weak coupling limit, one can write a more generic matrix
model in the weak 't Hooft coupling limit (\cite{shiraz} for details):
\be\label{eq:pf-weak-coup}
Z(\beta) = \int [dU] \exp[S^1_{\text{eff}} (U)]
\ee
where,
\be\label{eq:mostgeneric-acn}
S^1_{\text{eff}} = \sum_{\vect n} \frac{a_{\vect n}(\lambda,
  \beta,N)}{N^k} \prod_{i=1}^k \Tr U^{n_i}  
\ee
with the integers $n_i$ obeying $\sum_i n_i =0$ and the coefficients
$a_{\vect n}$, is a term with $k$ traces, making their first
appearance at $(k - 1)$ loops in perturbation theory and consequently
having a planar contribution starting with $\l^{k-2}$.

One can analyze this partition function in the large $N$ limit and
obtain different phases of the system. The main goal of this paper
is to find phase space distribution corresponding to each saddle 
point or phase. Before we elaborate the importance of phase space
distribution (see section (I.C)), we briefly discuss how one 
obtains a phase space distribution corresponding to different saddle points.

\subsection{Phase diagram from integral equation : Eigenvalue
  analysis}

The Haar measure and the effective action are invariant under unitary
transformation hence one can go to a diagonal basis where
\begin{eqnarray}
U\to e^{i\theta_{n}}\delta_{nm}.
\end{eqnarray}
In this basis the Haar measure and the traces can be written as
\begin{eqnarray}
\int
[DU] = \prod_{i}\int_{-\pi}^{\pi}d\theta_{i}\prod_{i<j} \sin^{2}
\left(\frac{\theta_{i}-\theta_{j}}{2}\right)\quad\quad\quad 
\Tr[U^{n}] = \sum_{j=1}^N e^{i\,n\theta_{j}} .
\end{eqnarray}
In $N \to \infty$ limit the partition function can be written in
terms of continuous variables $\theta(x)$ defined as,
$$
\theta(x) =\theta_i, \quad x= \frac{i}{N},
$$
\begin{eqnarray}
Z=\int [\cD \theta]\, e^{N^{2}\seff{\theta}}
\end{eqnarray}
where, $\seff{\theta}$ corresponds to the effective action along with
the contribution of the determinant factor coming from the Haar
measure of the unitary group. For the zero 't Hooft coupling case we have the full effective action as:
\begin{eqnarray}
\sum_{n=1}^{\infty}\frac{a_{n}(\beta)}{n}\int dx\cos n\theta(x)\int dy\cos n\theta(y)+\frac{1}{2}\int dx\Xint- dy\ln \sin^{2}\bigg(\frac{\theta(x)-\theta(y)}{2}\bigg)
\end{eqnarray}

The above unitary matrix models can be analyzed using standard
techniques in the large $N$ limit\footnote{See \cite{Marino:2004eq}
  for a review.}. In the large $N$ limit, the dominant contribution to the partition
function comes from the extremum condition,
\begin{eqnarray}
\frac{\delta S_{\rm eff}(\theta)}{\delta\theta(x)}=0
\end{eqnarray}
which can be written as,
\begin{eqnarray}\label{eq:eigenvalueeq}
\Xint- dy\, \cot\left(\frac{\theta(x)-\theta(y)}{2} \right)
= V'(\theta(x))
\end{eqnarray}
where $V(\theta)$ is potential obtained by taking the large $N$ limit of
$S^1_{\text{eff}}[U]$ only in the zero 't Hooft coupling case. Introducing the eigenvalue density
\be 
\sigma(\theta)={1\over
  N}\sum_{i=1}^N\delta(\theta-\theta_i) =\frac{\partial x}{\partial \theta}
\ee
 the above saddle point equation can be rewritten in terms of the density $\sigma(\theta)$ as:
\begin{eqnarray}
\Xint- d\theta'\sigma(\theta')\, \cot\left(\frac{\theta-\theta'}{2} \right)
= V'(\theta)\,,
\end{eqnarray}
with the additional implicit normalization condition
\begin{eqnarray}
\int d\theta\,\sigma(\theta)=1 .
\end{eqnarray}

Thus, given the effective action (or $V'(\theta)$) one has to solve
this {\it integral equation} to find the saddle point configurations for
$\s$. The solution to this equation depends on the parameters of the
theory and the condition that $\s(\theta)>0$. As we vary the values of
the parameters of the theory, the system jumps from one phase to another 
phase, which demonstrates the phase transition in the system.

\subsection{Phase diagram from algebraic equation}

We follow a method, which is not so well studied, to find different
phases of the system and their dependence on the parameters without
solving the integral equation (\ref{eq:eigenvalueeq}), which in
general is difficult to handle. We solve \ds \ equation, derived for
$U(N)$ matrix model on $S^3$, at large $N$ and obtain the complete
phase diagram. This method is advantageous because, in the large $N$
limit \ds \ equation becomes algebraic, hence easy to solve. In fact,
solution of \ds \ equation also gives the eigenvalue density 
$\s(\theta)$ for different phases.  One can compute free energy of
different phases in this method as well. This technique was applied to
compute phase diagram of single or multi plaquette model in
\cite{Friedan:1980tu}. This is an useful technique to solve simple
models like $\s$ model, Gross-Neveu model, random matrix model
etc. \cite{spenta-ds}. \cite{spenta-ds} also used this equation to
write down the complete string equation of $U(\infty)$ lattice gauge
theory.

In standard Quantum field theory, the invariance property of the
measure can be used to construct functional equations that correlation
functions of the theory satisfy. These functional equations are called
\ds \ equations. Exploiting the fact that the Haar measure $[dU]$ of a
unitary group is invariant under left multiplication and introducing
a {\it ``Resolvent''} $R(z)=N^{-1}Tr[(1-zU)^{-1}]$, one can write down
an algebraic equation for $R(z)$ in large $N$ limit. This is the \ds \
equation in unitary matrix model. This equation can be simplified
further using the fact that product of single trace operators can be
factorized in large $N$ limit i.e. $\langle g\,f\rangle=\langle
g\rangle\langle f\rangle$, ($f,g$) being single trace operators).

\subsection{Phase diagram from Young diagram and free fermions}

The integral over unitary matrices in equation (\ref{eq:pf-weak-coup})
can as well be done in a different way. Expanding the exponential in
terms of character $\chi$ of conjugacy classes of a symmetric group
and using the orthogonality relation between the characters of $U(N)$ in different
representations, it is possible to write down the partition function in
terms of character $\chi$ and a sum over representations of unitary
group of rank $N$ \cite{Kazakov:1995ae, Marino:2001re}. Since
representations of a unitary group can be described by Young diagrams,
this is, therefore, equivalent to finding the Young tableaux distribution
which dominates the partition function in the large $N$ limit.

It was suggested in \cite{douglas} that the large $N$ limit of unitary
matrix models admits a phase space description. This follows from the
fact that eigenvalues of the holonomy matrix behave like position of free
fermions whereas, the number of boxes in the Young tableaux are like
momentum. This means eigenvalue distribution is like position
distribution and arrangements of boxes in the Young diagram is like
momentum distribution for $N$ non-interacting fermions. An exact phase
space description of different saddle points of a unitary matrix model
was constructed in \cite{Dutta:2007ws}. More precisely,
\cite{Dutta:2007ws} found phase space distribution for three different
phases, which appear in a ``restricted'' class \footnote{The phase space
  description holds for a special class of unitary matrix model where
  the effective action depends on $\Tr U \Tr U^{\dagger}$ and its
  power.} of unitary matrix models. 

These three phases are following.
\begin{figure}[h]
\begin{subfigure}{.3\textwidth}
\centering
\includegraphics[width=2cm,height=2cm]{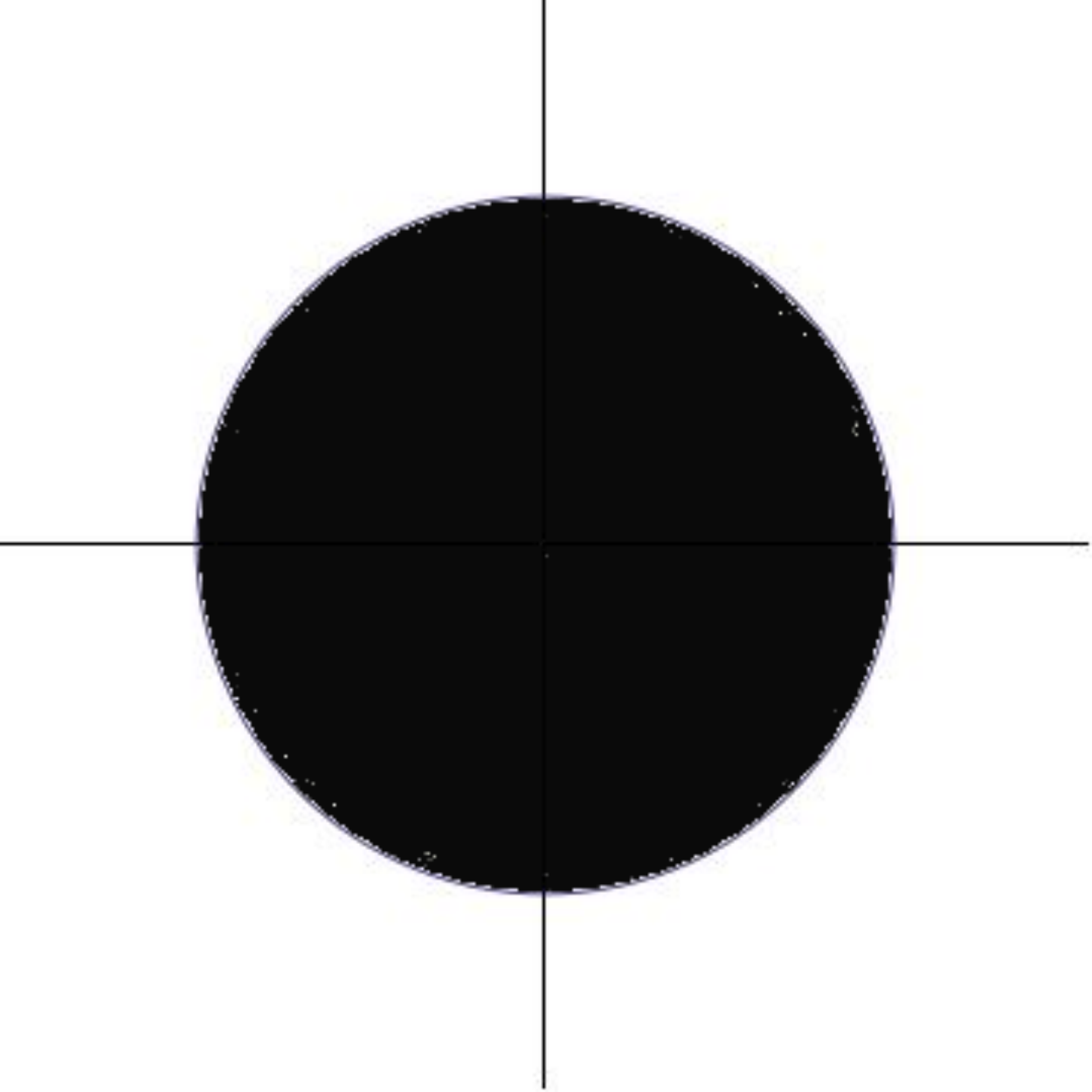}
\caption{Trivial representation/constant eigenvalue distribution
  $\leftrightarrow$ Thermal \ads.}
\end{subfigure}%
\begin{subfigure}{.5\textwidth}
\centering
\includegraphics[width=2cm,height=2cm]{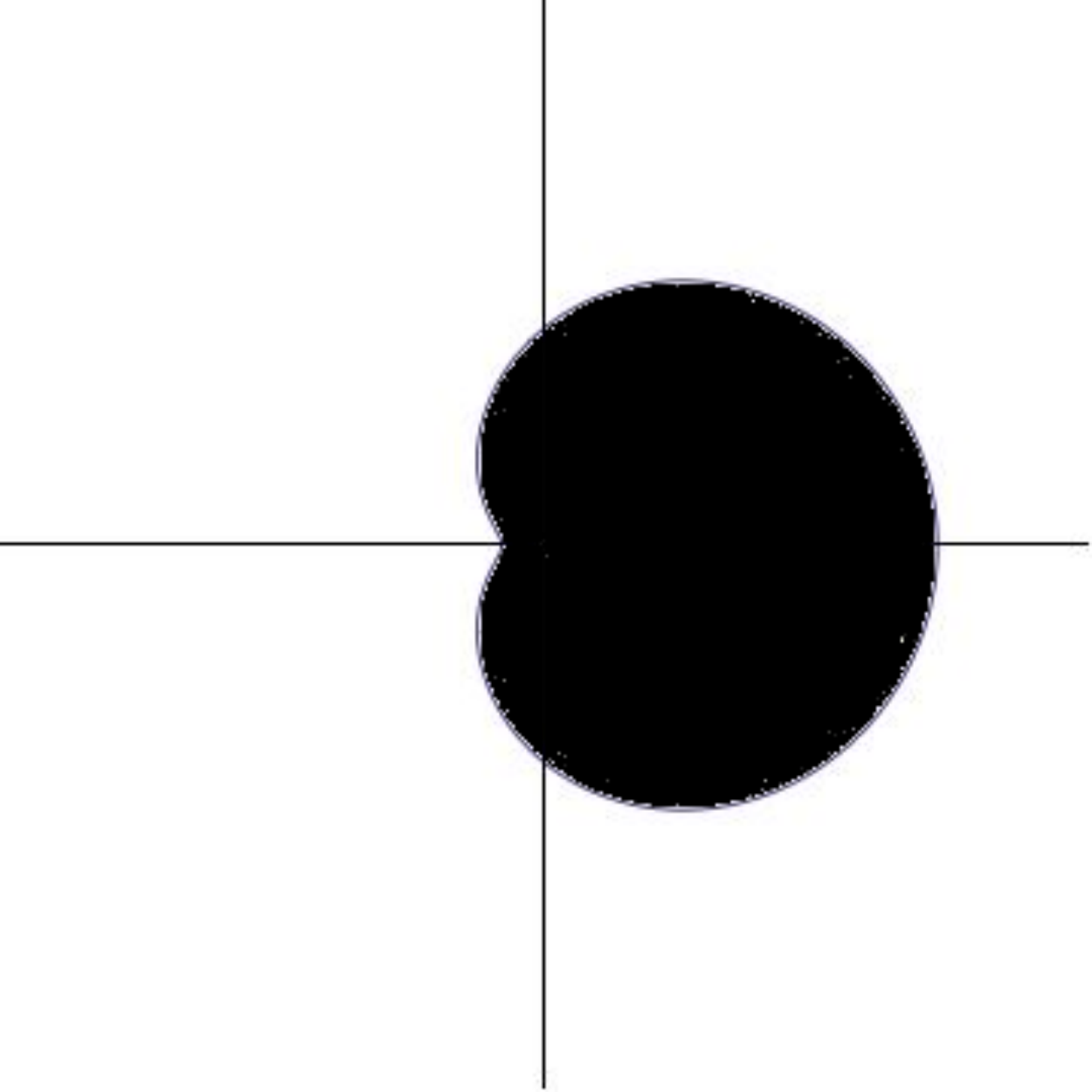}
\caption{Representation with finite number of rows empty - symmetric
  distribution of eigenvalues from $-\pi$ to $\pi$ $\leftrightarrow$
  small black hole.}
\end{subfigure}
\begin{subfigure}{.6\textwidth}
\centering
\includegraphics[width=2cm,height=2cm]{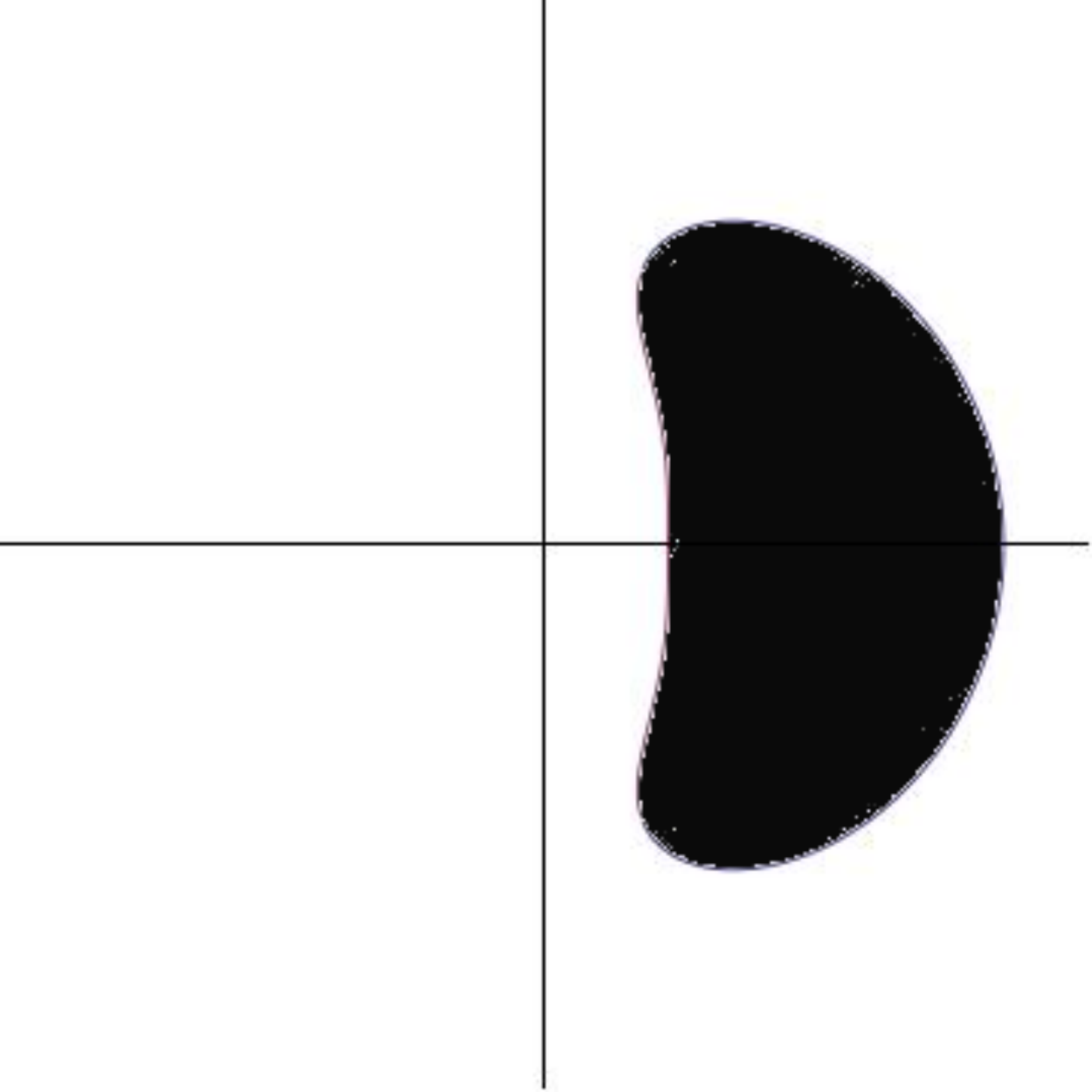}
\caption{Representation with all the rows are filled - eigenvalue
  distribution is symmetric and has a finite gap $\leftrightarrow$ big
  black hole.}
\end{subfigure}
\caption{Phase space distribution for different phases of a class of
  unitary matrix model.}
\label{fig:intro}
\end{figure}
\begin{itemize}
\item The first phase corresponds to a trivial representation : {\it
    no} box in the Young diagram. This saddle point correspond to
  trivial (constant) distribution of eigenvalues.
\item The second saddle point is nontrivial. It corresponds to a
  distribution of boxes in a Young diagram where a finite number of
  rows are empty. This corresponds to a symmetric distribution of
  eigenvalues between $-\pi$ and $\pi$, without any gap.
\item For the third phase all the $N$ rows of the Young diagram are
  filled. This phase has one gap in eigenvalue distribution. This is
  called {\it one-gap phase}\footnote{This phase is also called
    one-cut phase, for a reason which will be discussed in the main
    section.}.
\end{itemize}
Phase space distribution corresponding to these three phases are given
in the figure \ref{fig:intro}.

The shape of the phase space regions is not modified in functional
form when one includes perturbative corrections in terms of an
effective action involving only the relevant winding number one modes
(powers of $\Tr U \Tr U^{\dagger}$) or adds a chemical potential. This
free fermi picture is very intriguing. One can ask why there should be
a free fermion picture for large $N$ saddle points of an interacting
theory. We do not think the answer to this question is well
understood. On the other hand, it has been discussed in \cite{spenta-AG}
that different 
phases of dual gravity theory are identified with different phases on the
weak coupling side in the context of the AdS/CFT correspondence (see figure \ref{fig:intro}). However, it is not clear what is the phase on the gravity side corresponding to two gap solution 
on the weak coupling side. Free fermi phase space distribution could be of
great help to understand a dual description of weakly coupled saddle
points. However, we must emphasize here that there is no direct way 
to construct the phases on the string theory (or gravity) side from the phase space distribution, but the 
nature of transition from one phase to another phase and other characteristics on the weak 
coupling side may help to understand the properties of phases on the 
gravity side. Hence, it is interesting and important
to understand first if the free fermi description of a large N matrix
model is universal, in the sense, if all the large N saddle points
have an underlying free fermi picture.

\cite{Dutta:2007ws} did not consider one important phase which appears
in the class of matrix model they studied. This phase pops up in the
same regime where the third (one-gapped) phase appears. The eigenvalue
distribution corresponding to this phase has two gaps. We call this
phase as {\it two-gap solution} or {\it two-cut solution}. In the
context of unitary model we are considering, this phase has more free
energy than the third one and hence always thermodynamically
disfavoured. However, that does not stop this branch of solution to
exist. In fact single-plaquette model considered in
\cite{Friedan:1980tu, Jurkiewicz:1982iz} has a stable two-gapped or
two-cut phase for a large range of parameters. Large $N$ Chern-Simons
theory with a Gross-Witten-Wadia potential also exhibits a stable
two-gap solution \cite{shiraz-CS}. Therefore, it is interesting and
important to understand the phase space distribution for two-gap
branch. To find the distribution in phase space we need to first
compute the momentum distribution {\it i.e} how boxes are distributed
in a corresponding Young diagram. {\it The goal of this paper is to
  find the Young tableaux distribution and phase space distribution
  corresponding to two-gap solution.}

While analyzing different saddle points in terms of dominant unitary
representations, we observe that one can classify different solutions
depending on number of discontinuities (order $N$ jumps) in $h_i$, where,
$h_i$ is related to $\l_i$, number of boxes in $i^{th}$ row of a Young
diagram by $h_i=\l_i+N-i$ and $h_i\geq h_{i+1}, \ \forall \
i=[1,N]$. More precisely, we classify the distribution into two
different classes,
\begin{enumerate}
\item $\frac{h_i-h_{i+1}}{N} \sim 0, \ \forall \ i=[1,N]$. This
  implies number of boxes in a Young diagram smoothly decreases from 
  $h_1-N+1$ to $h_N$ in the large $N$ limit.
\item $\frac{h_{i'}-h_{i'+1}}{N} \sim \cO(1)$ for some
  $i'\in[1,N]$. This corresponds to an $\cO(N)$ jump in the number of
  boxes at $i=i'$. There can be more than one such jumps.
\end{enumerate}
The first class was considered in \cite{Dutta:2007ws}. This class has
two subclasses depending on whether all the rows of a diagram are
filled up or a finite number of rows are empty. The first subclass
corresponds to one-gap solution whereas, the second subclass maps to
no-gap distribution in eigenvalue side. In this paper we show that the
second class with one $\cO(N)$ jump (or one discontinuity) in $h_i$
corresponds to two-gaps in eigenvalue distribution. Having found the
Young tableaux distribution we immediately identify that there exists
a relation between eigenvalue distribution and Young tableaux
distribution which gives rise to phase space distribution for two-gap
solution, like other two phases.

Organization of this paper is following. In section
\ref{sec:resolvent} we review the basics of the \ds \ equation in
quantum field theory and derive the same for a class of unitary matrix
model. In section \ref{sec:zero-coup} we discuss the phase diagram of
free theory solving the \ds \ equation for $a_{1}$ model, explicitly, where we
have only one coupling $a_{1}$ and the rest $a_{n}$, for $n>1$ are
assumed to be zero. We obtain all the three phases, namely no-gap,
one-gap and two-gap phases and their free energies. We also 
discuss how to generalize our procedure to find different phases for a 
generic model. The phase diagram for weakly coupled theory
is discussed in section
\ref{sec:abmodel}. Section \ref{sec:YTside} explains momentum
distribution or Young tableaux distribution corresponding to different
saddle points of the theory. In section \ref{sec:identification} we
find relation between the Young Tableaux and Eigenvalue distributions
which enables us to draw the phase space distribution for different
saddle points in section \ref{sec:phsespace}.  In Appendix
\ref{sec:ds-eqn} we derive the \ds equation for the most generic
Unitary matrix model relevant for our case. Then in Appendix
\ref{app:ds-eq-most-generic} we discuss the resolvent and the spectral
densities for the most generic cases, while in Appendix \ref{app:app1}
we particularly concentrate on a model with only two cycles namely
$(a_{1}-a_{2})$ model, and describe its phase space in full
detail. Next we move onto the Young Tableaux side and discuss the
important properties of the resolvent function for the Young Tableaux
distribution in Appendix \ref{app:hprop}, and finally discuss a
possible approach to generalize the construction of the resolvent for
the $(a_{1}-a_{2})$ case in Appendix \ref{app:yt-a1a2}.

\section{\ds \ equation and unitary matrix model}\label{sec:resolvent}

\subsection{The Dyson-Schwinger equation}

Path integral quantization of any field theory gives us a prescription
to directly compute the correlation functions bypassing the
Hamiltonian construction, Hilbert space of states and equations of
motion. The underlying symmetries of the theory become manifest in
path integral formalism. For example, lets consider scalar field
theory for simplicity. In the Hamiltonian formalism the Lorentz
symmetry was not manifest, as we define the field momentum varying the
Lagrangian with respect to time derivative of the scalar field and
impose commutation relation between the field momentum and the field. Although
the correlation functions are Lorentz invariant but the construction
was not manifestly Lorentz invariant.

 Let us consider a $n$ point correlation
function $\< X\> = \< \Phi(x_1) \cdots \Phi(x_n)\>$,
\be
\< X \> = \frac1Z\int [{\cD}\Phi] \exp\lb -S[\Phi]\rb \Phi(x_1) \cdots
\Phi(x_n) . 
\ee
Since the translation invariance of measure $[\cD \Phi]$ is the
defining property of the path integral, we have the following identity
\begin{eqnarray}
  \frac1Z\int [{\cD}\Phi] \frac{\delta}{\delta\Phi(x)}\exp\lb
  -S[\Phi]\rb \Phi(x_1) \cdots 
  \Phi(x_n) .=0
\end{eqnarray}
This yields the following identity between the correlation functions
\begin{eqnarray}
  \left \langle\frac{\delta S[\phi]}{\delta\Phi(x)} \right \rangle =
  \sum_{i=1}^{n}
  \delta(x-x_{i})\langle\Phi(x_{1}).. \Phi(x_{i-1})\Phi(x_{i+1}).. \Phi(x_{n})
  \rangle
\end{eqnarray}
This is called the Dyson-Schwinger equation associated with a given
model in quantum field theory.

While the above identity was constructed using the invariance of the measure under translation, there exists more general identities constructed from invariance of the partition function under continuous symmetries. Global symmetries of classical Lagrangian give rise to some currents
which are conserved on-shell. This is Noether's theorem. We ask what
is the consequence of invariance of functional integral under a
symmetry. We shall see that the answer is a quantum generalization of
Noether's equation.

Consider a general quantum field theory of a field $\Phi(x)$ and its
variation 
\be
\Phi'(x) = \Phi(x) - i \omega_a G_a \Phi(x)
\ee
where $\omega$'s are infinitesimal parameters and $G_a$'s are
corresponding generators.
Since $\Phi$ is integration variable on the right hand side we can
first change it to $\Phi'$ and assuming that the measure is invariant
under the above transformation $i.e.$ $[\cD\Phi] = [\cD \Phi']$ we can
write
\ben
\<X\> &=& \frac1Z \int [\cD \Phi'] \exp \lb -S[\Phi']\rb \Phi'(x_1) \cdots
\Phi'(x_n) \nn\\
&=& \frac1Z \int [\cD \Phi] (X + \delta X) \exp\lB -S[\Phi] - \int
d^dx \pa_{\mu} j^{\mu}_a(x) \omega_a(x)\rB .
\een
Here we have used 
\be
S[\Phi'] = S[\Phi] + \int d^dx \pa_{\mu}j^{\mu}_a(x) \omega_a(x)
\ee
where we have considered $\omega_a$'s to be local and $j^{\mu}_a$'s are the
classical Noether's current. Expanding the exponential up to first
order in $\omega$ we find,
\be \label{dX}
\<\d X\> = \int d^dx \ \pa_{\mu} \< j^{\mu}_a(x) X\> \omega_a(x) .
\ee
The left hand side up to order $\o$ is given by,
\ben
\d X &=& X'-X = (\Phi(x_1) - i \omega_a(x_1) G_a \Phi(x_1)) \cdots
(\Phi(x_n) - i \omega_a(x_n) G_a \Phi(x_n)) - \Phi(x_1) \cdots
\Phi(x_n) \nn\\
&=& -i \int d^d x \ \omega_a(x) \sum_{i=1}^n \delta(x-x_i) \lb
\Phi(x_1) \cdots G_a \Phi(x_i) \cdots \Phi(x_n) \rb .
\een
Since eqn. (\ref{dX}) holds for any arbitrary $\o (x)$, we can write
\be
\pa_{\mu}\<j_a^{\mu}(x) \Phi(x_1) \cdots \Phi(x_n)\> = -i \sum_{i=1}^n
\d(x-x_i) \< \Phi(x_1) \cdots G_a \Phi(x_i) \cdots \Phi(x_n) \> .
\ee
This is Dyson-Schwinger equation associated with classical Noether
theorem.

\subsection{Unitary matrix model and Dyson-Schwinger equation
}\label{sec:mm-ds-eqn} 
Here we discuss the Dyson-Schwinger equation for the generalized class of Unitary matrix models that we discussed earlier. 
We start from eqn. (\ref{eq:mostgeneric-acn}). The partition function
is invariant under $U\ra U^{\dagger}$. Therefore expectation value of
any gauge invariant operator which is invariant under conjugation is
of interest. Any gauge invariant and conjugation invariant operator
can be generated by the functions $\Tr U^k$ and $\Tr U^{-k}$. On the
other hand, expectation of product of such operators can be factorized
in the large $N$ limit. Therefore, it is suffice to consider
expectation value of $\Tr U^k$ for any $k$. Now, $\< \Tr U^k\>$ is a
real number because,
\ben
\< \Tr U^{\dagger k}\> =\< \Tr U^{-k}\>  &=& \int [dU] \exp[S_{\text{eff}} (U)] \Tr
U^{\dagger k} \nn \\
&=& \int [dU^{\dagger}] \exp[S_{\text{eff}} (U^{\dagger})] \Tr
U^{\dagger k}, \quad \text{since measure and action are invariant}\nn
\\ 
&=& \int [dU] \exp[S_{\text{eff}} (U)] \Tr
U^{ k} \quad \text{replacing} \ U^{\dagger} \ \text{by} \ U \nn \\
&=& \< \Tr U^{k}\> .
\een
Therefore it is sufficient to consider the expectation values of $\<
\Tr U^{k}\>$ for $k>0$.

It is well known that the Haar measure on the group $U(N)$ is
invariant under action of any element of the group itself.  To write
down the Dyson-Schwinger equation for this model, let us define the
following holomorphic function
\be
R(z) =N^{-1}\langle Tr[(1-zU)^{-1}]\rangle .
\ee
Expanding the right hand side one can write,
\be\label{rzexpan}
R(z) = 1 + \frac1N \lb z \<\Tr U\> + z^2  \<\Tr U^2\> + z^3 \<\Tr
U^3\> +\cdots \rb
\ee
Since $N^{-1}\av{\Tr U^k}$ lies between $-1$ and $+1$ the right hand
side of eqn. (\ref{rzexpan}) is convergent for $|z|<1$. Therefore,
$R(z)$ is an analytic and holomorphic function in the interior of the unit disk.

We use the technique introduced in \cite{Friedan:1980tu} to derive the
Dyson-Schwinger equation in terms or $R(z)$. In large $N$ limit, it
turns out that, the Dyson-Schwinger equation is a purely algebraic.

Here we skip the detailed algebra and write down the final version of
Dyson-Schwinger equation. The explicit calculation has been provided
in appendix \ref{sec:ds-eqn}.

The equation is given by,
\begin{eqnarray}\label{DSfinal}
R(z)^{2}-R(z)+\sum_{\vec n}\tilde{a}_{\vec n}\sum_{i=1}^{k}\prod_{{j=1}\atop{j\neq
  i}}^{k}\frac{1}{|n_{j}|{!}}\partial_{z}^{|n_{j}|}R(z)\bigg|_{0}n_{i}F_{i}=0 
\end{eqnarray}
where,
\begin{eqnarray}\label{DSfinal1}
F_{i} &=&\frac{1}{2\pi i}\oint_{\it{C}}
\frac{R(w)}{w^{n_{i}}}\frac{1}{(w-z)}dw, \ \ \text{for} \ \ n_{i}> 0,\notag\\
&=&\sum_{k=0}^{|n_{j}|-1}z^{k}\frac{1}{(|n_{j}|-k){!}} \partial_{z}^{|n_{j}|-k}R(z)\bigg
  |_{0}+z^{|n_{j}|}R(z),
\ \ \text{for} \ \ n_{i}<0 
\end{eqnarray}
and 
\be
\tilde{a}_{\vec n}  = a_{\vec n}/N^{2}.
\ee
The contour $\it{C}$ being the unit circle. We note that this is an algebraic equation for $R(z)$ in the large $N$
limit. This is, in deed, a very powerful equation. For example, one can study the
phase structure of $\cN =4$ SYM theory from the analytic properties of
the resolvent. We elaborate this in the next two sections.

\section{Phase diagram at zero coupling}\label{sec:zero-coup}

In this section we study the phase structure of the free theory from
the analytic properties of the resolvent. Remember that the resolvent
$R(z)$ is analytic inside the unit circle in the complex $z$ plane. We
first discuss a simple case to understand how a complete phase
structure is captured in $R(z)$.

\subsection{Zero coupling : ``$a_1$ model'' - approximation 1
}\label{sec:amodel}

In the limit $\lambda =0$ the form of the effective action is given
by,
\be
\seff U = \sum_{n_1} \frac{a_{\{n_1,-n_1\}}(\beta)}{N^2} \Tr U^{n_1}\Tr U^{-n_1}
= \sum_{n_1} \tilde a_{\{n_1,-n_1\}}(\beta) \Tr U^{n_1}\Tr U^{-n_1} .
\ee
For further simplification let us assume all $\tilde a_{\{n_1,-n_1\}}
=0$ for $n_1\geq 2$. We shall see that this simple model captures the
important feature of the phase structure at zero coupling.

Denoting $\tilde a_{\{1,-1\}} = a_1$ the Dyson-Schwinger equation
(\ref{DSfinal}) becomes,
\begin{eqnarray}
R(z)^{2}-R(z)+a_1R'(0)\lb\frac{R(z)}{z}-\frac{1}{z}\rb-
a_1R'(0)\lb R'(0)+zR(z)\rb &=&0\notag\\ 
\Rightarrow \, R(z)^{2}
+\lb a_1R'(0)\lb\frac{1}{z}-z\rb-1\rb R(z)-
a_1R'(0) \lb\frac{1}{z}+R'(0)\rb&=&0 .
\end{eqnarray}
Redefining $\beta_1 = a_1R'(0)$ we have,
\be \label{eq:Rzeqn1-zerocoup}
R(z)^{2}
+\lb \beta_1\lb z^{-1}-z\rb-1\rb R(z)-
\beta_1 \lb z^{-1}+R'(0)\rb =0 .
\ee
This equation is equivalent to Dyson-Schwinger equation for one
plaquette model \cite{Friedan:1980tu}. Hence one can carry over the
solution presented in \cite{Friedan:1980tu} for this model.

\subsubsection{Different types of solutions }

Here we discuss the different type of solutions we obtain for $R(z)$ depending on the value of $\beta_{1}$ which satisfy the analytic properties of $R(z)$. To understand different types of solution we 
solve eqn. (\ref{eq:Rzeqn1-zerocoup}) and find
\be\label{Rzform-a1}
R(z)=\frac{1}{2} \left(1+\beta_1
  \left(z-\frac{1}{z}\right)+\sqrt{F(z)}\right) 
\ee
where,
\be F(z) = \left[\beta_1 \left(z+\frac{1}{z}\right)+1\right]^2 +4
\beta_1 \left(R'(0)-\beta_1
 \right) \, .
\ee
From eqn. (\ref{Rzform-a1}) it is clear that analytic properties of
$R(z)$ inside a unit circle depends on the function $F(z)$. If $F(z)$
has zeros of order one then $R(z)$ will have branch cuts. On the other
hand if all the zeros of the function $F(z)$ are of even order then
$R(z)$ does not have any branch cut. Therefore we can classify the
solutions into two types. No-cut solution and cut solutions.

To understand these two different types of solutions we need to check
the zeros of $F(z)$. $F(z)$ is a polynomial of degree four. Hence,
$F(z)=0$ has four solutions
\be \label{Fzsol1}
\begin{split}
  z_1&=-\frac{1-2 \sqrt{\beta _1} \sqrt{\beta _1-R'(0)}+\sqrt{1-4
      \beta _1 R'(0)-4 \sqrt{\beta _1} \sqrt{\beta_1 -R'(0)}}}{2
    \beta_1
  }\\
  z_2&= \frac{-1+2 \sqrt{\beta _1} \sqrt{\beta _1-R'(0)}+\sqrt{1-4
      \beta _1 R'(0)-4 \sqrt{\beta _1} \sqrt{\beta_1 -R'(0)}}}{2
    \beta_1
  }\\
  z_3&= -\frac{1+2 \sqrt{\beta _1} \sqrt{\beta _1-R'(0)}+\sqrt{1-4
      \beta _1 R'(0)+4 \sqrt{\beta _1} \sqrt{\beta_1 -R'(0)}}}{2
    \beta_1
  }\\
  z_4&= \frac{-1-2 \sqrt{\beta _1} \sqrt{\beta _1-R'(0)}+\sqrt{1-4
      \beta _1 R'(0)+4 \sqrt{\beta _1} \sqrt{\beta_1 -R'(0)}}}{2
    \beta_1 }\, .
\end{split}
\ee
Note that $z_1 z_2 =1$ and $z_3 z_4=1$. This means, if one root of
$F(z)$ lies inside the unit circle then the other root lies out side
the circle. Therefore, if all four roots are distinct then the
function $R(z)$ can not be analytic inside a unit circle, as it can
not be Taylor expanded about the root inside, unless all roots live on
the boundary of the unit circle (remember the domain of convergence of
$R(z)$ is $|z|\le 1$). Hence, there are three possibilities.
\begin{itemize}
\item \type {1} - {\bf No-cut
    solution}:\, Roots are pair wise same, $i.e.$ $F(z)$ has two zeros
  of order two. In this case $R(z)$ has no cut inside the unit
  circle. We denote this solution by \tp 1.
\item \type{2} - {\bf One-cut solution}:\, Two roots are same and
  other two are different and live on $|z|=1$ line. In this case
  $R(z)$ has a branch cut with end points of the cut lie on the unit
  circle.  We denote this solution by \tp 2.
\item \type{3} - {\bf Two-cut solution}:\ All four roots are different
  and lies on the unit circle.  We denote this solution by \tp 3.
\end{itemize}

Note that, although we have chosen the branch cuts lying inside the
unit circle but the branch points always lie outside the domain of
analyticity of $R(z)$. In fact, one can also choose the branch cuts
lying outside the circle. However, according to our choice here, we
name the solutions depending on the number of cuts inside the circle.

\subsection*{\type{1} : No-cut solution}\label{sec:no-cut}

From the above solutions (\ref{Fzsol1}) it is clear that when $R'(0)
=\beta_1$, $z_1$ becomes equal to $z_3$ and similarly $z_2$ is same as
$z_4$. Therefore, $F(z)$ will have two zeros of order 2, hence $R(z)$
does not have any branch cut inside the unit circle.

Since $\beta_1 = R'(0) a_1$, therefore for this branch we have
\be
R'(0) (1-a_1)=0.
\ee
For $a_1\neq 1$ the only solution we have $R'(0) =\beta =0$. In this
case $z_1$ and $z_3$ go to infinity and $z_2$ and $z_4$ approach to
$z=0$ and resolvent is given by,
\be\label{eq:res-no-cut1}
R(z) = 1.
\ee
We define spectral density $\sigma(\theta)$ by,
\be\label{eq:specden}
\sigma(\theta) = \frac1{2\pi}\lb 2\Re(R(e^{i \theta})) -1\rb.
\ee
The spectral density, which measures the density of eigenvalue
distribution of the holonomy matrix, defined in eqn.
(\ref{eq:holonomy}), is always positive definite. For $\b_1=0$, the
spectral density is given by,
\be\label{eq:specden-case1a}
\sigma(\theta) = \frac1{2\pi}.
\ee

For $a_1=1$ all the roots are finite and non-zero. The resolvent, in
this case, is given by,
\be\label{eq:res-no-cut2}
R(z) = 1 +\beta_1 z .
\ee
The spectral density for this branch is give by,
\be \label{eq:specden-case1b}
\s(\theta) = \frac1{2\pi}(1+2\b_1\cos \theta).
\ee

For $a_1=1$ case we also have,
\be
z_1+z_2 = z_3+z_4 = -\frac1{\beta_1}.
\ee
Since, $z_1z_2=1$ we can write $z_1 = r_1 e^{i\theta_1}$ and $z_2=
\frac1{r_1}e^{-i\theta_1}$ (similar argument can be given for $z_3$
and $z_4$ also). Thus we get,
\be
\lb r_1+\frac1{r_1}\rb \cos\theta_1+i \lb r_1-\frac1{r_1}\rb \sin
\theta_1 = -\frac1{\beta_1}.
\ee
The right hand side has no imaginary part. Hence, there are two
possibilities: either $\theta_1 = \pi$\footnote{Note that $\theta_1$
  can not be $0$ as the right hand side is negative and $r_1$ is
  positive.} or $r_1=1$. In the first case all four roots lie on
negative real axis and
\be
r_1+\frac1{r_1} =\frac1{\beta_1}.
\ee
This equation has real positive solution only when $0\leq\beta_1\leq
\frac12$. 

In the second case, when $r_1=1$, all roots lie on the boundary of the
unit circle and
\be
2 \cos\theta_1 = -\frac1{\beta_1}.
\ee
This equation is solvable for $\theta_1$ when $\beta_1>\frac12$.
However, for $\b_1>1/2$ the spectral density (\ref{eq:specden-case1b})
becomes negative for some values of $-\pi\leq \theta\leq \pi$. Hence,
$\b_1>1/2$ solution is not physical.

\begin{figure}[h]
\begin{subfigure}{.5\textwidth}
  \centering
  \includegraphics[width=6.5cm,height=4.35cm]{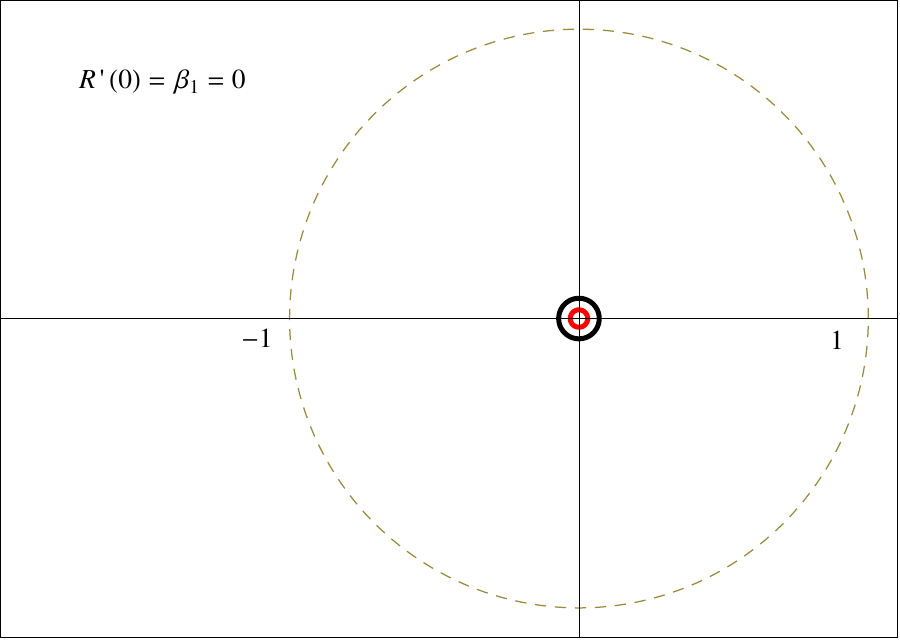}
  \caption{$\b_1=0$. Two roots are at z=0, other two are at $\infty$}
\end{subfigure}%
\begin{subfigure}{.5\textwidth}
  \centering
  \includegraphics[width=6.5cm,height=4.35cm]{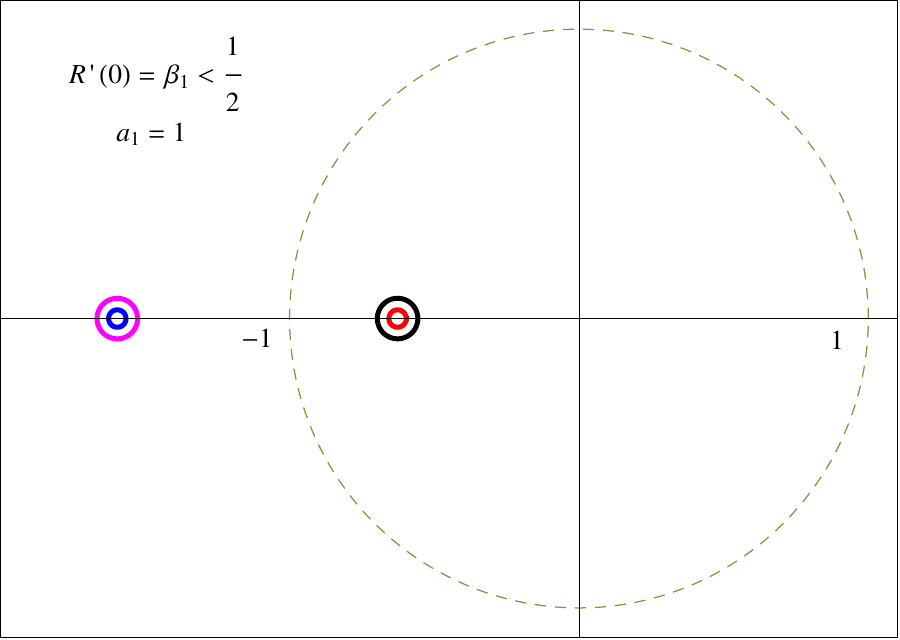}
  \caption{$0<\b_1<1/2$. All the roots on negative real axis.}
\end{subfigure}
\begin{subfigure}{.5\textwidth}
  \centering
  \includegraphics[width=6.5cm,height=4.35cm]{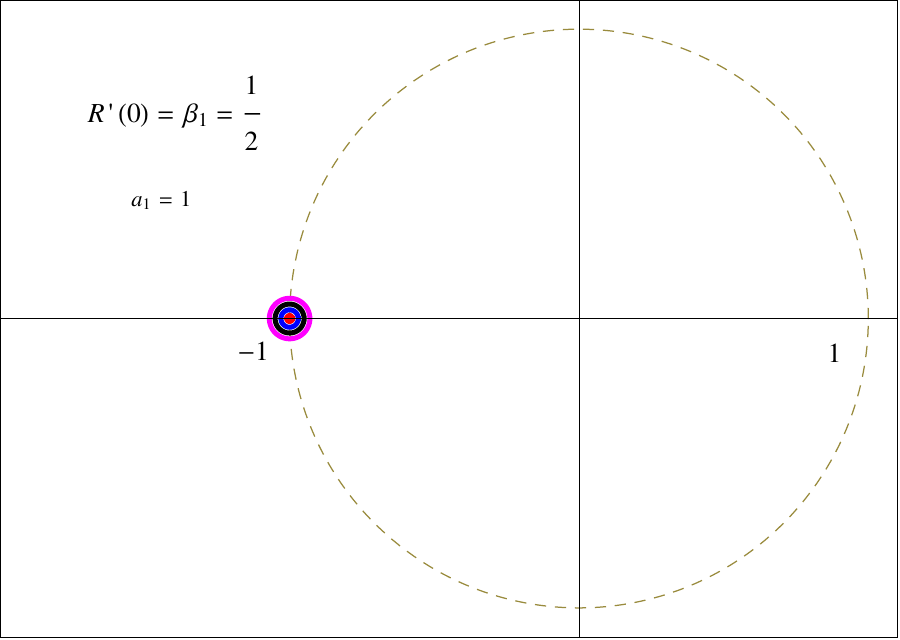}
  \caption{$\b_1=1/2$. All the roots are at $z=-1$.}
\end{subfigure}%
\begin{subfigure}{.5\textwidth}
  \centering
  \includegraphics[width=6.5cm,height=4.35cm]{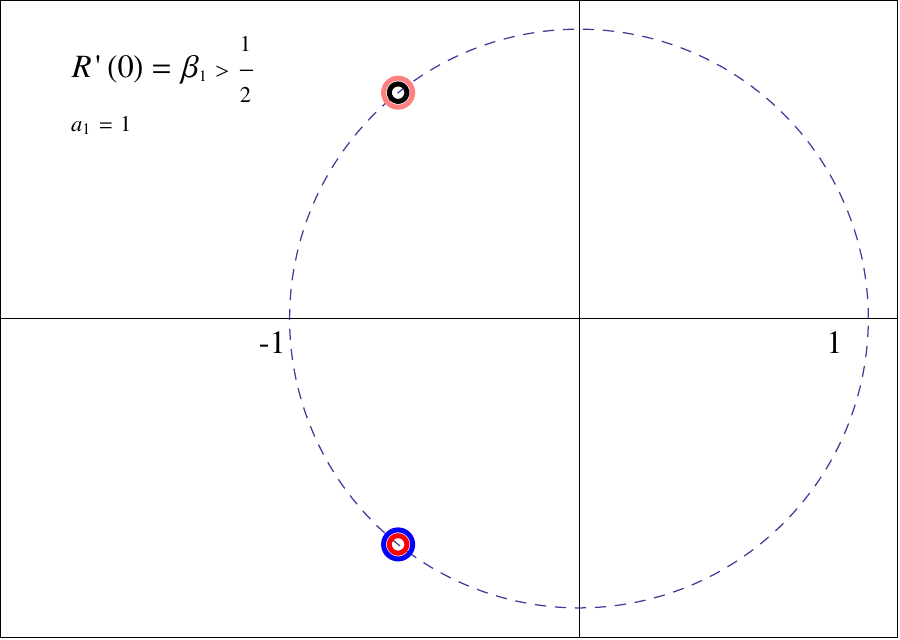}
  \caption{$\b_1>1/2$. All the roots on the circle. For this branch
    spectral density becomes negative.}
\end{subfigure}%
\caption{Position of the roots of $F(z)$ in the complex $z$ plane for
  no-cut solution \tp 1}
\label{fig:roots-Fz-nocut}
\end{figure}

Thus, we see that no-cut solution (\tp 1) exists for $0 \leq
\beta_1\leq 1/2$ and for this solution the roots are lying on negative
real axis in complex $z$ plane. See figure \ref{fig:roots-Fz-nocut}.

\subsection*{\type{2} : One-cut solution} \label{sec:one-cut}

\begin{figure}[h]
\centering
\includegraphics[width=7.5cm,height=5cm]{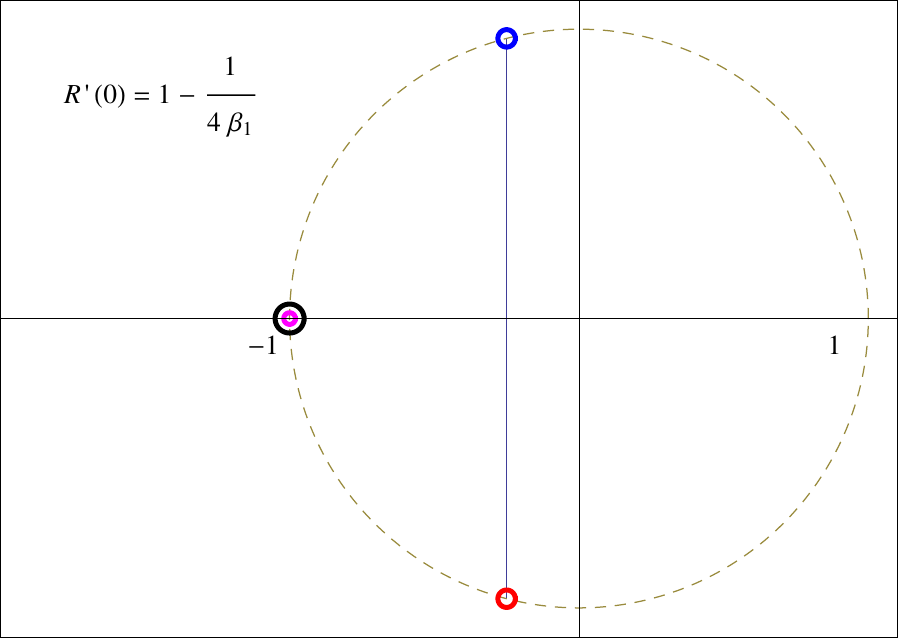}
\caption{Position of the roots of $F(z)$ in the complex $z$ plane for
  one-cut solution.}
\label{fig:roots-Fz-onecut}
\end{figure}

%
From the position of zeros of $F(z)$ given by eq. (\ref{Fzsol1}) we
see that \tp 2 occurs when 
\be\label{eq:one-cut-cond} 
R'(0) = 1 - \frac1{4\beta_1} .  
\ee
In this case, 
\be
z_1= -\frac1{\beta_1} \lB (1-\beta_1) + \sqrt{1-2\beta_1}\rB,
\quad \ z_2= -\frac1{\beta_1} \lB (1-\beta_1) -
\sqrt{1-2\beta_1}\rB, 
\ee
and $z_3 = z_4 =-1$. The roots are plotted in figure
\ref{fig:roots-Fz-onecut}. 
Since $z_1 z_2=1$, writing $z_1 = r_1
e^{i\theta_1}$ and $z_2= \frac1{r_1}e^{-i\theta_1}$ we have
\be
\lb r_1+\frac1{r_1}\rb \cos\theta_1+i \lb r_1-\frac1{r_1}\rb \sin
\theta_1 = -\frac2{\beta_1} +2 .
\ee
These two roots $z_1$ and $z_2$ lie on the boundary of unit circle,
hence $r_1=1$ and,
\be\label{onecutcondition-a1}
\cos \theta_1 = 1-\frac1{\beta_1}.
\ee
Therefore this branch of solution exists for $\beta_1\geq
\frac12$. The resolvent, for this branch, is given by
\be \label{eq:res-one-cut}
R(z) = \frac12\lB 1+\beta_1 (z-z^{-1}) + \beta_1(1+z^{-1})
\sqrt{z^2 + 2(\beta_1^{-1} -1)z +1}\rB .
\ee
The condition for one-cut solution can also be written in terms of
$a_1$ from eqn. (\ref{eq:one-cut-cond}),
\be\label{eq:one-cut-cond-2}
a_1 = \frac{4\beta_1^2}{4\beta_1 -1}.
\ee
Since $\beta_1\geq \frac12$ this branch exists for $a_1\geq 1$.

The spectral density corresponding to this branch is given by,
\be\label{eq:specden-case2}
\sigma(\theta) = \frac{2\beta_1}{\pi} \sqrt{\sin^2\frac{\theta_0}{2} -
  \sin^2\frac{\theta}{2}} \cos\frac{\theta}2, \qquad \sin^2
\frac{\theta_0}{2} = \frac1{2\beta_1}.
\ee
There is a gap in eigenvalue distribution {\it i.e.} eigenvalue
distribution is zero between $|\theta|>\theta_0$. That is why, \tp 2
phase is also called ``gapped'' phase.

\subsection*{\type{3} : Two-cut solution} \label{sec:two-cut}

\begin{figure}[h]
\centering
\includegraphics[width=7.5cm,height=5cm]{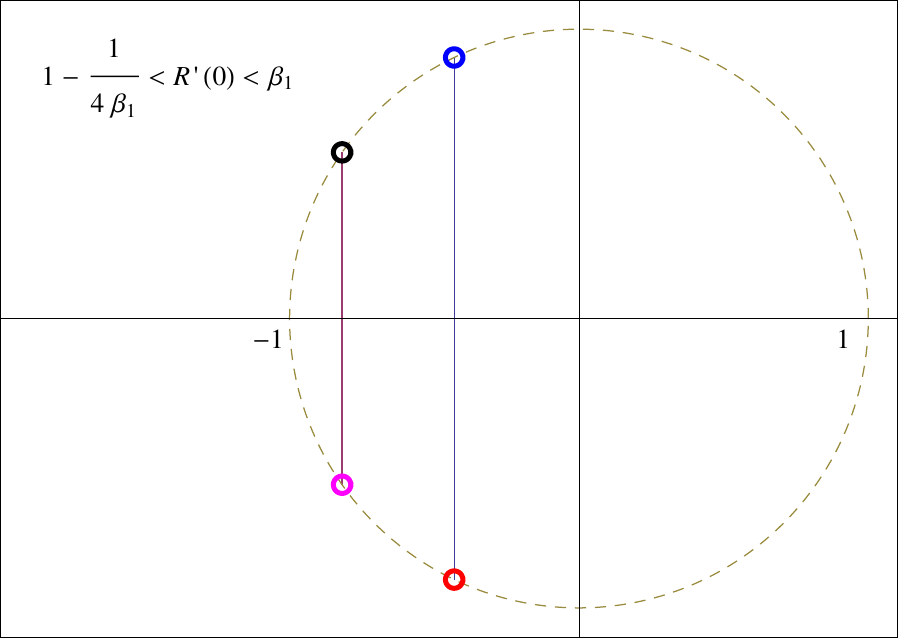}
\caption{Position of the roots of $F(z)$ in the complex $z$ plane for
  two-cut solution.}
\label{fig:roots-Fz-twocut}
\end{figure}

%
Two-cut solution exists when all the roots of $F(z)=0$ are
distinct. This happens when
\be\label{eq:2cut-cond1}
 R'(0)\geq 1-\frac1{4\beta_1}
\ee
and all four roots lie on the boundary of unit circle (figure
\ref{fig:roots-Fz-twocut}).
The resolvent is given by,
\be \label{eq:res-two-cut}
R(z) =\frac12 \lB 1 + \beta_1 \lb z-\frac1z\rb +\beta_1
\sqrt{(z-2\cos\theta_1 + \frac1z) (z-2\cos\theta_2+\frac1z)}\rB
\ee
where,
\be\label{eq:theta1-theta2}
\cos\theta_1= \frac12\lB -\frac1{\beta_1} + 2 \sqrt{1-\frac1{a_1}}\rB,
\quad \cos\theta_2= \frac12\lB -\frac1{\beta_1} - 2 \sqrt{1-\frac1{a_1}}\rB.
\ee
From this resolvent we find,
\be
R'(0) = \beta_1 \lB 1-\frac14 (\cos\theta_1 -\cos\theta_2)^2\rB,
\ee
which implies that two cut solution exists when 
\be\label{eq:2cut-cond2}
R'(0) < \beta_1 .
\ee
Combining eqn. (\ref {eq:2cut-cond1}) and (\ref{eq:2cut-cond2}) we
find that the two-cut solution exists for,
\be \label{eq:2cut-lim1}
1< a_1 < \frac{4\beta_1^2}{4\beta_1-1}
\ee
or equivalently,
\be \label{eq:2cut-lim2}
1< a_1 < \frac{1}{4R'(0)(1-R'(0))}.
\ee

The spectral density for this branch is given by,
\be\label{eq:specden-case3}
\sigma(\theta) = \frac{2\b_1}{\pi}\sqrt{\lb \sin^2\frac{\theta_1}{2} -
  \sin^2\frac{\theta}{2} \rb \lb \sin^2\frac{\theta_2}{2} -
  \sin^2\frac{\theta}{2} \rb }.
\ee
Thus we see that there are two distinct gaps in eigenvalue
distribution for two-cut solution.

Here we have obtained only the limiting conditions on the parameters
(eqn. \ref{eq:2cut-lim1} or \ref{eq:2cut-lim2}) for two cut
solution. However exact dependence can be found from the condition
that the number of eigenvalues lying between $\theta_1$ and $\theta_2$
is zero, {\it i.e.}
\be
\int_{\theta_1}^{\theta_2} d\theta \s(\theta) =0.
\ee
This equation gives a relation between $\theta_1$ and $\theta_2$ and
hence, between $a_1$ and $\b_1$ using eqn. (\ref{eq:theta1-theta2}).
However, the exact relation is not required for our purpose. Hence we
shall not find this relation in this paper.

\subsubsection{Free energy for different phases}

Free energy, in the large $N$ limit, is given by
\be
F(T)= - T \ln Z(T)
\ee
where, 
\be
Z(T) = \int [dU] \exp\lB a_1(T) \Tr U \Tr U^{\dagger}\rB.
\ee
To calculate the free energy we use the following trick. Temperature
dependence of partition function comes from the temperature dependence
of $a_1$, as the holonomy matrix is independent of
temperature. Differentiating the partition function with respect to
temperature we obtain,
\be
\frac{d Z(T)}{d T} = \int [dU]  \exp\lB a_1(T) \Tr U \Tr
U^{\dagger}\rB \Tr U \Tr 
U^{\dagger} \frac{d a_1(T)}{d T} .
\ee
In the large $N$ limit it becomes,
\be
\begin{split}
 \frac{d Z(T)}{d T} &= N^2 Z(T) (R'(0))^2 \frac{d a_1(T)}{d T} \\
\Rightarrow \frac{d}{d T} \ln Z(T) &=N^2 (R'(0))^2 \frac{d a_1(T)}{d
T} .
\end{split}
\ee

For \tp 1, we have either $R'(0)=0$ or $a_1$ constant. Hence Free
energy is zero. Whereas, in \tp 2, for single-cut solution we
have\footnote{Note that $R'(0)$ is also function of temperature. Here
  we have not written the argument for brevity.}
\be
a_1(T) = \frac1{4R'(0)(1-R'(0))}
\ee
hence, 
\be
\frac{da_1}{d T} = - \frac1{4 (R'(0))^2} \frac{1-2 R'(0)}{(1-R'(0))^2}
\frac{dR'(0)}{d T} .
\ee
Therefore, we find,
\be \ln Z(T) = N^2 \lB \frac14 \frac1{1-R'(0)} + \frac12
\ln(R'(0)-1)\rB + \cC, \quad \text{where} \ \cC \ \text{is integration
  constant.}  \ee
At $\b_1 = 1/2$ ($i.e.$ $R'(0)=1/2$) the free energy should vanish
hence we get,
\be
\cC = -\frac12+\frac12 \ln(-2).
\ee
Substituting the value of $\cC$, the free energy for one-cut solution
becomes,
\be 
F_{\mathbb{T 2}}(T) =
-N^2 T \lB \frac1{4(1-R'(0))} -\frac12 + \frac12 \ln2(1-R'(0))\rB=
-N^2 T \lB \beta_1 -\frac12 - \frac12 \ln(2\beta_1)\rB.  
\ee
This expression matches with \cite{Dutta:2007ws}.

For two cut solution we have $ 1< a_1 <
\frac{1}{4R'(0)(1-R'(0))}$. Since it is difficult to find the exact
relation between $R'(0)$ and $a_1$, hence it is difficult to calculate
free energy for two-cut solution. However, we consider a naive
parametric relation between $R'(0)$ and $a_1(T)$ and find free energy
for this parametrization.  We take,
\be
a_1(T) = 1+x \lB \frac{(2R'(0)^2-1)^2}{4R'(0)(1-R'(0))}\rB , \qquad
0<x<1 .
\ee
For $x=0$ we get back the no-cut branch corresponding to $a_1=1$ and
for $x=1$ we get back the one-cut branch. For this parametrization
free energy is given by,
\be
F_{\mathbb{T 3}} (T) =
-N^2 T  x \lB \frac1{4(1-R'(0))} -\frac12 + \frac12 \ln2(1-R'(0))\rB .
\ee
Since $0<x<1$, we find free energy of two cut solution is greater than
that of one cut solution and hence thermodynamically disfavoured. 

\subsubsection{Phase diagram for $a_1$ model}

\begin{figure}[h]
\centering
\includegraphics[width=7cm,height=6cm]{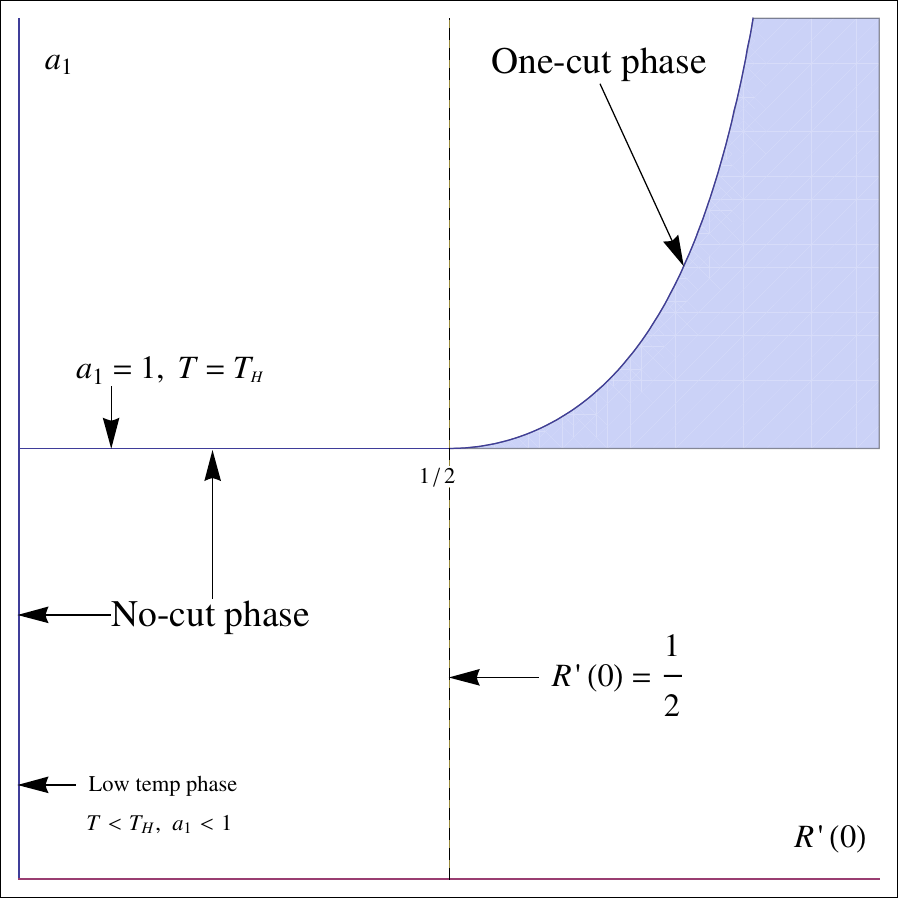}
\caption{Phase diagram for $a_1$ model in $(R'(0),a_1)$ plane.}
\label{fig:phase-dig-a1model}
\end{figure}

Let us summarize the complete phase diagram as a function of
temperature for $a_1$-model. $a_1(T)$ is a monotonically increasing
function of temperature with $a_1(0)=0$. Therefore, the system jumps
from one phase to other as we vary temperature. In figure
\ref{fig:phase-dig-a1model} we draw the complete phase diagram for
$a_1$-model.
\begin{itemize}
\item At low temperature, when, $a_1<1$, the minimum action
  configuration is given by a constant resolvent and hence a constant
  eigenvalue distribution.  The free energy is zero for this
  configuration (to order $N^2$). This phase is represented by a line
  $(\Rp=0, a_1<1)$ in the phase diagram.
\item As we increase temperature to $T=T_H$, $a_1$ becomes equal to
  $1$ ($a_1(T_H)=1$).  There is a continuous family of minimum action
  configurations (labeled by a parameter $\Rp=\b_1<1/2$) for which the
  resolvent is linear in $z$, given by eqn. (\ref{eq:res-no-cut2}).
  Eigenvalue density for this phase (eqn. (\ref{eq:specden-case1b}))
  is symmetrically distributed between $-\pi$ and $\pi$.  All these
  configurations also have zero free energy.
\item For $T>T_H$ {\it i.e.} $a_1>1$, there is a new saddle point,
  with resolvent having a cut inside the unit circle is given by
  eqn. (\ref{eq:res-one-cut}) and eigenvalue distribution function has
  a gap. The free energy for this configuration is negative. This
  branch is given by the upper boundary line of the shaded region in
  the phase diagram.

\item For $T>T_H$ there exists another branch whose resolvent
  (eqn. (\ref{eq:res-two-cut})) has two cuts inside the unit circle in
  $z$ plane and eigenvalue distribution has two gaps. But this phase
  is thermodynamically disfavoured as it has more free energy than one
  gap phase. This phase lies somewhere in the shaded region in the
  phase diagram depending on the temperature.
\end{itemize}

Thus we see that there is a first order 
phase transition at $a_1=1$, which corresponds to a 
temperature $T=T_H$.

However, the model we have considered so far is the simplest one. But
one can generalize the solution discussed in this section for a more
generic class of model. In the next subsection we shall consider the
next term in the effective action of free theory and discuss the
solution of Dyson-Schwinger equation.

\subsection{Zero coupling : ($a_1, a_2$) model - approximation
  2}\label{sec:aprox2}

In this section we keep the term proportional to $\Tr U^2 \Tr U^{-2}$
in the effective action and analyze the structure of the resolvent $R(z)$ and the function $F(z)$, from which the phase structure can be obtained. This term is suppressed by an $N^2$ factor
with respect to the first term. We consider this term to understand
how to find the solution to Dyson-Schwinger equation for a general
class of action. Our discussion in this section is general and valid
for a general class of action.

In presence of this term the Dyson-Schwinger equation becomes,
\begin{eqnarray}
&&R^{2}(z)+\lB\beta_{1}\lb \frac{1}{z}-z\rb+\beta_{2} \lb
\frac{1}{z^{2}}-z^{2}\rb -1\rB
R(z)
-\beta_{1}\lb \frac{1}{z}+R'(0)\rb -\beta_{2}\lb
\frac{R'(0)}{z}+\frac{1}{z^{2}}+\frac{R''(0)}{2}+R'(0)z\rb =0,
\notag\\
&& \text{where}, \ \ \  \beta_{2}=\tilde{a}_{+2,-2}R''(0).
\end{eqnarray}
This is a quadratic equation of $R(z)$ and it has two solutions. Let
us write the solutions in the following form
\begin{eqnarray}\label{Rzform}
R(z)=\frac{1}{2}\lB
1+\beta_{1}\lb z-\frac{1}{z}\rb+\beta_{2}\lb z^{2}-\frac{1}{z^{2}}\rb
\pm\sqrt{F(z)}\rB.   
\end{eqnarray}
The function $F(z)$ is given by
\be
\begin{split}\label{eq:Fz-a1a2-twocut}
F(z)&=\bigg[\beta_{1}\lb\frac{1}{z}-1\rb
+\beta_{2}\lb\frac{1}{z^{2}}-z^{2}\rb-1\bigg]^{2}  
+4\bigg[\beta_{1}\lb\frac{1}{z}+R'(0)\rb+\beta_{2}\lb\frac{R'(0)}{z}+\frac{1}{z^{2}}
+\frac{R''(0)}{2}+R'(0)z\rb\bigg]\\
&=\bigg[\beta_{1}\lb\frac{1}{z}+z\rb+\beta_{2}\lb\frac{1}{z^{2}}+z^{2}\rb
+1\bigg]^{2} -4\lb\beta_{1}z+\beta_{2}z^{2}\rb \lB\frac{\beta_{1}}{z}
+\frac{\beta_{2}}{z^{2}}\rB +4\beta_{1}R'(0)+4\beta_{2}\lB\frac{R'(0)}{z}
+\frac{R''(0)}{2}+R'(0)z\rB  .
\end{split} 
\ee
Note that $F(z)$ is invariant under $z\to\frac{1}{z}$.

\subsubsection{Different types of solutions - a generic discussion}

Although, we refer to the expressions for $R(z)$ and $F(z)$ obtained
for the particular model, but our argument holds in general.

From invariance of function $F(z)$ under $z\ra 1/z$ symmetry we
immediately realize that for any multiplicity one root $\lambda$,
there exists another multiplicity one root $1/\lambda$, since $z$ and
$1/z$ satisfy the same equation. This means that roots of multiplicity
one always occur in distinct pairs. Each distinct pair has product
equal to 1. From analyticity condition of $R(z)$ we know that none of
the multiplicity one root can lie inside the unit circle, while from
the product of pair of roots we know that if one root lies inside, the
other one must lie outside the unit circle. This means the
multiplicity one roots must lie on the unit circle itself for any
$F(z)$ (which is always of even order). Therefore, the generalized form
for $F(z)$ can be written as follows.
\begin{eqnarray}
F(z)=C\prod_{i=i}^{n}\left(z+a_{i}+\frac{1}{z}\right)
\end{eqnarray}
$C$ being some constant. A pair of multiplicity one root means there
exists a $a_{i}\neq a_{j}$, for $i\neq j$ and multiplicity 2 means
there exists some $a_{i}=a_{j}$, for $i\neq j$ and $a_{i}\neq a_{k}$
for $k\neq j\neq i$. 

For one-cut solution to exist ({\it i.e.} $R(z)$ has one branch cut
inside the unit circle) there must be one $a_{i}$ which occurs only
once. From the analyticity condition of $R(z)$, we find
$a_{i}=-2\cos\theta$ and hence the roots are equal to
$e^{i\theta},e^{-i\theta}$. Now, let there exist at least one of these
$a_{i}=\tilde{a}_{1}$ (more can occur). Then $F(z)$ can be factorized as
\begin{eqnarray}
F(z)=\lb z+\tilde{a}_1+\frac{1}{z}\rb f(z),
\end{eqnarray}
The function $f(z)$ will also have the symmetry $z\to{1/z}$. A generic
form can be written as,
\be
f(z) = \cdots+ a_p z^p +a_{p-1}z^{p-1} +\cdots a_0 + \cdots +
\frac{a_{p-1}}{z^{p-1}} + \frac{a_p}{z^p} +\cdots. 
\ee
However, $f(z)$ may have more pairs of multiplicity 1 factors of form
$z+a_{i}+\frac{1}{z}$. Now using the condition $R(0)=1$ we find a
relation for $\tilde{a}_{1}$ or in other word $\cos\theta$ in terms of
$\beta_i$ s in the action, (the other coefficients $a_{0},..,a_{p}$ also being given in terms of $\beta_{i}$ s and $R^{n}(0)$). For existence of at least one pair of
multiplicity 1 root, there must exist a $\cos\theta$. If for the
choice of parameters we see that there exists no $\cos\theta$ i.e. its
bound is broken, then there can be no pair of roots of multiplicity
one. This means the only possible solution $F(z)$ is that all the
roots are of multiplicity 2 (or of higher even order). Thus if one cut
solution is not possible then only possible solution is a no-cut.

For example, in the previous subsection we have seen that to have one
cut solution we must need $\beta>1/2$. Therefore, for $\beta<1/2$ we
can have only no-cut solution.

However, one can have no-cut solution even if the condition for
one-cut solution is satisfied. But in that case, the roots of
multiplicity 2 (minimum) will live on the unit circle and hence $R(z)$
will violate the condition of positivity of spectral density.

We present the discussion on phase diagram for this model in appendix
\ref{app:app1}.

\section{Phase diagram at weak coupling : A phenomenological
  model} \label{sec:abmodel}

The \ds \ equation can be written for the most generic class of weakly
coupled theory given by eqn. (\ref{eq:mostgeneric-acn}). However,
finding the solution to that equation and hence phase diagram is
difficult. One can, instead, construct an analytically solvable
phenomenological effective action which captures all the essential
features of the original theory \cite{spenta-AG}.

In the last section we have seen that in large the $N$ limit the order
parameter which distinguishes different phases is $\Rp= \left< \Tr
  U\right>$. Consequently, one can also imagine integrating out all
the ${\Tr}U^n$ (with $n \neq \pm 1$) and obtaining an effective action
purely in terms of ${\Tr}U{\Tr}U^{\dagger}$.  However, this is not
easy to carry out explicitly. Therefore we consider a phenomenological
models of the form \cite{spenta-AG}
\be\label{genZ}
Z = \int [dU] e^{N^2 \seff x} \ , \quad x={1 \over N^2} \Tr U
\Tr U^{\dagger} \ ,
\ee
where 
\be\label{eq:sefftrunc}
\seff U= a_1(\lambda, T) \Tr U \Tr
U^{\dagger} + {b_1(\lambda, T) \over N^2} (\Tr U  \Tr
U^{\dagger})^2 
+ {c_1(\lambda, T) \over N^4} \ (\Tr U \Tr
U^{\dagger})^3 + \ \cdots 
\ee
where $S(x)$ is convex and $S'(x)$ is concave.  This
phenomenological model captures all the essential features of
thermodynamic  properties and phase transition of the unitary matrix
model we are considering. The simplest of such model is the so-called
$(a,b)$ model \cite{spenta-AG} in which one keeps only the first two
coefficients in $\seff x$ given in eqn. (\ref{eq:sefftrunc}).
\be\label{Zab}
Z(a_1,b_1) = \int [dU] \exp \lB a_1\Tr  U \Tr U^{\dagger} + {b_1
\over N^2} \lb \Tr U \Tr U^{\dagger} \rb^2 \rB ,
\ee
where $a_1$ and $b_1$ are functions of temperature $T$ and $\lambda$
and $b_1>0$. 

In this section, we shall discuss the properties of phase diagram for
$(a,b)$ model. However, our discussion can be extended to more general
classes (\ref{eq:sefftrunc}). We present the generic discussion in
appendix \ref{app:ds-eq-most-generic}.

The Dyson-Schwinger equation (\ref{DSfinal}) for this 
model is given by,
\begin{eqnarray}
  R(z)^{2}-R(z)+a_1R'(0)\lB \lb\frac{R(z)}{z}-\frac{1}{z}\rb-
  \lb R'(0)+zR(z)\rb\rB + 2 b_1 R'(0)^3 \lB \lb\frac{R(z)}{z}-\frac{1}{z}\rb-
  \lb R'(0)+zR(z)\rb \rB=0
\end{eqnarray}
where $\tilde a_{1,-1}=a_1$ as before and $\tilde a_{1,1,-1,-1} =
b_1$. Redefining $\beta_{ab} = a_1R'(0) + 2 b_1 R'(0)^3$ we have,
\be \label{eq:Rzeqn1}
R(z)^{2}
+\lb \bab \lb z^{-1}-z\rb-1\rb R(z)-
\bab \lb z^{-1}+R'(0)\rb =0
\ee

This equation is exactly same as zero coupling case (approximation 1)
except $\beta_1$ replaced by $\bab$. Therefore, solution of this
equation is given by,
\be\label{Rzform1}
R(z)=\frac{1}{2} \left(1+\bab
  \left(z-\frac{1}{z}\right)+\sqrt{F(z)}\right) 
\ee
where,
\be F(z) = \left[\bab \left(z+\frac{1}{z}\right)+1\right]^2 +4
\bab \left(R'(0)-\bab
 \right) \ee

\subsection{ Different phases}

\subsubsection*{\type 1: No-cut solution}

The phase structure can be constructed as before. For no-cut solution
there are two possibilities : $\bab=0$ and $R'(0) = \bab$. The
resolvent, for these two cases, are given by $R(z)=1$ and $R(z) = 1+
\bab z$ respectively.

$\bab=0$ phase exists at any temperature, as before. This phase has
{\it zero} free energy and spectral density is constant for this
phase.

The other branch of no-cut solution exists when $R'(0) = \bab$. This
implies that,
\be
\bab^2 = \frac{1-a_1}{2b_1}.
\ee
For this branch, two zeros of $F(z)$ are lying at the same point
inside the unit circle and on the negative real axis. The other two
zeros lie outside the real axis (same as $a_1$ model). This branch
exists when $\bab \leq \frac12$, {\it i.e.},
\be
\bab^2 = \frac{1-a_1}{2b_1} \leq \frac14.
\ee

The spectral density for \tp 1 is given by,
\be
\s(\theta) = \frac1{2\pi} (1+2\bab \cos\theta).
\ee

\subsection*{Free energy}

One can calculate free energy of this branch following the
prescription given in the previous section. 
\ben 
\frac{d Z(T)}{d T} &=& \int [dU] \exp\lB a_1 \Tr U \Tr
U^{\dagger} + \frac{b_1}{N^2} (\Tr U \Tr U^{\dagger})^2 \rB \lb \Tr U
\Tr U^{\dagger} \frac{d a_1}{d T} + \frac{1}{N^2} (\Tr U \Tr
U^{\dagger})^2 \frac{d b_1}{d T}
\rb .  
\een
In the large $N$ limit we find,
\ben
  \frac{d}{d T} \ln Z(T) &=&  N^2  R'(0)^2 \lb \frac{d a_1}{d T} +
    R'(0)^2 \frac{d b_1}{d T}\rb.\nonumber\\
&=& N^2 \lB \frac{1-a_1}{2b_1} \frac{d a_1}{d T} + \lb
\frac{1-a_1}{2b_1} \rb^2 \frac{d b_1}{d T}\rB \nonumber\\
&=& N^2 \frac d{d T} f(a_1, b_1) 
\een
where,
\be
\frac{\pa f(a_1, b_1)}{\pa a_1} =  \frac{1-a_1}{2b_1}, \quad \text{and} \
\frac{\pa f(a_1, b_1)}{\pa b_1} = \lb
\frac{1-a_1}{2b_1} \rb^2.
\ee
Solving these two equations we find,
\be
f(a_1, b_1) = -\frac{(1-a_1)^2}{4 b_1}. 
\ee
Thus we find,
\ben
\ln Z = - \frac{(1-a_1)^2}{4b_1} + \cC .
\een
$\cC$ can be fixed by demanding that at $R'(0)=0$ $i.e.$ at $a_1=1$,
the free energy is zero. Thus we get $\cC = 0$. Hence
the free energy of \type 1 phase is given by,
\be\label{eq:free-en-ab-nocut}
F_{\mathbb{T1}} = -T \ln Z = N^2 T \frac{(1-a_1)^2}{4b_1}.
\ee
Thus we see free energy is positive for $b_1>0.$

\subsubsection*{\type 2 : One-cut solution}

For one cut solution we have the following condition,
\be
R'(0) =1 -\frac1{4\bab}
\ee
and the solution exists for $\bab\geq \frac12$. Plugging the value of
$\bab$ in terms of $a_1$ and $b_1$ we can write,
\be 
a_1 x+ 2b_1 x^3 \bab^2-1=0, \quad \text{where}\ \ x= \frac{4\bab
  -1}{4\bab^2}.  
\ee
The resolvent is given by,
\be
R(z) =  \frac12\lB 1+\bab (z-z^{-1}) + \bab(1+z^{-1})
\sqrt{z^2 + 2(\bab^{-1} -1)z +1}\rB .
\ee
Hence, the spectral density for this branch is given by,
\be
\s(\theta) = 4\bab\cos\frac{\theta}2 \sqrt{\sin^2\frac{\theta_1}2
  -\sin^2\frac{\theta}2}
\ee
where,
\be
\sin^2 \frac{\theta_1}{2} = \frac1{2\bab} .
\ee

\subsection*{Free energy}

The free energy of this branch can also be calculated in the similar
way as mentioned before. Variation of $\ln Z$ is given by,
\ben \label{dLnZ-ab1cut}
  \frac{d}{d T} \ln Z(T) &=&  N^2  R'(0)^2 \lb \frac{d a_1}{d T} +
    R'(0)^2 \frac{d b_1}{d T}\rb.
\een
For this phase,
\be
R'(0)=1-\frac1{4(a_1 R'(0) +2b_1 R'(0)^3)}.
\ee
Solving this equation we find,
\be
a_1= \frac1{4 \Rp (1-\Rp)} -2b_1 \Rp.
\ee
Differentiating both sides we find,
\be
\frac{da_1}{dT}=\frac{\frac{dR'(0)}{dT} \left(2 R'(0) \left(1-8 b_1
      \left(R'(0)-1\right)^2 
   R'(0)^2\right)-1\right)-8 \frac{db_1}{dT} \left(R'(0)-1\right)^2
R'(0)^4}{4 
   \left(R'(0)-1\right)^2 R'(0)^2}.
\ee
Substituting $\frac{da_1}{dT}$ in eqn. (\ref{dLnZ-ab1cut}) we find,
\ben
\frac1{N^2} \frac{d}{d T} \ln Z(T) &=& \lb\frac{-16 b_1 R'(0)^5+32 b_1 
  R'(0)^4-16 b_1 R'(0)^3+2 R'(0)-1}{4 
   \left(R'(0)-1\right)^2} \rb \frac{d\Rp}{dT} -\Rp^4
 \frac{db_1}{dT}\notag\\
&=& \frac d{d T}g(\Rp, b_1)
\een
where,
\be
\frac{\pa g}{\pa \Rp} = \lb\frac{-16 b_1 R'(0)^5+32 b_1
  R'(0)^4-16 b_1 R'(0)^3+2 R'(0)-1}{4 
   \left(R'(0)-1\right)^2} \rb, \quad \frac{\pa g}{\pa b_1} = -\Rp^4.
\ee
Solving these two equation we find,
\be
g(\Rp, b_1) = \frac{1}{4-4 R'(0)}+\frac{1}{2} \log
\left(R'(0)-1\right) - \Rp^4 b_1 .
\ee
Thus we find free energy
\be
F=-N^2 T \lB \frac{1}{4-4 R'(0)}+\frac{1}{2} \log
\left(R'(0)-1\right) - \Rp^4 b_1 \rB + \cC
\ee
Constant $\cC$ can be fixed by demanding that the free energy of
this branch should match with the free energy of no-cut solution at
$\Rp=1/2$. This implies,
$$
\cC = N^2 T \lb \frac12 -\frac12 \ln(-2)\rb.
$$
Hence we find free energy of \type 2 branch is given by,
\be
\begin{split}
 F_{\mathbb{T2}}&=  -N^2 T\lB  \frac{1}{4-4 R'(0)} -\frac12 +\frac{1}{2} \log
\left(1-R'(0)\right) - \Rp^4 b_1 \rB \\
&=  -N^2 T \lB \bab -\frac12
-\frac12 \ln (2\bab) - b_1\lb 1-\frac1{4\bab} \rb^4\rB .
\end{split}
\ee

\subsubsection*{\type 3 : Two-cut solution}

Two-cut solution, in the similar way, exists when
\be
 1-\frac1{4\bab}<R'(0)<\bab .
\ee
One can again calculate the free energy of this class for a particular
parametrization between $\Rp$ and $a_1, b_1$. It turns out that the
free energy of this branch is higher than the free energy of one-cut
solution and hence this phase is thermodynamically disfavoured.

\subsection{Phase diagram and its dual description}

Finally we summarize the phase diagram for $(a,b)$ model. However, the
basic structure of the phase diagram is same for a generic class of
phenomenological model governed by the action (\ref{eq:sefftrunc}).

\begin{itemize}
\item At low temperature the saddle point is characterized by $\bab=0$
  which has constant resolvent and uniform eigenvalue
  distribution. This phase corresponds to thermal $AdS$ in the dual
  gravity setup. This phase has zero free energy.
\item Then there is a saddle point with resolvent linear in $z$ when
  $\bab = \Rp$ {\it i.e.}
\be \bab^2 = {1-a_1 \over 2 b_1} \le {1 \over 4}\
.  \ee 
This is the unstable saddle point corresponding to the small black hole
(in the phase where it is to be viewed as an excited string state) and
has positive free energy given by eqn. (\ref{eq:free-en-ab-nocut}).
This saddle point exists in a temperature range
$T_c \le T \le T_H$.
\item There is the saddle point which obeys 
\ben \Rp=1-{1\over 4\bab}
  \quad \text{or}\ a_1 x+2b_1 x^3 \bab^2-1=0, \ \ \text{where} \
  x={4\bab-1\over 4\bab^2} 
\een
This equation has two real solutions for $\bab$ above a temperature
$T_0$. Let us denote these two values by $\bab^S$ and $\bab^B$ with
$\bab^B>\bab^S$\footnote{At $T=T_0$, $\bab^B=\bab^S$. As we increase
  temperature $\bab^B$ starts increasing and $\bab^S$ decreases.}. The
two values or phases correspond to the Big Black Hole (BBH) and the
Small Black Hole (SBH) (in the actual black hole regime) in the dual
side. Both these branches have gap in eigenvalue distribution. The BBH
($\bab^B$) solution exists for all temperatures greater than the
minimum $T_0$ for which this solution exists. While the SBH ($\bab^S$)
solution exists in the interval $T_o\leq T \leq T_c$. At $T_c$ which
corresponds to $\bab^S={1\over 2}$, this solution goes over into the
ungapped solution. This is called Gross-Witten-Wadia transition
\cite{gww}. 

\item There exists a temperature $T_1> T_0$ where the stable saddle
  points $\bab=0$ and $\bab^B$ exchange dominance. This temperature
  corresponds to the Hawking-Page temperature where the BBH has a
  lower free energy than thermal AdS in the semi-classical gravity
  path integral. The free energy for $\bab^B$ phase becomes negative
  for $T>T_1$.

Phase diagram in presence of charge or 
chemical potential has been discussed in \cite{Basu:2005pj,
  Yamada:2006rx, Harmark:2006di}.

\item There exists another phase in the same regime of parameters where
  the last phase appears. This phase has eigenvalue distribution with
  two gaps. But this phase has higher free energy than that of one-gap
  phase, hence thermodynamically disfavoured. The gravity dual to this
  phase is not well understood.
\end{itemize}


\section{Distribution in momentum space}\label{sec:YTside}

In this section we shall discuss how different saddle points of weakly
coupled gauge theories on $S^3$ are captured in terms of different
representation of $U(N)$ group in the large $N$ limit. Since different
representations correspond to different Young diagrams, therefore, we
find how boxes in Young diagrams are distributed corresponding to
different saddles of the theory. The advantage of studying the phases
in terms of large $N$ Young diagram is that one can directly find out
the phase space distribution of different saddle points. 
As we have already studied, eigenvalues of the holonomy matrix behave like position variables of 
fermions. At the same time, the representations of $U(N)$ also have an
interpretation in the language of non-interacting fermions with the
number of boxes of the Young tableaux being like the momentum
\cite{douglas}. Therefore knowing the momentum distribution is
important for drawing the phase space distribution.

Our goal, therefore, is to find out the momentum distribution
corresponding to no-cut, one-cut and two-cut phases.

\subsection{Partition function and Young tableaux}

We start with the partition function at {\it zero coupling}
\begin{eqnarray}
Z=\int{\cal{D}}U\, \exp\bigg[\sum_{n=1}^{\infty}
\frac{a_{n}}{n}\Tr[U^{n}]\Tr[U^{\dagger n}]\bigg] .
\end{eqnarray}
This can be expanded as follows
\begin{eqnarray}
\int{\cal{D}}U\,\sum_{\vec{k}}\frac{1}{z_{\vec{k}}}
\prod_{j}a_{j}^{k}\Upsilon_{\vec{k}}(U)\Upsilon_{\vec{k}}(U^{\dagger}) ,
\end{eqnarray}
where we use the following notations
\begin{eqnarray}
z_{\vec{k}}=\prod_{j}k_{j}{!}j^{k_{j}}\quad\quad\quad
\Upsilon_{\vec{k}}(U)=\prod_{j=1}^{\infty}(\Tr[U^{j}])^{k_{j}} .
\end{eqnarray}
$\Upsilon_{\vec{k}}(U)$ can be rewritten in terms of the characters of
the conjugacy class of the permutation group $S_{K}$ as follows
\begin{eqnarray}
\Upsilon_{\vec{k}}(U)=\sum_{R}\chi_{R}(C(\vec{k}))\Tr_{R}[U] .
\end{eqnarray}
Here $\chi_{R}(C(\vec{k}))$ is the character of the conjugacy class
$C(\vec{k})$ of the permutation group $S_{K}$ and $K=\sum_ j j k_{j}$
and $R$ denotes the specific representation of $U(N)$. Finally, using
the following orthogonality relation between the character of
representations of $U(N)$
\begin{eqnarray}
\int {\cal{D}}U\, \Tr_{R}[U] \Tr_{R'}[U^{\dagger}]=\delta_{RR'}
\end{eqnarray}
we obtain the following form for the partition function:
\begin{eqnarray}
  Z=\sum_{\vec{k}}\frac{\prod_{j}a_{j}^{k_{j}}}{z_{\vec{k}}}\sum_{R}
  \lB\chi_{R}(C(\vec{k}))\rB^2 .
\end{eqnarray}
This is an exact expression for the partition function for any
temperature and $N$ of the free gauge theory. However, it should be
noted that the answer is completely explicit.

A particular representation of an unitary group $U(N)$ can be labeled
by a Young diagram with maximum $N$ number of rows and arbitrary
numbers of boxes in each row up to a constraint that number of boxes
in a particular row can not be greater than the number of boxes in a
row before . Therefore, the sum of representations of $U(N)$ can be
cast as sum of different Young diagrams. If $K$ is the total number of
boxes in a particular Young diagram with $\lambda_i$ number of boxes
in $i^{th}$ row and $\sum_i \l_i =K$, then sum over representations
can be decomposed as
\begin{eqnarray}
\sum_{R} \rightarrow\sum_{K=1}^{\infty}
\sum_{\{\l_{i}\}}\delta\lb\sum_{i=1}^{N}\l_{i}-K\rb .
\end{eqnarray}
The partition function can be written as,
\ben\label{sec:pf-final}
Z =\sum_{K=1}^{\infty} \sum_{\vec {\l}}
\delta\lb\sum_{i=1}^{N}\l_{i}-K\rb
\sum_{\vec{k}}\frac{\prod_{j}a_{j}^{k_{j}}}{z_{\vec{k}}} \delta\lb K-
\sum_i i k_i\rb \lB \chi_{\vec{\l}}(C(\vec k))\rB^2.
\een
Note that the total number of boxes is same as the order of the
permutation group $S_K$.  The characters of the conjugacy class are
determined recursively by the Frobenius formula \cite{fulton-harris,
  hamermesh, lasalle}. Explicit expressions for the most general case
are not simple. We shall discuss about that in the next subsection.

\subsection{Character of permutation group}\label{character}

We discuss in detail the Frobenius character formula
\cite{fulton-harris} for the $\chi_{\vec{\l}}(C(\vec k))$ of the
permutation group.  Let $C({\vec{k}})$ denote the conjugacy class of
$S_K$ which is determined by a collection of sequence of
numbers
\begin{eqnarray}
\vec{k}=(k_{1},k_{2},\cdots), \qquad \text{with}, \ \
\sum_{i}ik_{i}=K 
\end{eqnarray}
$C({\vec{k}})$ consists of permutations having $k_{1}$ number of
1-cycles, $k_{2}$ number of 2-cycles and so on.  We introduce a set of
independent variables, $x_{1},\cdots,x_N$, and for a given Young
diagram $(\vec{\l})$ we have $\lambda_{1}\geq\lambda_{2}\geq \cdots
\geq \lambda_{N}\geq 0$. Then we define a power series and the
Vandermonde determinant as follows.
\begin{eqnarray}
  P_{j}(x)=\sum_{i=1}^{N}x_{i}^{j}\quad\quad
  \quad\Delta(x)=\prod_{i<j}(x_{i}-x_{j}).
\end{eqnarray}

For a set of non-negative integers, $(n_{1},\cdots ,n_{N})$, we define
\begin{equation}
  \lB f(x)\rB_{(n_{1},..,n_{N})}=\textrm{coefficient of }
  x_{1}^{n_{1}}\cdots x_{N}^{n_{N}}\textrm{ in }f(x) 
\end{equation}
Now for our given partition, $\vec \lambda$, of a Young diagram we
define,
\begin{eqnarray}
h_i = \l_i + N -i \qquad \forall \quad i =1, \cdots, N
\end{eqnarray}
with
\be\label{hmonotonicity}
h_1> h_2> \cdots > h_N \geq 0.
\ee
%
Then the character corresponding to a conjugacy class $C(\vec k)$ and
a representation characterized by $\vec \l$ is given by
\begin{eqnarray}
  \chi_{\vec \l }(C_{\vec{k}})=\bigg[\Delta(x).\prod_{j} \lb P_{j}(x)
  \rb^{k_j}\bigg]_{(h_{1},\cdots, h_{N})} 
\end{eqnarray}
This is the most generic formula for character. Using this formula one
can in principle write down the partition exactly. However it is still
complicated to solve the model and find the saddle points in large $N$
limit in presence of all $a_i$'s. In the next section, we shall
extremize the partition function in large $N$ limit to find out
different saddle points of the system in presence of one cycles
only. This is same as {\it $a_1$ model } discussed in
sec. \ref{sec:amodel}. Young Tableau distribution for {\it $a_1$ model}
has already been discussed in \cite{Dutta:2007ws}. Authors in
\cite{Dutta:2007ws} found Young Tableau or momentum distribution
corresponding to no-cut and one-cut solution discussed in the previous
section. Here, we extend the discussion and find momentum distribution
for two-cut solution.

In appendix \ref{app:yt-a1a2} we extend our calculation to find out
the saddle points in presence of both 1-cycle and 2-cycle. This is
similar to the model we considered in section \ref{sec:aprox2}.

Before we extremize the partition function to find the saddle point
equation let us simplify the formula for character for one cycle. This
expression will be useful in the next section.

\subsubsection{Character in presence of 1-cycles}

We consider a particular conjugacy class $C(\vec k) = (k,0,0,
\cdots)$, {\it i.e.} there is only $k$ 1-cycles.  In this case,
$k_{1}=k$, and character is given by
\begin{eqnarray}
  \chi_{\vec h}(C(k))=\bigg[\Delta(x) \lb
  \sum_{i=1}^{N}x_{i}\rb^{k}\bigg]_{(h_{1},\cdots,h_{N})}. 
\end{eqnarray}
The Vandermonde determinant can be written as
\begin{eqnarray}\label{eqn:vand}
  \sum_{\sigma\in S_{K}}(-1)^{\sigma} x_{N}^{\sigma(1)-1} \cdots
  x_{1}^{\sigma(N)-1}
\end{eqnarray}
Where $\sigma(i)$ is an element of the permutation group $S_{K}$. More over
\begin{eqnarray}\label{eqn:P}
  \lb\sum_{i=1}^{N}x_{i}\rb^{k}=\sum_{\vec r}
  \frac{k ! }{r_1 {!} r_2 ! \cdots r_N !} x_1^{r_1} x_2^{r_2}
  \cdots x_N^{r_N},
\end{eqnarray}
where the sum is performed over a set of non-negative integers, $\vec
r = (r_{1},..,r_{N})$ such that $r_1+r_2+\cdots +r_N=k$.

Multiplying eqn. (\ref{eqn:vand}) with eqn. (\ref{eqn:P}) and picking
up the coefficient of the $\prod_{i=1}^{N}x_{i}^{h_{i}}$ we find
\begin{eqnarray}
  \chi_{\vec h}(C(k)) = \sum_{\sigma\in S_{K}} (-1)^{\sigma}
  \sum_{\vec r} \frac{k !}{r_1 {!} r_2 ! \cdots r_N !}
  \prod_i\delta(r_i +\sigma(N-i+1) -1 -h_i)\, .
\end{eqnarray}
Note that $\sum_i r_i = k$ is automatically satisfied once we consider
$\sum_i \l_i = k$. Performing the summation $r_i$ we can write,
\be
\chi_{\vec h}(C(k)) = \sum_{\sigma\in S_{K}} (-1)^{\sigma} \frac{k
  !}{\prod_i (h_i +1 - \sigma(N-i+1))!}.
\ee
Here the sum is only over those $\sigma$ such that
$h_{N-i+1}-\sigma(i)+1\ge 0$. This can be rewritten using the property
of the determinant as
\begin{eqnarray}\label{eq:character-1c}
\chi_{\vec h}(C_{k})=\frac{k{!}}{\prod_{i}^{N}h_{i}{!}}\prod_{i<j}(h_{i}-h_{j}).
\end{eqnarray}

For completeness here we present a generic formula for the character
in presence of $k_1$ number of 1-cycles, $k_2$ number of 2-cycles up
to $k_{\o}$ number of $\o$-cycles. The expression for the character is
obtained following the same argument given above.
\begin{eqnarray}
  &&\chi_{\vec{k}}(C(\vec{k}))=\sum_{\vec{r}_{i}}
  \textrm{sign}(\sigma)\frac{k_{1}{!}..k_{\omega}{!}}{\prod_{j}r^{1}_{j}{!} 
    r^{2}_{j}{!}..r^{\omega}_{j}{!}}\delta^{j}(\sum_{i=1}^{\omega}ir^{i}_{j}+
  \sigma(N+1-j)-1-h_{j})\delta(\sum
  r^{1}_{j}-k_{1})...\delta(\sum r^{\omega}_{j}-k_{\omega})\notag\\ 
  &&=\frac{k_{1}{!}..k_{\omega}{!}}{\prod_{j}^{N}(l_{j}-\sum_{i=2}^{\omega}
    ir_{j}^{i})r_{j}^{2}{!}..r_{j}^{\omega}{!}}\prod_{p<q}(h_{p}-\sum_{i=2}^{\omega}
  ir_{p}^{i}-h_{q}+\sum_{i=2}^{\omega}ir_{q}^{i})\delta(\sum
  r_{j}^{2}-k_{2})... \delta(\sum r_{j}^{\omega}-k_{\omega}).
\end{eqnarray}

\subsection{Zero coupling : ``$a_1$ model'' - approximation 1}

We first look at a model where all the terms are zero apart from the
first term $a_{1}$ ($a_i=0$, for $i\geq 2$). 
\begin{eqnarray}\label{eq:a1pf}
Z=\int{\cal{D}}U\, \exp\bigg[
a_{1}\Tr[U]\Tr[U^{\dagger}]\bigg] .
\end{eqnarray}

Which is essentially same
as considering $\vec k = (k,0,0, \cdots)$. The partition function
(\ref{sec:pf-final}) takes the following form
\begin{eqnarray}
  Z &=& \sum_{K=1}^{\infty} \sum_{\vec {\l}}
  \delta\lb\sum_{i=1}^{N}\l_{i}-K\rb
  \sum_{k=1}^{\infty}\frac{a_{1}^{k}}{k!} \ \delta\lb K-
  k\rb \lB \chi_{\vec{\l}}(C( k))\rB^2 \notag\\
  &=& \sum_{k=1}^{\infty} \sum_{\vec {\l}} \frac{a_{1}^{k}}{k!} \ \delta\lb 
  \sum_{i=1}^{N}\l_{i} -
  k\rb \lB \chi_{\vec{\l}}(C( k))\rB^2.
\end{eqnarray}

Using the expression for character given in
eqn. (\ref{eq:character-1c}) we can explicitly write the partition
function for {\it a model}
\ben
Z &=& \sum_{k=1}^{\infty} \sum_{\vec {\l}} \frac{a_{1}^{k}}{k!} \ \delta\lb 
  \sum_{i=1}^{N}\l_{i} -
  k\rb \lB \frac{k{!}}{\prod_{i}^{N}h_{i}{!}}\prod_{i<j}(h_{i}-h_{j})
  \rB^2 .
\een
This expression is exact for all $N$. However, we shall consider large $N$
limit of this model and see how one can capture all the phases
obtained from the analysis of eigenvalue density.

\subsection*{ The large $N$ limit and saddle point equation}

In the limit $N\to\infty$ we introduce, following \cite{dk},  a
set of continuous variables
\begin{eqnarray}
  \lambda(x)=\frac{\lambda_{i}}{N},\qquad
  h(x)=\frac{h_{i}}{N},\qquad \text{where}, \quad x=\frac{i}{N}.
\end{eqnarray}
Since $i$ runs from $1$ to $N$, therefore in large $N$ limit
$x\in[0,1]$. The function $\l(x)$ or equivalently $h(x)$ captures the
profile or shape of Young diagram in large $N$ limit. The relation
between $\l(x)$ and $h(x)$, in continuum limit is given by
\be
  h(x)=\lambda(x)+1-x .
\ee
The condition $h_1>h_2>\cdots>h_N$ implies a strict monotonicity for
$h(x)$, {\it i.e.}
\be\label{eq:hmonotoni}
h(x)>h(y) \quad \text{for} \ \ y>x, \quad \text{which implies} \ \
\frac{\pa h(x)}{\pa x} < 0. 
\ee
In this continuum limit all the summation over $i$ is replaced by an
integration
\be
\sum_{i=1}^{N} \ra N \int_0^1 dx .
\ee
The total number of boxes $k$ in a Young diagram is given by
\begin{eqnarray}
k=\sum_{i=1}^N \lambda_{i}\to \ \ k=
N^{2}\bigg[\int_{0}^{1}dx\,h(x)-\frac{1}{2}\bigg]=N^{2}k', 
\end{eqnarray}
where $k'$ is a order 1 number. In the large $N$ limit, as we shall
see in the next section, the dominant contribution to the partition
function comes from the representation for which the total number of
boxes in the corresponding Young diagram is of the order of $N^2$.

The partition function given in eqn. (\ref{eq:a1pf}) can then be written
as
\begin{eqnarray}
Z = \sum_{\vec h} \exp \lB \ln k{!}  + \ln a_{1}  + \sum_{i\ne
  j}|h_{i}-h_{j}| - 2 \sum_{i} \ln h_{i}{!} \rB .
\end{eqnarray}
In the large $N$ limit using Sterling's approximation we find,
\begin{eqnarray}\label{eq:pfconti}
Z=\int[ {\cal{D}}\, h(x)] e^{-N^{2}\seff{h}}
\end{eqnarray}
where,
\begin{eqnarray}
-\seff{h}=\int_{0}^{1} dx \ \Xint-_{0}^{1}dy\, \ln|h(x)-h(y)|-2\int dx\,
h(x)\ln h(x)+k' \ln a_{1}k'+ k'+1,
\end{eqnarray}
with
$$
k'=
\int_{0}^{1}dx\,h(x)-\frac{1}{2} .
$$

From the expression of partition function given in
eqn. (\ref{eq:pfconti}) it is clear that in $N\ra \infty$ limit the
dominant contribution comes from those configurations for which
$\seff{h}$ is extremum. Therefore, varying $S_{\text{eff}}$ with
respect to $h(x)$ we obtain the following saddle point equation.
\begin{eqnarray}
\Xint-\frac{dy}{h(x)-h(y)}=\ln h(x)-\frac{1}{2} \ln [a_{1}k'] .
\end{eqnarray}

Following \cite{dk} we introduce the density of boxes in the
Young diagram defined by
\begin{eqnarray}
u(h)=-\frac{\partial x(h)}{\partial h}.
\end{eqnarray}
The density function $u(h)$ by definition satisfies the normalization
\be\label{eq:hnorma}
\int^{h_{U}}_{h_{L}}dh \,u(h)=1 .
\ee
The density function has a support between $h_L=h(1)$ and
$h_U=h(0)$. Moreover, from the monotonicity of  $h(x)$
(eqn. \ref{eq:hmonotoni}) it is easy to show that the density function
is positive definite. Also, it  obeys the normalization
(\ref{eq:hnorma}), hence we find
\be\label{eq:constraint}
u(h)\leq 1.
\ee

In terms of the Young Tableaux density the saddle equation becomes
\begin{eqnarray}\label{eq:inteq}
\Xint- ^{h_{U}}_{h_{L}} dh' \frac{u(h')}{h-h'}=\ln\bigg[\frac{h}{\xi}\bigg]
\end{eqnarray}
where $\xi^{2}=a_{1}k'$. 

Note that the parameter
$\xi$  sitting on the right hand side of the saddle equation depends
on $k'$ where $k'$ is given by
\be\label{eq:k'}
k' = \int_0^1 h(x) dx -\frac12 = \int_{h_L}^{h_U} h \ u(h) dh -\frac12,
\ee
itself depends of the density function $u(h)$. We have to solve this
equation self-consistently.

The general interacting unitary matrix model given by
eqn. (\ref{eq:mostgeneric-acn}) for weakly coupled gauge theory can
also be solved using the above technique. One can in principle write a
saddle point equation at large $N$ for the generic action. However,
the analysis becomes technically much more involved.  But the
essential phase structure of the theory is in any case captured by
models involving only ${\rm Tr}U\Tr U^{\dagger}$ as given in the eqn.
(\ref{eq:sefftrunc}). In particular, as we have seen, the $(a,b)$
model (eqn. (\ref{Zab})) does a good job in getting the detailed form
of the phase structure. In \cite{Dutta:2007ws} the authors obtained
the saddle equation for Young tableaux distribution function for the
class of phenomenological model (\ref{eq:sefftrunc}). The basic
structure of the saddle equation remains same as
eqn. (\ref{eq:inteq}), only the definition of the parameter $\xi$
changes. As we mentioned in the introduction, the basic shape of the
phase space distribution does not change under addition of
perturbative correction, therefore, here we shall not carryout the
computation of saddle equation for weak coupling model. We refer
\cite{Dutta:2007ws} for a detailed discussion.

\subsection*{Different classes of solutions}

We now discuss the possible classes of solution which satisfy the
integral equation (\ref{eq:inteq}) for Young Tableaux density. We
classify different saddle points depending on the fact whether the
variable $h$ is continuous from $h_L$ to $h_U$. Depending on this
there are primarily two classes.

\subsubsection*{ \class 1 ( \cl 1) : {\bf $h(x)$ is continuous} }

For this class the function $h(x)$ is continuous and monotonically
decreasing for $x\in[0,1]$. $h$ changes continuously from $h(1)=h_L$
to $h(0)=h_U$. In other words, $(h_i -h_{i+1})/N \ra 0$ $\forall \ i =
1, \cdots , N$. Therefore, for this branch the difference in number of
boxes between two consecutive rows is of the order 1 in the limit
$N\ra
\infty$. This class has two subclasses.\\

\noindent
{\class{1 a} (\cl{1a}) :}
In this case\footnote{We consider $h_L=p$ and $h_U=q$.}
\be\label{eq:ansatz-class1a}
0\leq u(h) < 1; \qquad h\in[p,q].
\ee
A typical Young diagram corresponding to such distribution has been
plotted in figure \ref{fig:YTC1a}.
\begin{figure}[h]
\ben
\begin{Young}
&&&&&&&&&&&&\cr
&&&&&&&&&&&&\cr
&&&&&&&&&&&&\cr
&&&&&&&&&&&\cr
&&&&&&&&&&\cr
&&&&&&&&&\cr
&&&&&&&&\cr
&&&&&&&\cr
&&&&&&\cr
&&&&&\cr
\end{Young}\notag
\een
\caption{A typical Young diagram for \cl{1a}.}
\label{fig:YTC1a}
\end{figure}

\begin{figure}[h]
\centering
\includegraphics[width=7cm,height=6cm]{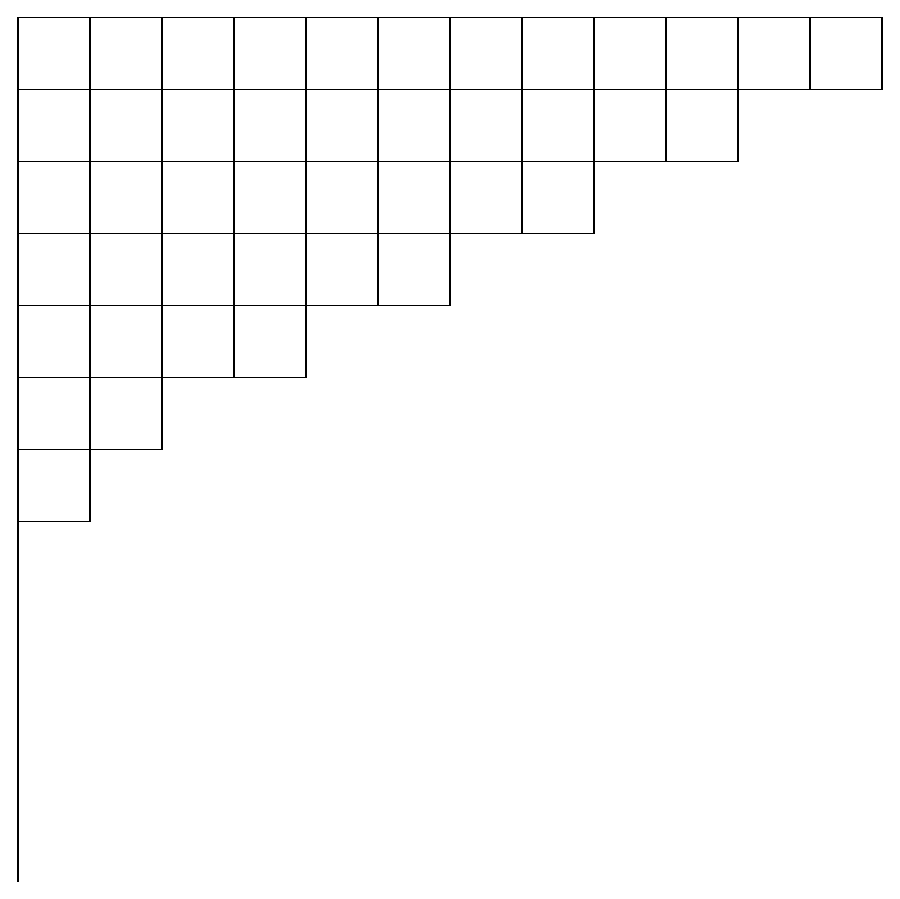}
\caption{A typical Young Tableaux for \cl{1b}}
\label{fig:YTC1b}
\end{figure}

\noindent
{\class{1b} (\cl{1b}):} 
 For this class distribution is given by
\ben \label{eq:ansatz-class1b}
u(h) &=& 1; \qquad \ \ \ \ h\in [0,p]\notag\\
&=& \tilde u(h); \qquad h\in [p,q]
\een
with $\tilde u(h)<1$.  For this branch $u(h)=1$ for $0<h<p$. Therefore
a finite number of rows in the particular Young diagram are empty. The
distribution has been depicted in figure \ref{fig:YTC1b}.
These two subclasses have been considered in \cite{Dutta:2007ws}. In
this paper we introduce a new class of solution where the support $h$
is discontinuous.

\subsubsection*{\class 2 : $h(x)$ {\bf is discontinuous}}

\begin{figure}[h]
\centering
\includegraphics[width=9cm,height=6cm]{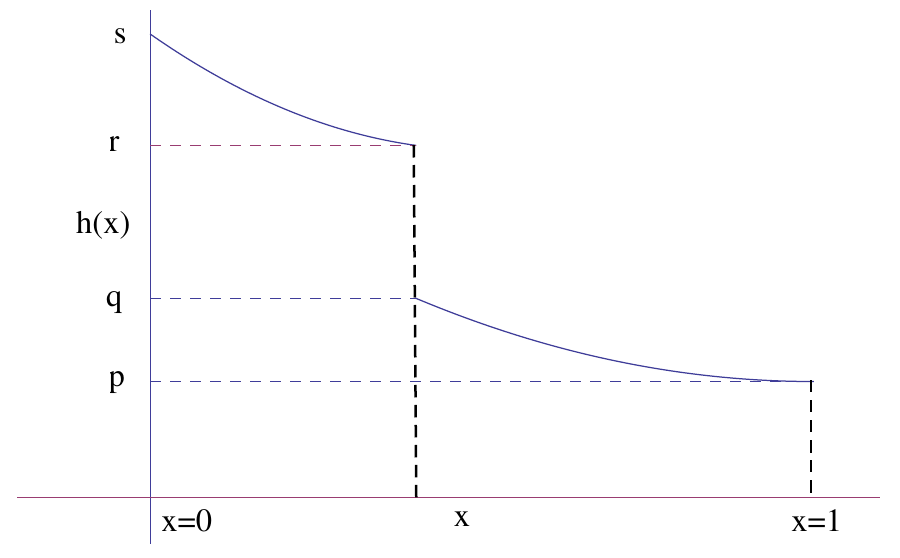}
\caption{$h(x)$ has a discontinuity for \cl{2}.}
\label{fig:h-discont}
\end{figure}

In this case the function $h(x)$ is discontinuous at some point
$\zeta$ between 0 and 1 (see figure \ref{fig:h-discont}).
\be \label{eq:ansatz-class2}
{{\lim}_{\epsilon\ra 0} } \lB h(\zeta-\epsilon)
-h(\zeta+\epsilon)\rB  \sim
{\cal O}(1), \qquad 0<\zeta<1 .
\ee

\begin{figure}[h]
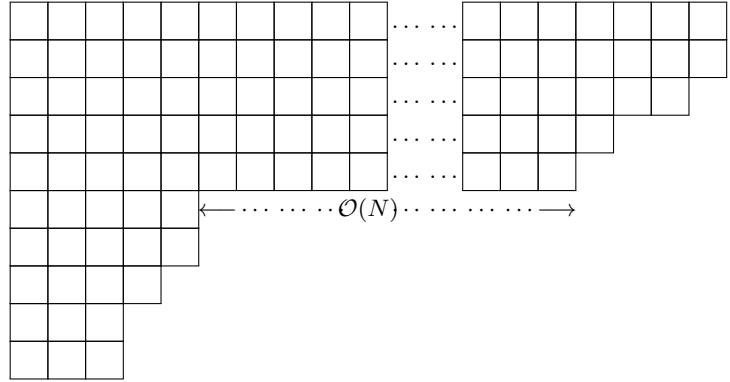

  \centering
\ytableausetup
{mathmode, boxsize=1.5em}
\begin{ytableau}
\, & \, & \, &\,&\,&\,&\,&\,& \, &\, &  \none[\dots] &\none[\dots] & \, &\,
& \, &\, & \, &\, & \\
\, & \, & \, &\,&\,&\,&\,&\,& \, &\, &  \none[\dots] & \none[\dots] & \, &\,
& \, &\, & \, &\, &\\
\, & \, & \, &\,&\,&\,&\,&\,& \, &\, & \none[\dots] &\none[\dots] & \, &\,
& \, &  \, &\, & \\ 
\, & \, & \, &\,&\,&\,&\,&\,& \, &\, & \none[\dots] & \none[\dots] &
\, & \, &\, &\\  
\, & \, & \, &\,&\,&\,&\,&\,& \, &\, & \none[\dots] & \none[\dots] &
\, & \, &\\  
\,& \, &\, & \, &\, & \none[\longleftarrow] &\none[\cdots]
&\none[\cdots]&\none[\cdots] &\none[\cO(N)]
&\none[\cdots]&\none[\cdots]& \none[\cdots]& \none[\cdots]&
\none[\longrightarrow]\\  
\,& \, &\, & \, & \\
\,& \, &\, & \\
\,& \, &\\
\,& \, & \\
\end{ytableau}
\caption{A typical Young diagram for \cl{2}.}
\label{fig:YTC2}
\end{figure}

The distribution $u(h)$ is therefore given by,
\ben\label{eq:ansatz-u-2}
u(h) &=& \tilde u_1(h),  \qquad p<h<q , \quad \tilde u_1(h)<1 \notag\\
&=& \tilde u_2(h),  \qquad r<h<s, \quad \tilde u_2(h)<1
\een
with 
\be
\int_q^r u(h) dh =0.
\ee
A typical representation corresponding to such a Young diagram has
been plotted in figure \ref{fig:YTC2}.

As we vary the order parameter $\xi$ the constraint
(\ref{eq:constraint}) will come into play and the system will jump
from one branch of solution to the other.

\subsection*{Finding densities for different classes}

To find Young tableaux distribution $u(h)$ we define a resolvent
$H(h)$ given by
\begin{eqnarray}\label{eq:defH}
H(h)=\int^{h_{U}}_{h_{L}}dh'\frac{u(h')}{h-h'}.
\end{eqnarray}
This function has the following properties:
\begin{enumerate}[(a)]
\item $H(h)$ is an analytic function of $h$ in complex $h$ plane with a
  branch cut along positive real interval where $u(h)$ has support.
\item $H(h)$ is real for real positive $h$ outside the support.
\item In the limit $h\ra \infty$, $H(h)$ has following asymptotic
  expansion
\be
H(h\ra \infty)\sim \frac{1}{h}+\lb k'+\frac{1}{2}\rb\frac{1}{h^{2}}.
\ee
This can be verified easily by expanding the resolvent for large $h$.
\item 
${\lim}\atop {\e \ra 0}$ \ $H(h+i\epsilon)+H(h-i\epsilon)=2 \ln \lB\frac{h}{\xi}\rB$
  for real $h$ which comes from the definition of a complex function
  with branch cut on the real line.
\item $u(h)={\lim \atop {\e \ra 0}}-\frac{1}{2\pi
    i}[H(h+i\epsilon)-H(h-i\epsilon)]$ for $h$ inside the support.
\end{enumerate}

\subsubsection{\class{1a} }

\begin{figure}[h]
\ben
\includegraphics[width=8cm,height=6cm]{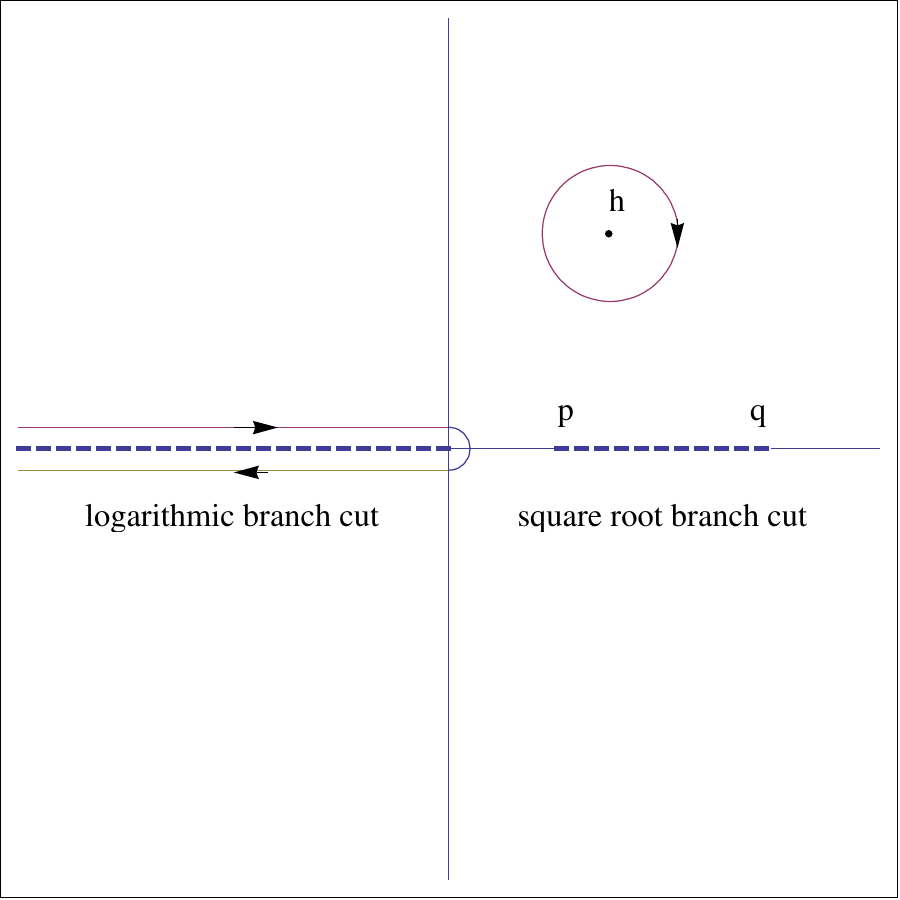}\notag
\een
\caption{Contour for eqn. (\ref{eq:H-C1a}).}
\label{fig:contour-1}
\end{figure}

Using standard techniques \cite{Kazakov:1995ae} we solve integral
equations (\ref{eq:inteq}) with logarithmic kernel.  There is a closed
expression for resolvent $H(h)$ in terms of a contour integral
\cite{Marino:2004eq}. It follows from the properties of $H(h)$ listed
above.
\begin{eqnarray}\label{eq:H-C1a}
  H(h)=-\sqrt{(h-p)(h-q)}\oint_{\cC}\frac{ds} {2\pi
    i}\frac{\ln(s/\xi)}{(s-h)\sqrt{(s-p)(s-q)}} 
\end{eqnarray}
where contour $\cC$ is an anticlockwise contour around the branch cut
$[p,q]$. By deforming the contour one can take the contour to be a
clockwise loop about $h$ and about the logarithmic branch cut from
$-\infty$ to $0$ and back (see figure \ref{fig:contour-1}).
Carrying out the contour integration we obtain,
\begin{eqnarray}\label{eq:H-1a}
H(h)=\ln\bigg[\frac{2 h^{2}-(\sqrt{q}-\sqrt{p})^{2} h+2q p-2(h+q
  p)\sqrt{(h-p)(h-q)}}{\xi (\sqrt{p}+\sqrt{q})^{2}}\bigg].
\end{eqnarray}
From the asymptotic expansion of $H(h)$ one finds,
\begin{eqnarray} \label{eq:pqvalue1a}
\sqrt{p} &=&\sqrt{\xi}+\frac{1}{\sqrt{2}}, \notag\\
\sqrt{q}&=&\sqrt{\xi}-\frac{1}{\sqrt{2}} \\
\text{and} \quad k'&=&\sqrt{q p}+\frac{1}{4} \notag
\end{eqnarray}
which implies,
\begin{eqnarray}\label{eq:condition-a1-1a}
a_{1}=\frac{4\xi^{2}}{4\xi-1} .
\end{eqnarray}
Since $p,q$ are real and positive therefore, this branch exists for
$\xi\geq \frac12$ (from eqn. (\ref{eq:pqvalue1a})). From
eqn. (\ref{eq:condition-a1-1a}), we therefore, conclude that this
class of solution exists for $a_1\geq 1$. 

Hence, this branch (\class{1a}) corresponds to 1-cut branch
(\type 2) discussed in section \ref{sec:one-cut}.

From the discontinuity of resolvent we find that the Young Tableaux
density $u(h)$, for this branch, is given by
\be
u(h) = \frac2\pi \cos^{-1} \lB \frac{h+\xi -1/2}{2\sqrt{\xi h}}\rB,
\quad h\in[p,q]. 
\ee
Equivalently we can write a quadratic equation for $h$,
\be \label{eq:h-eqn-class-1a}
h^2 -\lb 1+2\xi \cos(\pi u(h)) \rb h + \lb \xi -\frac12 \rb^2 =0.
\ee
If $h_1$ and $h_2$ are two roots of this solution then we find
\be\label{eq:root-relation-1a}
h_1+h_2 = 1+2\xi \cos(\pi u(h)), \quad h_1 h_2 = \lb \xi -\frac12
\rb^2.
\ee
For this branch, we also note that
\be\label{eq:pq-root-1a}
p+q = 1+2\xi, \quad \text{and} \quad p q = \lb \xi -\frac12 \rb^2 =
h_1 h_2 . 
\ee

\subsubsection*{An alternate way to compute the resolvent}

This is an important section. We understand that finding an exact
expression for resolvent is important to get distributions of boxes in
Young diagram. In the last section we performed the contour
integration explicitly and found $H(h)$. However, the expression for
resolvent becomes complicated, as we shall see later, when variable
$h$ becomes discontinuous between $h_L$ and $h_U$ and hence solving
the contour integration turns out to be difficult. In this section we
explain how one can obtain the above result for resolvent from its
asymptotic properties without doing any contour integration.

We take the following ansatz for resolvent $H(h)$ following
\cite{lensspaces}
\begin{eqnarray}\label{eq:generic-ans-H}
H(h)=2\ln\lB\frac{g(h)-\sqrt{g( h)^{2}-f(h)^{2}}}{\kappa \ v(h)}\rB
\end{eqnarray}
where $g(h), \ f(h)$ and $v(h)$ are polynomials of $h$ and $\kappa$ is
constant.

If $H(h)$ has single branch cut between $p$ and $q$ then, $g^2-f^2$
can at most be polynomial of degree 2, hence $g$ and $f$ are
polynomial of degree 1 or less\footnote{$g^2-f^2$ can have a form like
  $w(h)(h-p)(h-q)$. In that case we absorb extra factor of $w(h)$ in
  $v(h)$ by re-definition.}.

From the property (d) of $H(h)$ given above one can show that,
\begin{eqnarray}\label{eq:property-d}
  {\lim \atop \e \ra 0} \lB H(h+i\epsilon)+H(h-i\epsilon)\rB &=& 2\ln\lB
  \frac{g(h)-i\sqrt{f^{2}(h)-g^{2}(h)}}{\kappa \ v(h)}\rB + 2 \ln\lB
  \frac{g(h)+i\sqrt{f^{2}(h)-g^{2}(h)}}{\kappa \ v(h)}\rB \notag\\
  &=& 2 \ln\lB\frac{f^{2}(h)}{\kappa^2 v^2(h)}\rB=2
  \ln\bigg[\frac{h}{\xi}\bigg]. 
\end{eqnarray}
Therefore we find,
\be f(h) = \frac\kappa {\sqrt{\xi}} \sqrt{h}\ v(h).  \ee
From the asymptotic expansion of $H(h)$, given in property (c), we
find
\ben
g(h) = h+\frac{2\xi-1}{2}, \quad v(h)= \frac{2\xi}{\kappa},\quad
\text{and} \ f(h) = 2 \sqrt{\xi \ h}
\een
and 
\be
\xi = k'+\frac14.
\ee
Substituting the values of these functions in
eqn. (\ref{eq:generic-ans-H}) we obtain the expression for $H(h)$
\be 
H(h)=2\ln\lB \frac{1 -2 h-2 \xi+\sqrt{4 h^2-4 h (2 \xi +1)+(1-2
    \xi )^2}} {4 \xi } \rB 
\ee
which matches with eqn. (\ref{eq:H-1a}). Since the resolvent has a
branch cut between $p$ and $q$, therefore
\be
 g^2-f^2=h^2-4 h (2 \xi +1) +(1-2
    \xi )^2 = (h-p)(h-q).
\ee
Solving this equation one finds eqn. (\ref{eq:pqvalue1a}). Thus, we
recover all the essential relations including the expression for the
resolvent without solving the contour integration.

\subsubsection{ \class {1b}}

\begin{figure}[h]
\centering
\includegraphics[width=8cm,height=6cm]{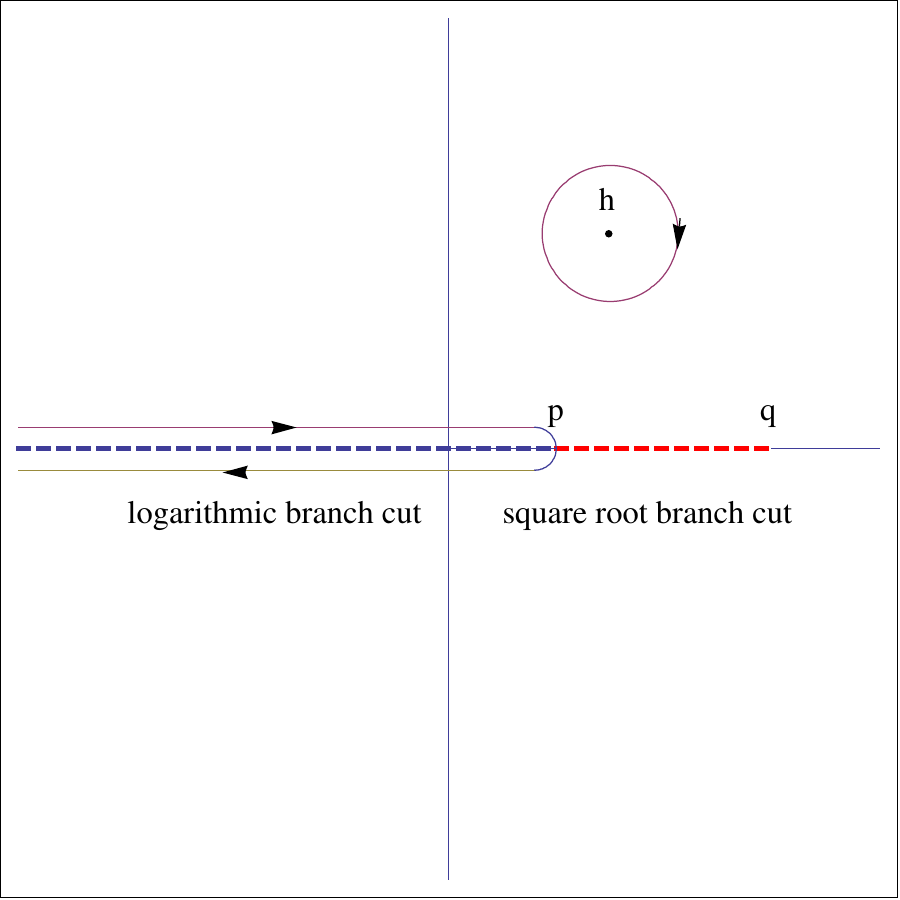}
\caption{Contour for Solution \class{1b} }
\label{cont_b2}
\end{figure}

Using the ansatz (\ref{eq:ansatz-class1a}) for $u(h)$, for this class
the saddle point equation becomes,
\be \label{eq:sadeqbra2}
\Xint-_{p}^{q} dh' {\tilde u(h') \over h - h'}
= \ \ln \lB {h \over \xi} \rB - \ln \lB {h \over h- p } \rB
{\hskip .5cm} \text{ where,} \ h \in [p, q] \ .
\ee
The resolvent as defined in eqn. (\ref{eq:defH}) becomes (with $h_L=0$
and $h_U=q$),
\be
H(h) = \ln \lB {h \over h-p} \rB + \int_{p}^{q} dh'  {\tilde
u(h') \over h  -  h'} \ .
\ee 
Again, one can write an expression for $H(h)$ depending on its
analytic properties. It is given by,
\be\label{Hbra2}
H(h) = \ln \lB {h \over h-p} \rB - \sqrt{(h - p) \ (h - q)}
\oint_{\cC} {ds \over 2 \pi i} { \ln \lb s/\xi \rb - \ln \lB s/( s
- p)\rB \over (s - h) \sqrt{(s - p) \ (s - q)}}\ .
\ee
The contour is shown in fig \ref{cont_b2}. 
However, we shall not perform this contour integration to find $H(h)$,
rather we shall follow the technique defined in the last section.

Let us write
\be 
H(h) = \ln \lB {h \over h-p} \rB + H_1(h), \qquad H_1(h) =
\int_{p}^{q} dh' {\tilde u(h') \over h - h'} \ .  
\ee
We take the ansatz for $H_1(h)$ as before,
\begin{eqnarray}\label{eq:generic-ans-H1}
H_1(h)=2\ln\lB\frac{g(h)-\sqrt{g( h)^{2}-f(h)^{2}}}{\kappa \ v(h)}\rB .
\end{eqnarray}
It follows from eqn. (\ref{eq:sadeqbra2}) that $H_1(h)$ satisfies,
\begin{eqnarray}\label{eq:property-d-H1}
  {\lim \atop \e \ra 0} \lB H_1(h+i\epsilon)+H_1(h-i\epsilon)\rB =2
  \ln\bigg[\frac{h-p}{\xi}\bigg]. 
\end{eqnarray}
Hence, from eqn. (\ref{eq:property-d-H1}) and asymptotic properties of
$H(h)$ we find that,
\be
g(h) = h + \frac12 (2\xi -1- p), \quad v(h) = \frac{2\xi}{\kappa},
\quad f(h) = 2\sqrt{\xi(h-p)}
\ee
and
\be \label{eq:k'-1b}
4\xi (1-p) -(1-p)^2 = 4 k'.
\ee
Plugging the values of these functions in to the ansatz we find,
\be 
H(h) = \ln\lB \frac{h \left(1-2 h-2 \xi +p+2
    \sqrt{(h-\frac12-\frac{p}2)^2-\xi (2 h-3 p+1) 
      + \xi ^2}\right)^2}{16 \xi ^2 (h-p)}\rB.   
\ee
Since, $H(h)$ has a square root branch cut between $p,\ q$, we find
that the term inside the square root can be written as $(h-p)(h-q)$, with
\be
p= 1-2\xi, \quad q=1+2\xi .
\ee
Hence, further simplifying $H(h)$ we finally get,
\be
H(h) = \ln \lB \frac{h \left(h-1-\sqrt{(h-1)^2 -4 \xi ^2}\right)}{2
  \xi ^2} \rB
\ee
which matches with the expression given in \cite{Dutta:2007ws}.

Since $p=1-2\xi$ and $p>0$, therefore this branch exists for
$\xi\leq\frac12$. From eqn. (\ref{eq:k'-1b}) we find, 
\be\label{k-vr2}
k' = \xi^2 \ .  
\ee 
From the definition $a_1k'=\xi^2$ we obtain
\ben\label{a1-vr2}
\text{ either} \ \xi &=&0 \notag\\
\text{ or}, a_1 &=& 1.  
\een 
The former implies the uniform distribution 
\be\label{unfrm} 
u(h)=1 \
\ \ \ h \in [0,1].  
\ee 
This is therefore a saddle point for any value of $a_1$. This is in
fact the density corresponding to the trivial representation $n_i=0$.
The latter corresponds to a family of saddle points labeled by
$0<\xi<1/2$ which exists only at $a_1=1$.

Therefore, we see that \class{1b} corresponds to no-cut
branch \type{1} discussed in section (\ref{sec:no-cut})

From the discontinuity of the resolvent we find the Young Tableaux
distribution for this branch is given by,
\be \label{eq:usol2b}
\tilde u(h) =  {1 \over \pi}  \cos^{-1} \lB {h-1 \over 2
\xi} \rB \ .
\ee
Hence, $h$ can be written in terms of $\pi \tilde u(h)$ as,
\be
h= 1+2\xi \cos\pi \tilde u(h).
\ee

\subsubsection{ \class 2 : \bf{ discontinuous support}}

For solution class 2 $h(x)$ is discontinuous and $u(h)$ is non-zero
for $h\in[q,p],[r,s]$. The saddle point equation, for this class, is
given by (using ansatz (\ref{eq:ansatz-u-2})),
\be
\Xint -_p^q \frac{\tilde u_1(h')}{h-h'} dh'+\Xint -_r^s \frac{\tilde
  u_2(h')}{h-h'} dh' = \ln\lB\frac{h}{\xi}\rB, \qquad p<q<r<s.
\ee
We define the resolvent for this case 
\begin{eqnarray}
H(h)=\int^q_p dh'\frac{\tilde u_1(h')}{h-h'}+\int^s_r dh' \frac{\tilde
  u_2(h')}{h-h'}.
\end{eqnarray}
The resolvent has the following properties.
\begin{enumerate}[(a)]

\item It is a analytic function of $h$ in the complex plane with two
  branch cuts from $p$ to $q$ and $r$ to $s$ on positive real axis.

\item It is real for other values of the positive real $h$.

\item 
$H(h)\sim \frac{1}{h}+(k'+\frac{1}{2})\frac{1}{h^{2}}$ as $h\to
  \infty$.

\item 
  ${\lim \atop \e \ra 0} \lB H(h+i\epsilon) + H(h-i\epsilon)\rB =2
  \ln\lB\frac{h}{\xi}\rB$ for real $h\in[p,q],[r,s]$, which follows
  from the definition of a complex function with a branch cut on the real
  line.

\item $\tilde u_{1(2)}(h)= {\lim \atop \e \ra 0} -\frac{1}{2\pi
    i}[H(h+i\epsilon)-H(h-i\epsilon)]$ for $h\in[p,q] ([r,s])$.

\end{enumerate}
Derivation of properties (d) and (e) has been discussed in appendix
\ref{app:hprop}.

\begin{figure}[h]
\centering
\includegraphics[width=8cm,height=6cm]{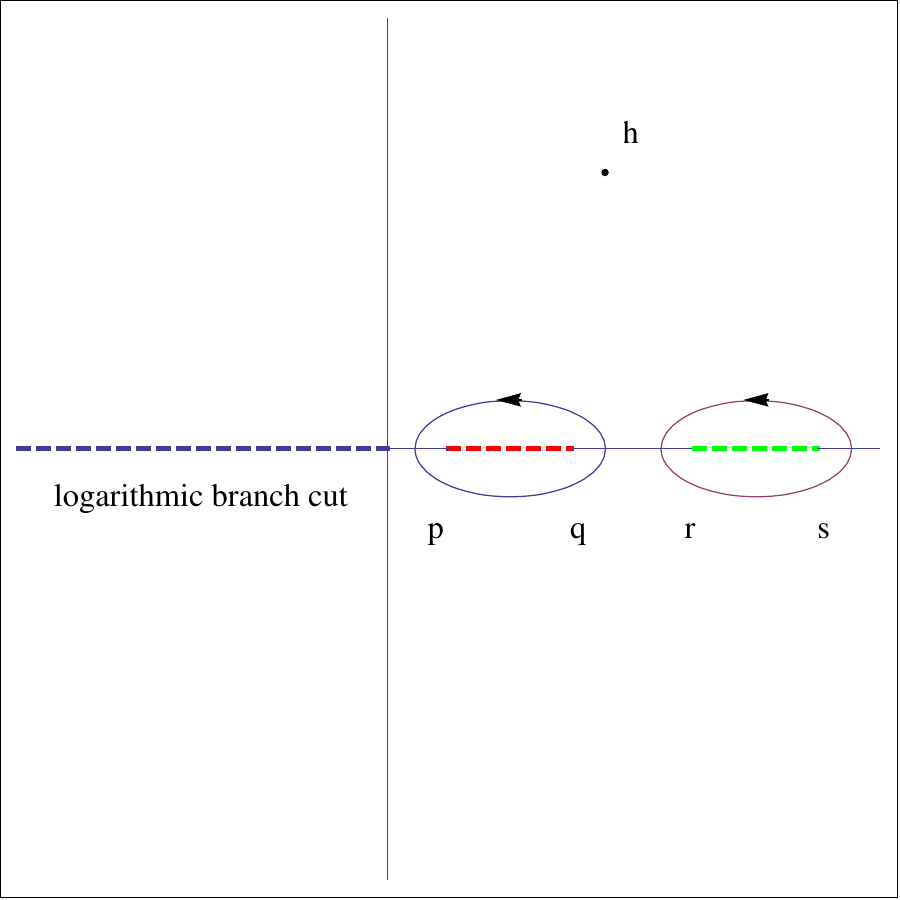}
\caption{Contour for Solution Class 2}
\label{cont-2}
\end{figure}

The resolvent $H(h)$ for this class can be constructed analogously
\begin{eqnarray}
H(h)=-\sqrt{(h-p)(h-q)(h-r)(h-s)}\oint_{\cC}\frac{dz}{2\pi i}
\frac{\ln(z/\xi)}{(z-h)\sqrt{(z-p)(z-q)(z-r)(z-s)}} 
\end{eqnarray}
where the contour $\cC$ is defined anticlockwise around each of the
branch cuts $[p,q]$ and $[r,s]$. See figure \ref{cont-2}.

The contour can be deformed and can be taken around the point $h$,
infinity and along the branch cut of the logarithm, as before. But the integral
obtained in this manner is extremely tedious. Instead we use an
alternate method.

\subsubsection*{Alternate method to find resolvent in \class 2}

We take the following ansatz for $H(h)$
\begin{eqnarray}
H(h)=2\ln\lB\frac{g(h)-\sqrt{g(h)^{2}-f(h)^{2}}}{\kappa \ v(h)}\rB.
\end{eqnarray}
Since $H(h)$ has two branch cuts between $[p,q], \ [r,s]$ therefore,
$\sqrt{g(h)^{2}-f(h)^{2}} = \sqrt{(h-p)(h-q)(h-r)(h-s)}$ up to overall
normalization. Hence, $g(h)$ is a quadratic polynomial and $f(h)$ is
quadratic at most.

From property (d) of $H(h)$ we find,
\begin{eqnarray}\label{property1}
  H(h+i\epsilon)+H(h-i\epsilon)&&=2
  \ln\lB\frac{g(h)-i\sqrt{f(h)^{2}-g(h)^{2}}}{\k \ v(h)}\rB +
  2 \ln\lB \frac{g(h)+i\sqrt{f(h)^{2}-g(h)^{2}}}{\k \ v(h)}\rB \notag\\  
  &&=2 \ln\lB\frac{f(h)^{2}}{\k^{2} v(h)^2}\rB = 2
  \ln\lB\frac{h}{\xi}\rB
\end{eqnarray}
which implies
\be
f(h) = \frac\kappa {\sqrt{\xi}} \sqrt{h}\ v(h).
\ee

From the asymptotic expansion we find that,
\be
g(h) = h^2 + \lb \xi -\frac12 \rb h + b, \quad v(h) = \frac{2\xi}{\k} h
\ee
From the condition, $\sqrt{g(h)^{2}-f(h)^{2}} =
\sqrt{(h-p)(h-q)(h-r)(h-s)}$ we find that,
\ben\label{eq:pqrs-relation}
\begin{split}
&  p q r s = b^2, \\
& p+q+r+s = 2\xi+1, \\
& p(q r+q s+r s) +q r s = -(2\xi -1)b , \\ 
& p(q+r+s)+q(r+s)+r s = 2b+\lb \xi -\frac12\rb^2 .
\end{split}
\een
Solving the first equation we find,
\be
b = - \sqrt{p q r s}.
\ee
We choose the negative solution to get the correct branch. Plugging
the value of $b$ in the third equation above we find,
\be
p(q r+q s+r s) +q r s = \sqrt{p q r s}(2\xi -1).
\ee
Since $p,\ q,\ r,\ s$ are greater than zero, this implies this branch
of solution exists for $\xi\geq \frac12$.

\begin{figure}[h]
\be
\includegraphics[width=8cm,height=5cm]{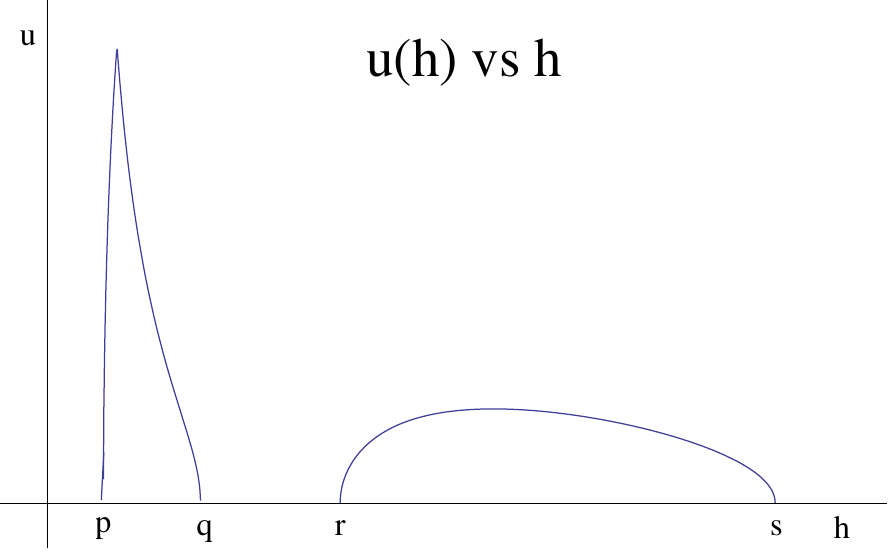}\notag
\ee
\caption{$u(h)$ vs. $h$ for \class 2}
\label{fig:uplot-2}
\end{figure}

Now from the definition of $u(h)$ we have,
\begin{eqnarray}
u(h)&&=-\frac{1}{2\pi i}[H(h+i\epsilon)-H(h-i\epsilon)]\notag\\
&&=-\frac{1}{2\pi i}\lB 2 \ln\lb
\frac{g(h)-i\sqrt{f(h)^{2}-g(h)^{2}}}{\k \ u(h)} \rb - 2 \ln\lb
\frac{g(h) + i \sqrt{f(h)^{2}-g(h)^{2}}}{\k \ u(h)}\rb \rB \notag\\
&&=-\frac{1}{\pi i} \ln\lb \frac{g(h)-i\sqrt{f(h)^{2} -
    g(h)^{2}}}{g(h)+i\sqrt{f(h)^{2}-g(h)^{2}}}\rb.
\end{eqnarray}
This expression can be written in terms of inverse trigonometric
function as
\begin{align}
  \label{density-C2}
  u(h) =&
 \left\{ \begin{array}{rl} 
                    & \frac{2}{\pi} \tan^{-1}\lB
\frac{\sqrt{f(h)^{2}-g(h)^{2}}}{g(h)}\rB\,, \quad \text{for}\ \ g(h) >0\\
                    & 2-\frac{2}{\pi} \tan^{-1}\lB
\frac{\sqrt{f(h)^{2}-g(h)^{2}}}{|g(h)|}\rB \,, \quad \text{for}\ \
g(h) <0 \, 
                    \end{array} \right.\\
= & \quad \frac{2}{\pi} \cos^{-1} \lB \frac{|g(h)|}{f(h)}\rB \, .
\end{align}
Plugging the values of $g(h)$ and $f(h)$ we obtain 
\be
u(h)=\frac2{\pi}\cos^{-1} \lB \frac{|b+h \left(h+\xi
    -\frac{1}{2}\right)|}{2 h^{3/2} \sqrt{\xi }} \rB
\ee
which gives a quartic equation for $h$,
\ben \label{eq:h-eqn-class-2}
h^4 -(1+2\xi \cos(\pi u(h)))h^3 + \lb \lb \xi -\frac12\rb^2 +2b \rb
h^2 + (2\xi-1) b h +b^2 =0.  \een
A typical distribution $u(h)$ has been plotted in figure
\ref{fig:uplot-2}. We note that the distribution is zero between $q$
and $r$.

Let us denote the solution of this equations by $h_1,\ h_2,\ h_3$ and
$h_4$. It follows from the properties of quartic equations that,
\ben \label{eq:h1h2-relation}
\begin{split}
&h_1+h_2+h_3+h_4 = 1+2\xi \cos(\pi u(h))\\
& h_1(h_2+h_3+h_4)+h_2(h_3+h_4)+h_3 h_4 = \lb \xi -\frac12\rb^2 +
2b\\
& h_1 (h_2 h_3 +h_2 h_4 +h_3 h_4) + h_2 h_3 h_4 = -(2\xi-1)b\\
& h_1 h_2 h_3 h_4 = b^2.  
\end{split}
\een

Comparing these relations with eqn. (\ref{eq:root-relation-1a}) and
(\ref{eq:pq-root-1a}) we see that in this case also sum over roots are
same as $1+2\xi \cos(\pi u(h))$. Also it is interesting to note that,
if we set $r=s=0$ then we get back  \class {1a}. For $r=s=0$,
$b=0$ and we find $p+q = 1+2\xi$ and $p q = \lb \xi -\frac12
\rb^2$. All other equations in (\ref{eq:pqrs-relation}) are trivially
satisfied.

Finally, comparing the coefficient of $\frac1{h^2}$ term in asymptotic
expansion of $H(h)$ we find that,
\ben \label{eq:b-k'}
&&\xi +\frac14 -2b = k'+\frac12 \notag\\
&\Rightarrow& b = \frac{\xi}{2} -\frac18 -\frac{\xi^2}{2a_1}.
\een
Substituting the value of $b$ in the last relation or
eqn. (\ref{eq:pqrs-relation}) we find,
\be\label{eq:h1h2-and}
p(q+r+s)+q(r+s)+r s =h_1(h_2+h_3+h_4)+h_2(h_3+h_4)+h_3 h_4 = \xi^2 \lb
1-\frac1{a_1}\rb. 
\ee
Since $p,\ q,\ r,\ s > 0$, the above relation implies that this branch
is possible for $a_1> 1$. Also, since $b=-\sqrt{p q r s}<0$, from
eqn. (\ref{eq:b-k'}), we see that $a_1$ has an upper bound as well
\be
a_1 < \frac{4\xi^2}{4\xi -1}.
\ee
Thus, we identify this branch to \type 3 branch in the eigenvalue
side.\\

\noindent
{\bf A dictionary :}\ \ Comparing the equations in section
\ref{sec:amodel} and this section we find that different parameters on
both sides are related in the following way.
\be
\xi =\b_1, \qquad \Rp =\frac{k'}{\xi}.
\ee
It is easy to check that, using the above dictionary one can obtain
the equations between the parameters (saddle point conditions) in one
side from the equations on other side for $a_1$ model.

\section{Relation between the Young Tableaux and Eigenvalue
  distributions} \label{sec:identification}

The analysis of the different phases of gauge theories in terms of
dominant representations (or Young diagrams) of $U(N)$ groups is very
different from the usual eigenvalue analysis reviewed discussed in
section \ref{sec:amodel}.  However, it was observed in
\cite{Dutta:2007ws} that there is nevertheless, a simple relationship
between the saddle point configurations $u(h)$, in both the high and
low temperature phases, with the corresponding saddle point eigenvalue
densities.

We shall first review the relationship between \type 1 and
\class{1b} and \type 2 and \class{1a} as
described in \cite{Dutta:2007ws}.

\subsection{Identification between \type 1 and  \class{1b}}

Consider first the low temperature saddle point $u(h)=1, \ \xi =0$. The
corresponding saddle point for the eigenvalue density is
$\sigma(\theta)={1\over 2\pi}, \ \b_1=0$. From the form of these two
distributions, we notice that they are functional inverses of each
other. Therefore, we can make the identification
\be\label{eq:rel1} 
\begin{split}
u &= {\theta \over \pi} \\
{h \over 2 \pi} &=
\sigma(\theta)\ .  
\end{split}
\ee

This case is rather trivial. Lets consider the phases for $\b_1<1/2$
and $\xi < 1/2$. The Young diagram density is given by,
\be
u(h)={1 \over \pi}  \cos ^{-1} \lB {h-1 \over 2
\xi} \rB \quad h \in [p, q],  \qquad u(h) = 1,
\quad h \in [0,p]; \ \ 2\xi \leq 1.
\ee
Using the identification defined above we find,
\be \label{eq:h-theta-eqn-1b}
h= 1+2\xi \cos\theta
\ee
and hence,
\be
\sigma(\theta)={1\over
2\pi}(1+2\xi \cos\theta)
\ee
which matches with the eigenvalue distribution
(\ref{eq:specden-case1b}) with $\b_1=\xi$.

\subsection{Identification between \type 2 and  \class{1a}}

In this case the identification is more subtle. In this case we modify
the identification (\ref{eq:rel1}) to,
\be\label{eq:identification-onecut}
\begin{split}
  u &=
{\theta \over \pi} \qquad \text{and}\\
 \sigma(\theta) &=\frac{\sqrt{(h_1+h_2)^2-4h_1h_2}}{2\pi} 
= \frac{h_1-h_2}{2\pi} 
\end{split}
\ee 
where, $h_1$ and $h_2$ are two roots of eqn. 
(\ref{eq:h-eqn-class-1a}) with $h_1>h_2$. Using the identification
between $u$ and $\theta$ one can write from
eqn. (\ref{eq:h-eqn-class-1a})
\be \label{eq:h-theta-eqn-1a}
h^2 -(1+2\xi \cos \theta)h + \lb \xi -\frac12\rb^2=0.
\ee
Calculating the roots of this equation we find
\be
\frac{\sqrt{(h_1+h_2)^2-4h_1h_2}}{2\pi} = \frac{2\xi}{\pi} \sqrt{\sin^2\frac{\theta_0}{2} -
  \sin^2\frac{\theta}{2}} \cos\frac{\theta}2, \qquad \sin^2
\frac{\theta_0}{2} = \frac1{2\xi}.
\ee
Thus we find,
\be
\s(\theta) =\frac{2\xi}{\pi} \sqrt{\sin^2\frac{\theta_0}{2} -
  \sin^2\frac{\theta}{2}} \cos\frac{\theta}2,
\ee
which matches with eqn. (\ref{eq:specden-case2}) with $\b_1=\xi$.

\subsection{Identification between \type 3 and \class 2}

Following the identification between \type 2 and \class {1a} we now
generalize the identification between \type 3 and \class 2. Note that
for \class 2, $h$ satisfies a quartic equation
(\ref{eq:h-eqn-class-2}) with four roots denoted by $h_1, h_2, h_3$
and $h_4$. Since, in the limit $b=0$ \class 2 boils down to \class
{1a}, we expect our identification in this section will also reduce to
eqn. (\ref{eq:identification-onecut}).

We consider the following identification,
\be
\label{eq:identification-twocut}
\begin{split}
  u &=
{\theta \over \pi} \qquad \text{and} \\
 \sigma(\theta) &=\frac{\sqrt{(h_1+h_2+h_3+h_4)^2-
4\lb h_1 (h_2+h_3+h_4)+h_2(h_3+h_4)+h_3h_4\rb}}{2\pi} \, .
\end{split}
\ee
Using the above identification between $u$ and $\theta$ we find from
eqn. (\ref{eq:h-eqn-class-2}) 
\ben \label{eq:h-theta-eqn-2}
h^4 -(1+2\xi \cos\theta)h^3 + \lb \lb \xi -\frac12\rb^2 +2b \rb
h^2 + (2\xi-1) b h +b^2 =0.  \een
Finally, from the properties of roots of the above equation and
eqn. (\ref{eq:h1h2-and}),
\be 
\sigma(\theta) = \frac{2 \xi}{\pi} \sqrt{\lb \sin^2\frac{\theta_1}{2} -
  \sin^2\frac{\theta}{2} \rb \lb \sin^2\frac{\theta_2}{2} -
  \sin^2\frac{\theta}{2} \rb}
\ee
where,
\be
\theta_{1} = \cos^{-1}\lB -\frac{1}{2\xi} + \sqrt{1-\frac1{a_1}}\rB,
\quad
\theta_{2} = \cos^{-1}\lB -\frac{1}{2\xi} - \sqrt{1-\frac1{a_1}}\rB .
\ee
Therefore, this expression exactly matches with eqn. (\ref{eq:specden-case3})
with $\b_1=\xi$.

\section{Fermionic phase space}\label{sec:phsespace}

The identification between the set $(h, u(h))$ and $(\theta,
\sigma(\theta))$ at the saddle points has a very natural
interpretation, as pointed out in \cite{Dutta:2007ws}, in terms of a
free fermionic picture. The fermionic picture follows from the
following two facts.
\begin{itemize} 
\item The eigenvalues $\theta_i$s of holonomy matrix $U(N)$ behave
  like coordinates of free fermions.
\item The number of boxes $n_i$s (or $h_i$s) in a Young diagram being
  like momentum of non-interacting fermions \cite{douglas}.
\end{itemize} 
The eigenvalue distribution $\s(\theta)$ tells how the position
coordinates of $N$ fermions are distributed whereas, $u(h)$ describes
the distribution of momenta of fermions.

\cite{Dutta:2007ws} considered a phase space distribution which gives
rise to these individual distributions. In the classical (i.e. large
$N$) limit, one can describe this system of $N$ fermions in terms of
an incompressible fluid occupying a region of the two dimensional
phase space.

Therefore, different saddle points are described by some configuration
or region $\Gamma$ in phase space with phase space density function
$\o(h,\theta)$ defined in a two dimensional plane spanned by $h$ and
$\theta$. $h$ is a radial direction and $\theta$ is angular. The phase
space density has the property
\ben
\o(h, \theta) &=& \frac1{2\pi}, \quad \text{for}
(h,\theta) \in \Gamma,\notag\\ 
&=& 0 , \quad \text{otherwise}.
\een
Momentum and position distributions are defined in terms of phase
space density as follows,
\ben\label{defden} 
u(h) &=&
\int_{-\pi}^{\pi} \o(h,\theta)d\theta \notag\\
\sigma(\theta) &=&
\int_0^{\infty} \o(h,\theta)d h 
\een 
where the first integral is at constant $h$ and the second at constant
$\theta$.  Also note that
\be \label{mesurefac}
\int
\o(h,\theta)dh d\theta =1\ .
\ee
It follows from the normalization of two partial
densities\footnote{Actual polar coordinate related to $h$ by
  $h={r^2\over 2}$.}.

Our next goal is to find out the boundary of the region $\Gamma$ for
different saddle points. We assume that the boundary is specified by
two curves $h_+({\theta})$ and \hmt\, where,
\be
h_+(\theta) +h_-(\theta) = \sum_i h_i(\theta), \qquad \hp \hmm =
\sum_{i,j\atop{i\neq j}} h_i h_j
\ee
where $h_i$'s are roots of the equations satisfied by $h$ for
different classes.

\begin{figure}[h]
\begin{subfigure}{.3\textwidth}
  \centering
  \includegraphics[width=5cm,height=5cm]{ps-1b1}
  \caption{$\xi=0$}
\end{subfigure}%
\begin{subfigure}{.3\textwidth}
  \centering
  \includegraphics[width=5cm,height=5cm]{ps-1b3}
  \caption{$0<\xi<1/2$}
\end{subfigure}%
\begin{subfigure}{.3\textwidth}
  \centering
  \includegraphics[width=5cm,height=5cm]{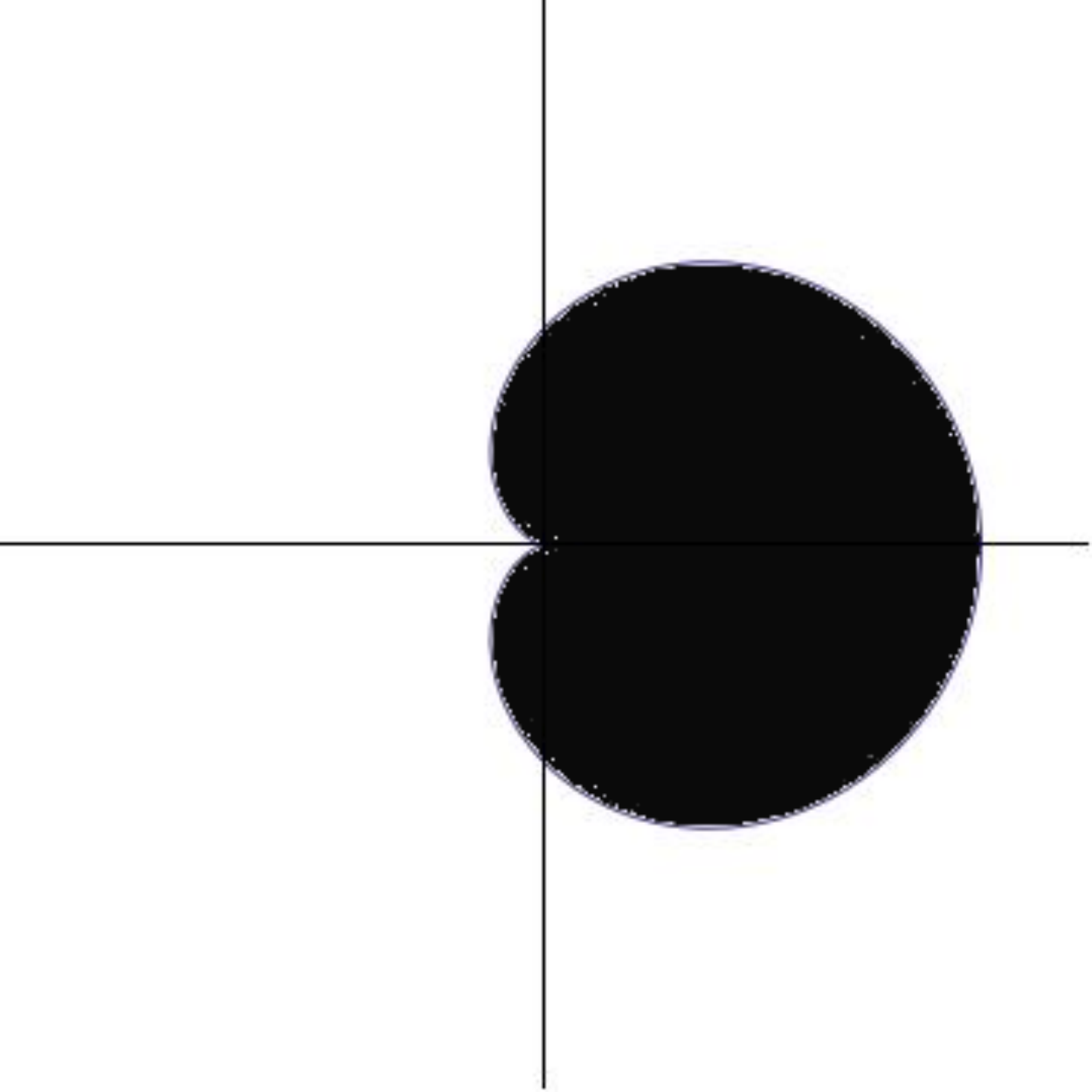}
  \caption{$\xi=1/2$}
\end{subfigure}%
\caption{Phase space distribution for \class {1b}.} 
\label{fig:ps-1b}
\end{figure}

For \class {1b} the governing equation is given by
eqn. (\ref{eq:h-theta-eqn-1b}). This is a linear equation and has one
solution. Therefore, $\hp=h_1$ and $\hmm=0$. This distribution in
phase space is therefore drawn in figure \ref{fig:ps-1b}.

For \class{1b} the governing equation is quadratic
(\ref{eq:h-theta-eqn-1a}). It has two solutions $h_1$ and
$h_2$. Therefore, $\hp+\hmm = h_1+h_2$ and $\hp \hmm = h_1 h_2$.
Hence,
\be\label{eq:ev-dis-ps}
\s(\theta)=\frac{\hp-\hmm}{2\pi}.
\ee
%
\begin{figure}[h]
\centering
\includegraphics[width=5cm,height=5cm]{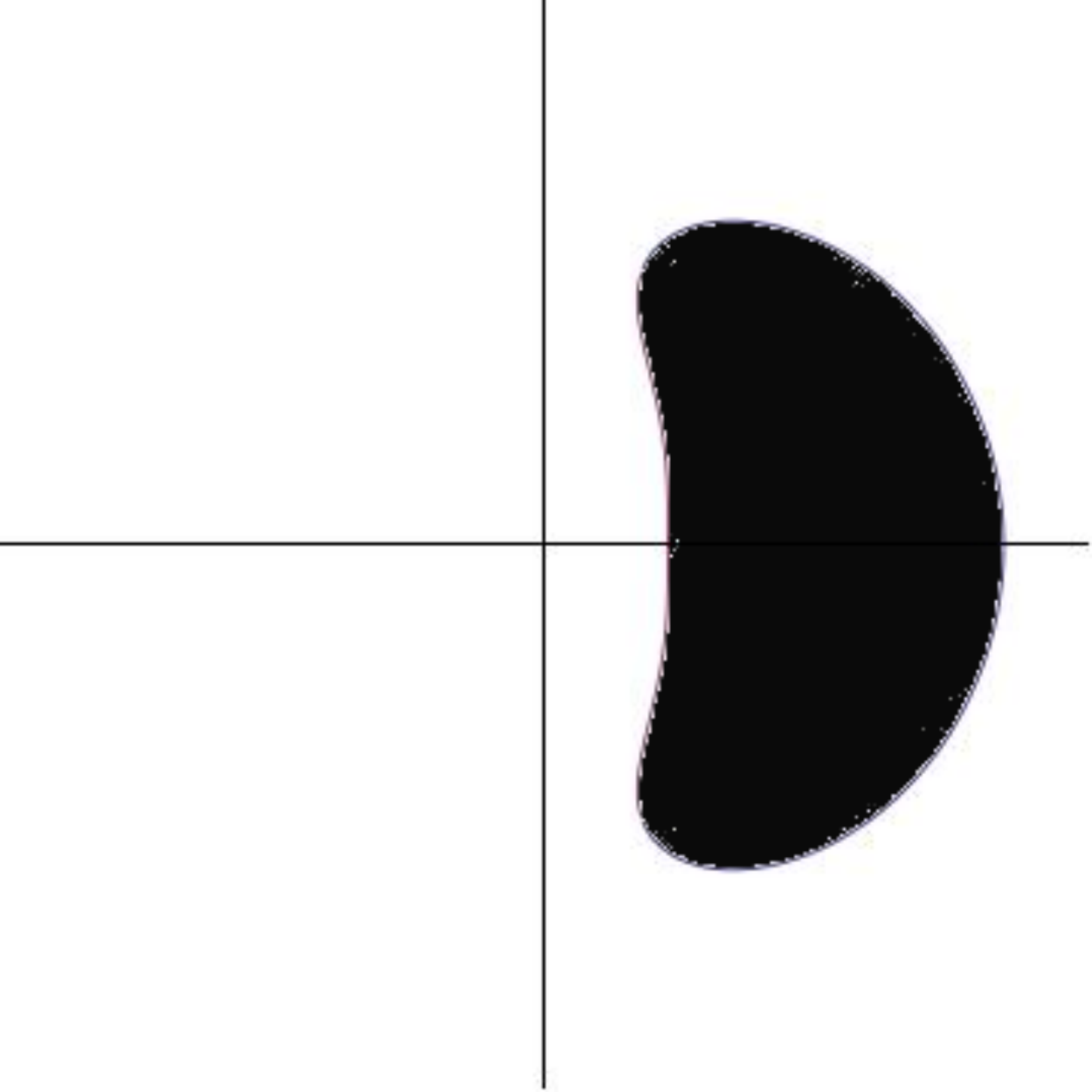}
\caption{Phase space distribution for \class {1a}}
\label{fig:ps-1a}
\end{figure}
The distribution is depicted in figure \ref{fig:ps-1a}.

\begin{figure}[h]
\centering
\includegraphics[width=5cm,height=5cm]{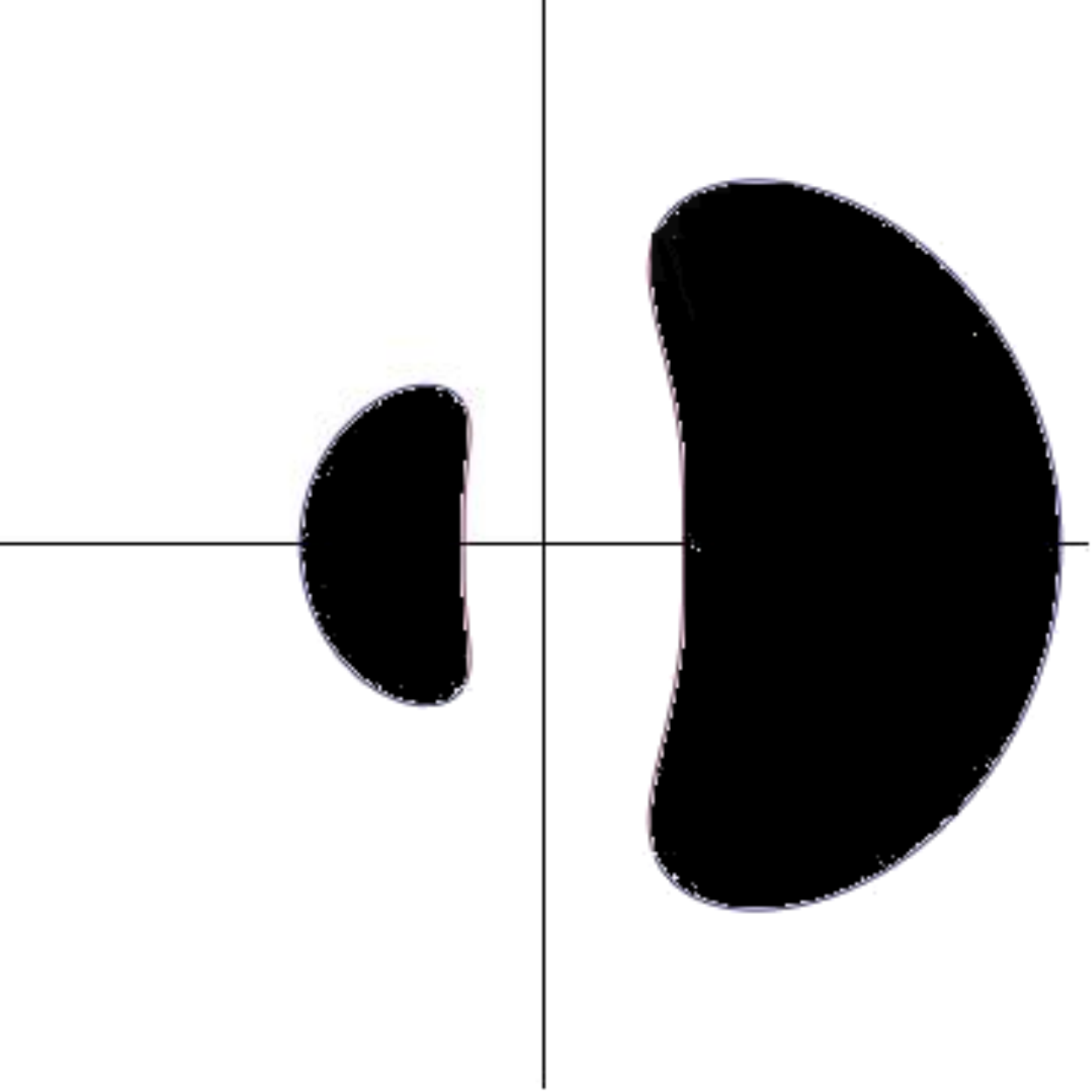}
\caption{Phase space distribution for \class 2. For the left island
  $\hp$ and $\hmm$ becomes negative. However, their difference is
  always positive. Hence we plot absolute value of $\hp$ and $\hmm$
  when they are real.}
\label{fig:ps-2}
\end{figure}

For  \class 2 the distribution is nontrivial. Here, we take,
\be
\hp+\hmm = h_1+h_2+h_3+h_4, \qquad \hp \hmm =
h_1(h_2+h_3+h_4)+h_2(h_3+h_4)+h_3 h_4.
\ee 
The equation which governs the shape of the fluid droplets is given
by,
\be
\hp^2 -(1+2\xi \cos \theta) \hp + \xi^2\lb 1-\frac1{a_1}\rb =0.
\ee
For this case also eigenvalue distribution is given by
eqn. (\ref{eq:ev-dis-ps}). The phase space distribution is given by
figure \ref{fig:ps-2}.

The topology of the phase space distribution for two-gap solution is
different than the other two branches. In fact, this topology is
robust and does not change under inclusion of weak coupling terms of
the form given in eqn. (\ref{eq:sefftrunc})\footnote{Note that, all
  the terms appearing in this model action correspond to one cycle in
  character expansion of partition function discussed in section
  \ref{sec:YTside}.}. This is because the equation satisfied by $h$
remains quartic even after addition of these terms. Only the
coefficients of different powers of $h$ change. In fact, the generic
$(a,b)$ type model given by eqn. (\ref{eq:sefftrunc}) can have at most
two-gap or two-cut solution and hence the topology of phase space
distribution is also fixed.  However, finding distribution for generic
weak coupling action is difficult \cite{winp}.
\\

\noindent {\bf Acknowledgement :} We wold like to thank Rajesh
Gopakumar for useful discussion and his valuable comments on the
manuscript. We express our sincere gratitude to N. Banerjee,
R. Banerjee, A. Bhand, A. Mangalsuli, S. Sarkar for useful
discussion. We appreciate the effort of S. K. Pathak for arranging
important articles required for this work. PD would like to
acknowledge the hospitality of IISER Pune where a part of this work
was done. Finally, we are grateful to people of India for their
unconditional support towards researches in basic sciences.

\appendix

\section{Derivation of Dyson-Schwinger equation for unitary matrix
  model}\label{sec:ds-eqn}

We use the technique introduced in \cite{Friedan:1980tu} to derive the
Dyson-Schwinger equation. Firstly the generic partition function that we start with is as follows:
\begin{eqnarray}
Z=\int [dU]\exp \lb\sum_{\vec n} \frac{a_{\vec n}}{N^{k}}
\prod_{i=1}^{k} \Tr [U^{n_{i}}]\rb
\end{eqnarray}
with $\sum_{i}n_{i}=0$ and the constants $a_{\vec{n}}$, are such that the action is invariant under $U\to U^{-1}$. Next let us look at the following expectation
value
\begin{eqnarray}
\< N^{-1} \Tr[X(1-zU)^{-1}] \> =\int [dU]\,N^{-1} \Tr[X(1-zU)^{-1}] \
\exp\lb\seff U\rb, 
\end{eqnarray}
where $X$ is a skew adjoint $N\times N$ matrix. We change the
integration variable $U \ra \ex^{t X} U$. 
Therefore we get,
\begin{eqnarray}
\< N^{-1} \Tr[X(1-zU)^{-1}] \> =\int [d\ex^{t X} U]\,N^{-1} \Tr[X(1-z
\ex^{t X} U)^{-1}] \
\exp\lb\seff{\ex^{t X} U}\rb, 
\end{eqnarray}
The Haar measure is invariant
under this transformation by definition.:
\be
[d \ \ex^{t X} U] = [dU] 
\ee
That is because under this transformation we have the action on the Haar measure:
\begin{eqnarray}
\prod_{i}\int^{\pi}_{-pi}d\theta_{i}\prod_{i<j}\sin^{2}\bigg(\frac{\theta_{i}-\theta_{j}+\theta'_{i}-\theta'_{j}}{2}\bigg)
\end{eqnarray}
Under redefinition of each of the integration variables $\theta_{i}\to\theta_{i}-\theta'_{i}$, the measure remains invariant i.e. $d\theta_{i}$ remains invariant and thus the Haar measure remains invariant. 

Now the right hand side is independent of $t$, therefore we can write,
\begin{eqnarray} \label{DS1}
&&\frac{d}{dt}\int [d U]
N^{-1}\Tr[X(1-z \ex^{tX}U)^{-1}] \exp \lb\sum_{\vec n} \frac{a_{\vec n}}{N^{k}}
\prod_{i=1}^{k} \Tr [\ex^{n_{i}tX}U^{n_{i}}]\rb\Bigg|_{0}=0\notag 
\end{eqnarray}
We use the following result to simplify the above expression.
\ben
&&\frac{d}{dt}\lB (1- z\ex^{t X} U)(1-z\ex^{t X} U)^{-1}\rB \Bigg
|_{t=0}= \frac{d}{dt} \lB\mathbb{1}\rB = 0  \nn \\
&\Ra & (-z X \ex^{tX} U) (1-z \ex^{t X} U)^{-1} + (1-z \ex^{t X} U)
\frac{d}{dt} (1-z \ex^{t X} U)^{-1}\Bigg |_{t=0} =0 \nn\\
&\Ra& \frac{d}{dt} (1-z  \ex^{tX}U)^{-1} \Bigg|_{t=0}= (1-z  U)^{-1} (z X U) (1-z  U)^{-1} 
\een
which implies,
\ben
&&\frac{d}{dt} \lB X(1-z \ex^{t X} U)^{-1} \rB\bigg|_{t=0}= X (1-z  U)^{-1} (z X U) (1-z
U)^{-1} \nn\\
&\Ra& \frac{d}{dt} \Tr\lB X(1-z \ex^{t X} U)^{-1} \rB\bigg|_{t=0} = \Tr\lB X (1-z  U)^{-1}
(z X U) (1-z U)^{-1}\rB
\een
Therefore from eqn. (\ref{DS1}) we get,
\ben\label{DS2}
&&\int [dU]\bigg[[N^{-1}\Tr[X(1-zU)^{-1}zXU(1-zU)^{-1}]\nn\\
&+&N^{-1}\ Tr[X(1-zU)^{-1}]\sum_{\vec n}a_{\vec n}\frac{1}{N^{k}}\sum_{i=1}^{k}
\prod_{{j=1}\atop{j\neq i}}^{k}\Tr[U^{n_{j}}]n_{i}\Tr[XU^{n_{i}}]\bigg]\exp[\seff{U}]=0
\een
Eqn. (\ref{DS2}) holds for any $X$. There are $N^2$ independent
$N\times N$ anti-hermitian matrices. We write eqn. (\ref{DS2}) for
each such independent $X$ and add up to get
\ben\label{DS3}
&&\sum_a\int [dU]\bigg[[N^{-1}\Tr[X^a(1-zU)^{-1}zX^aU(1-zU)^{-1}]\nn\\
&+&N^{-1}\ Tr[X^a(1-zU)^{-1}]\sum_{\vec n}a_{\vec n}\frac{1}{N^{k}}\sum_{i=1}^{k}
\prod_{{j=1}\atop {j\neq i}}^{k}\Tr[U^{n_{j}}]n_{i}\Tr[X^aU^{n_{i}}]\bigg]\exp[\seff{U}]=0
\een
The first term can be written as,
\ben\label{1sterm}
\sum_a\int [dU]\bigg[[N^{-1}\Tr[X^a(1-zU)^{-1}&z X^a U&(1-zU)^{-1}]\bigg]
\exp[\seff{U}] =
\sum_a\< N^{-1}\Tr[X^a(1-zU)^{-1}z X^a U(1-zU)^{-1}]\> \nn\\
&=& z N^{-1}\sum_a \sum_{{i,j,k,}\atop{l,m}} X^a_{ij} \lB (1-zU)^{-1}\rB_{jk}
X^a_{kl}U_{lm}\lB (1-zU)^{-1}\rB_{mi} 
\een
Using the following identity for appropriately normalized basis
$\{X_{a}\}$ of the skew adjoint matrices 
$$
\sum_{a}X^{a}_{ij} X^{a}_{kl}=N^{-1}\delta_{li}\delta_{jk}
$$
we have
\ben\label{1sterm2}
& \sum_a&\int [dU]\bigg[[N^{-1}\Tr[X^a(1-zU)^{-1}z X^a U(1-zU)^{-1}]\bigg]
\exp[\seff{U}] \nn\\
&=& \<N^{-1}\Tr[(1-zU)^{-1} ]N^{-1}\Tr[z U(1-zU)^{-1}]\> \nn\\
&=& \<N^{-1}\Tr[(1-zU)^{-1} ]N^{-1}\Tr[-\mathbb 1 +(1-z U)^{-1}]\>
\nn\\
&=& - \<N^{-1}\Tr[(1-zU)^{-1} ]\> + \<\lB N^{-1}\Tr[(1-zU)^{-1} ]\rB^2\>\nn\\
\een
Similarly the second term can be written as,
\ben
&\sum_a&\int [dU]\bigg[ N^{-1}\ Tr[X^a(1-zU)^{-1}]\sum_{\vec n}a_{\vec
  n}\frac{1}{N^{k}}\sum_{i=1}^{k} 
\prod_{{j=1}\atop{j\neq
  i}}^{k}\Tr[U^{n_{j}}]n_{i}\Tr[X^aU^{n_{i}}]\bigg]\exp[\seff{U}]\nn\\
&=& \< N^{-1}\ Tr[X^a(1-zU)^{-1}]\sum_{\vec n}a_{\vec
  n}\frac{1}{N^{k}}\sum_{i=1}^{k} 
\prod_{{j=1}\atop{j\neq
  i}}^{k}\Tr[U^{n_{j}}]n_{i}\Tr[X^aU^{n_{i}}]  \>\nn\\
&=&
N^{-1}\sum_{\vec n}\frac{a_{\vec n}}{N}\frac{1}{N^{k}} 
\sum_{i=1}^{k}\prod_{{j=1}\atop{j\neq
  i}}^{k} \< Tr[U^{n_{j}}]n_{i}Tr[U^{n_{i}}(1-zU)^{-1}]\>
\een
Thus we get,
\ben \label{DS4}
- \<N^{-1}\Tr[(1-zU)^{-1} ]\> + \<\lB N^{-1}\Tr[(1-zU)^{-1} ]\rB^2\> +
N^{-1}\sum_{\vec n}\frac{a_{\vec n}}{N}\frac{1}{N^{k}}  
\sum_{i=1}^{k}\prod_{{j=1}\atop{j\neq
  i}}^{k} \< Tr[U^{n_{j}}]n_{i}Tr[U^{n_{i}}(1-zU)^{-1}]\> =0\notag\\
\een
In the large $N$ limit one can use the factorization i.e. $\langle
f(U)\,g(U)\rangle=\langle f(U)\rangle\langle g(U)\rangle$, where
$f,\,g$ being single trace operators. Using this factorization
eqn. (\ref{DS4}) becomes,
\begin{eqnarray}
R(z)^{2}-R(z)+\lim_{N\to\infty}N^{-1}\sum_{\vec n}\frac{a_{\vec n}}{N}\frac{1}{N^{k}}
\sum_{i=1}^{k}\prod_{{j=1}\atop{j\neq i}}^{k}n_{i}\langle
Tr[U^{n_{j}}]\rangle\langle Tr[U^{n_{i}}(1-zU)^{-1}]\rangle=0
\end{eqnarray}
This is the Dyson-Schwinger equation for this matrix model. One can
again simplify the equation to write the moments of single trace
operators completely in terms of resolvent. From eqn. (\ref{rzexpan})
we can calculate the moments of the single trace of matrix $U$ from
the resolvent
\begin{eqnarray}
\langle
Tr[U^{n_{j}}]\rangle=\frac{N}{|n_{j}|{!}}\partial_{z}^{|n_{j}|}R(z)|_{0}=\frac{N}{2\pi
  i}\oint_{\cC} \frac{R(w)}{w^{n_{j}+1}}dw 
\end{eqnarray}
where the contour has been chosen around the point $w=0$. For $n_j>0$,
one can write,
\begin{eqnarray}
\<\Tr[U^{n_j}(\mathbb 1 -z U)^{-1}]\>&=&\<
\Tr[U^{n_{j}}(\mathbb 1+zU+(zU)^{2}+\cdots)]\>\nn\\
&=&\frac{N}{2\pi 
  i}\oint_{\cC}
\frac{R(w)}{w^{n_{j}+1}}(1+\frac{z}{w}+\frac{z^{2}}{w^{2}}+\cdots)dw\notag\\
&=&\frac{N}{2\pi i}\oint_{\cC}
\frac{R(w)}{w^{n_{j}+1}}\frac{1}{1-\frac{z}{w}}dw\notag\\ 
&=&\frac{N}{2\pi i}\oint_{\cC} \frac{R(w)}{w^{n_{j}}}\frac{1}{(w-z)}dw .
\end{eqnarray}
Since the point $z$ is inside the unit circle, as evident from the taylor expansion of the resolvent, the contour $\cC$ now encompasses the entire unit circle. 

For $n_{j}$ negative we have,
\begin{eqnarray}
U^{-|n_{j}|}(I+zU+(zU)^{2}+\cdots)
&=& \sum_{k=0}^{\infty}z^{k}U^{k-|n_{j}|}\notag\\
&=& \sum_{k=0}^{|n_{j}|-1}z^{k}U^{k-|n_{j}|}+\sum_{k=|n_{j}|}^{\infty}z^{k}U^{k-|n_{j}|}\notag\\
&=&\sum_{k=0}^{|n_{j}|-1}z^{k}U^{k-|n_{j}|}+z^{|n_{j}|} \sum_{k-|n_{j}|=0}^{\infty}z^{k-|n_{j}|}U^{k-|n_{j}|}\notag\\ 
&=& \sum_{k=0}^{|n_{j}|-1}z^{k}U^{k-|n_{j}|}+z^{|n_{j}|}(I-zU)^{-1}
\end{eqnarray}
Taking trace and expectation value on both sides we can write,
\begin{eqnarray}
\<\Tr [U^{-|n_j|} (\mathbb 1 - z
U)^{-1}]\> =
\sum_{k=0}^{|n_{j}|-1}z^{k}\frac{N}{(|n_{j}|-k){!}} \partial_{z}^{|n_{j}|-k}R(z)\bigg|_{0}+z^{|n_{j}|}NR(z). 
\end{eqnarray}
 Therefore, the Dyson-Schwinger equation becomes,
\begin{eqnarray} \label{eq:DSfinal1-app}
R(z)^{2}-R(z)+\lim_{N\to\infty}\sum_{\vec n}\frac{a_{\vec n}}{N^{2}}
\sum_{i=1}^{k}\prod_{{j=1}\atop{j\neq 
  i}}^{k}\frac{1}{|n_{j}|{!}}\partial_{z}^{|n_{j}|}R(z)\bigg|_{0}n_{i}F_{i}=0 
\end{eqnarray}
where,
\begin{eqnarray}
F_{i} &=&\frac{1}{2\pi i}\oint
\frac{R(w)}{w^{n_{i}}}\frac{1}{(w-z)}dw, \ \ \text{for} \ \ n_{i}> 0,\notag\\
&=&\sum_{k=0}^{|n_{j}|-1}z^{k}\frac{1}{(|n_{j}|-k){!}} \partial_{z}^{|n_{j}|-k}R(z)\bigg
  |_{0}+z^{|n_{j}|}R(z),
\ \ \text{for} \ \ n_{i}<0 
\end{eqnarray}
We see that this is an algebraic equation for $R(z)$ in the large $N$
limit. This is a very powerful equation. From the analytic properties
of the resolvent one can study the phase structure of $\cN =4$ SYM
theory.

Redefining $a_{\vec n}/N^{2}=\tilde{a}_{\vec n}$ for simplicity and
taking the large $N$ limit eqn. (\ref{eq:DSfinal1-app}) becomes,
\begin{eqnarray}
R(z)^{2}-R(z)+\sum_{\vec n}\tilde{a}_{\vec n}\sum_{i=1}^{k}\prod_{{j=1}\atop{j\neq
  i}}^{k}\frac{1}{|n_{j}|{!}}\partial_{z}^{|n_{j}|}R(z)|_{0}n_{i}F_{i}=0 .
\end{eqnarray}

\section{Dyson-Schwinger equation for most generic
  model}\label{app:ds-eq-most-generic} 

\subsection{Generic zero coupling}
The partition function for the generic zero coupling case is:
\begin{eqnarray}
Z=\int [dU]\exp\bigg(\sum_{n_{1}}\frac{a_{\{n_{1},-n_{1}\}}}{N^{2}}Tr[U^{n_{1}}]Tr[U^{-n_{1}}]\bigg)
\end{eqnarray}
For this generic zero-coupling model (restricting to order $l$) the resolvent takes the following
form:
\begin{eqnarray}
R(z)=\frac{1}{2}[1-\sum_{i=1}^{l}\beta_{i}[\frac{1}{z^{i}}-z^{i}]+\sqrt{F(z)}]
\end{eqnarray}
Where $F(z)$ can be rewritten after some algebra :
\begin{eqnarray}
  &&F(z)=\bigg[1+\sum_{i=1}^{l}\beta_{i}[\frac{1}{z^{i}}
  +z^{i}]\bigg]^{2}-4\bigg[\sum_{i=1}^{l}\frac{\beta_{i}}{z^{i}}\bigg]
  \bigg[\sum_{i=1}^{l}\beta_{i}z^{i}\bigg]\notag\\  
  &&\quad\quad\quad\quad+4\beta_{1}R'(0)
  +4\beta_{2}[\frac{R'(0)}{z}+\frac{R''(0)}{2}+R'(0)z]+..\notag\\ 
\end{eqnarray}
Where the dots mean similar terms with $\beta_{3}$ and so on. It is
evidently in $z\to\frac{1}{z}$ form. The no-cut solution, following
arguments from earlier is:
\begin{eqnarray}
R(z)=1+\sum_{i}^{l}\beta_{i}z^{i}
\end{eqnarray}
One can easily write down the generic form of the one-cut solution for
zero-coupling model including up to order $l$. The Resolvent in that
case assuming a one-cut solution will be:
\begin{eqnarray}
  R(z)=\frac{1}{2}[1-\sum_{i=1}^{l}\beta_{i}[\frac{1}{z^{i}}-z^{i}]
  +\frac{\beta_{l}}{z^{l}}(z^{2l-1}+\sum_{j=l}^{2l-2}\omega_{j}
  (z^{j}+z^{2l-1-j})+1)\sqrt{z^{2}-2\,\cos(\theta_{0})z+1}]  
\end{eqnarray}
Where $\beta_{l}=\tilde{a}_{+l,-l}R^{l}(0)/(l-1){!}$ and the
coefficients $\omega_{j}$ are obtained by setting $R(0)=1$ and
$\cos(\theta_{0})$ obtained analogously as earlier by equating the two
equations one obtains for a single $\omega_{j}$, (as one had done for
 $l=2$ case). One can also note that the following
equations can be obtained from the relation $R(0)=1$, and so the
coefficient of $z^{-i}$ for $i> 0$ is zero.
\begin{eqnarray}
  &&-\beta_{i}z^{-i}+\beta_{l}\sum_{j=l}^{2l-2}
  \sum_{m=0}^{2}\sum_{n=0}^{\infty}\omega_{j}z^{l-1-j}a_{m}
  z^{m}P_{n}(\cos(\theta_{0}))z^{n}\delta_{l-1-j+m+n,-i}\notag\\  
  &&+\beta_{l}\sum_{m=0}^{2}\sum_{n=0}^{\infty}a_{m}z^{m}P_{n}
  (\cos(\theta_{0}))z^{n}z^{-l}\delta_{m+n-l,-i}=0\notag\\ 
  &&\Rightarrow(-\beta_{i}+\beta_{l}\sum_{j=l}^{2l-2}
  \sum_{m=0}^{2}\omega_{j}a_{m}P_{j+1-l-m-i}(\cos(\theta_{0})))
  +\beta_{l}\sum_{m=0}^{2}a_{m}P_{l-m-i}(\cos(\theta_{0}))=0,\quad 
  1\ge i\ge l-1\notag  
\end{eqnarray}
Along with
\begin{eqnarray}
  &&-\beta_{l}+\beta_{l}=0 \notag\\
  &&\beta_{l}\omega_{l}+\beta_{l}\sum_{j=l}^{2l-2}
  \sum_{m=0}^{2}\sum_{n=0}^{\infty}\omega_{j}z^{l-1-j}a_{m}z^{m}
  P_{n}(\cos(\theta_{0}))z^{n}\delta_{l-1-j+m+n,0}\notag\\  
  &&+\beta_{l}\sum_{m=0}^{2}\sum_{n=0}^{\infty}a_{m}z^{m}
  P_{n}(\cos(\theta_{0}))z^{n-l}\delta_{m+n-l,0}=1\notag\\ 
  &&\Rightarrow\beta_{l}\omega_{l}+\beta_{l}\sum_{j=l}^{2l-2}
  \sum_{m=0}^{2}\omega_{j}a_{m}P_{j+1-m-l}(\cos(\theta_{0}))
  +\beta_{l}\sum_{m=0}^{2}a_{m}P_{l-m}(\cos(\theta_{0}))=1  
\end{eqnarray}
Where $a_{0}=1$, $a_{1}=-2\cos(\theta_{0})$, $a_{2}=1$. Here we have
used the formula for the generating function of the Legendre
polynomials. Let us now write down the spectral density function,
$\frac{1}{2\pi}(2\Re[R(z)]-1)$ for $z=e^{i\theta}$. This gives the following for the
above case:
\begin{eqnarray}
&&\beta_{l}\bigg(e^{i(l-1)\theta}+\sum_{j=l}^{2l-2}
\omega_{j}(e^{i(j-l)\theta}+e^{i(l-1-j)\theta})+e^{-il\theta}\bigg)
\sqrt{(e^{i\theta}-e^{i\theta_{0}})(e^{i\theta}-e^{-i\theta_{0}})}\notag\\  
&&\Rightarrow\beta_{l}\bigg(e^{i(l-\frac{1}{2})\theta}
+\sum_{j=l}^{2l-2}\omega_{j}(e^{i(j-l+\frac{1}{2})\theta}
+e^{i(l-\frac{1}{2}-j)\theta})+e^{-i(l-\frac{1}{2})\theta}\bigg)
e^{-i\frac{\theta}{2}}2e^{i\frac{\theta}{2}}\sqrt{\sin^{2}
  (\frac{\theta_{0}}{2})-\sin^{2}(\frac{\theta}{2})}\bigg)\notag\\    
&&\Rightarrow 4\beta_{l}\bigg(\cos((l-\frac{1}{2})\theta)+
+\sum_{j=l}^{2l-2}\omega_{j}\cos((j-l+\frac{1}{2})\theta)\bigg)
\sqrt{\sin^{2}(\frac{\theta_{0}}{2})-\sin^{2}(\frac{\theta}{2})}\notag\\  
\end{eqnarray}
One can now calculate the moments. To calculate them we use Cauchy
integral formula
\begin{eqnarray}
R^{n}(0)=\frac{n{!}}{2\pi i}\oint_{\cC} dz\frac{R(z)}{z^{n+1}}
\end{eqnarray}
The contour being taken around origin. Putting the form of $R(z)$ in this we obtain:
\begin{eqnarray}
  R^{n}(0)=\frac{n{!}}{2\pi i}\oint \frac{dz}{z^{n+1}}\frac{1}{2}
  [1-\sum_{i=1}^{l}\beta_{i}[\frac{1}{z^{i}}-z^{i}]+\frac{\beta_{l}}{z^{l}}(z^{2l-1}
  +\sum_{j=l}^{2l-2}\omega_{j}(z^{j}+z^{2l-1-j})+1)\sqrt{z^{2}-2\,
    \cos(\theta_{0})z+1}]\notag\\  
\end{eqnarray}
Using the formula for generating function of Legendre polynomials
again, and the residue theorem we obtain:
\begin{eqnarray}
  &&R^{n}(0)=\frac{n{!}}{2}[\beta_{n}+\beta_{l}\delta_{n,l-1} +\beta_{l}
  \sum_{j=l}^{2l-2}\sum_{m=0}^{2}\sum_{k=0}^{\infty}\omega_{j}a_{m}P_{k}
  (\cos(\theta_{0}))(\delta_{j+m+k-l-n,0}+\delta_{l-1-j-n+m+k,0})\notag\\
  &&\quad\quad\quad\quad\quad+\beta_{l}\sum_{m=0}^{2}
  \sum_{k=0}^{\infty}a_{m}P_{k}(\cos(\theta_{0}))\delta_{-l-n+m+k,0}]\notag\\ 
  &&\Rightarrow
  R^{n}(0)=\frac{n{!}}{2}[\beta_{n}+\beta_{l}\delta_{n,l-1} +
  \beta_{l}\sum_{j=l}^{2l-2}\sum_{m=0}^{2}\omega_{j}a_{m}P_{l+n-j-m}
  (\cos(\theta_{0})) 
  +\beta_{l}\sum_{j=l}^{2l-2}\sum_{m=0}^{2}\omega_{j}a_{m}
  P_{n+j+1-l-m}(\cos(\theta_{0}))\notag\\  
  &&\quad\quad\quad\quad\quad+\beta_{l}\sum_{m=0}^{2}
  a_{m}P_{l+n-m}(\cos(\theta_{0}))]\notag\\ 
  &&\Rightarrow \beta_{n}\frac{(n-1){!}}{\tilde{a}_{+n,-n}}
  =\frac{n{!}}{2}[\beta_{n}+\beta_{l}\delta_{n,l-1}+\beta_{l}
  \sum_{j=l}^{2l-2}\sum_{m=0}^{2}\omega_{j}a_{m}P_{l+n-j-m}
  (\cos(\theta_{0}))\notag\\   
  &&\quad\quad\quad\quad\quad+\beta_{l}\sum_{j=l}^{2l-2}
  \sum_{m=0}^{2}\omega_{j}a_{m}P_{n+j+1-l-m}(\cos(\theta_{0}))
  +\beta_{l}\sum_{m=0}^{2}a_{m}P_{l+n-m}(\cos(\theta_{0}))]\notag\\  
\end{eqnarray}
These form a set of $l$ linear simultaneous equation in $\beta_{n}$
after substituting for $\omega_{j}$. To have a solution the
determinant of this equation must be zero, yielding the relation
between $\tilde{a}_{+n,-n}$ and $\cos(\theta_{0})$.

\subsection{generic non zero coupling}
Moving onto addition of any generic term for the nonzero coupling case, which can be
dealt in a similar fashion. Let us consider a term, characterized by
an integer $l$ in the following way:
\begin{eqnarray}
  \sum_{c_{i},d_{j}}\bigg[a_{c_{i},d_{j},l}\prod_{i}^{n}(Tr[U^{i}])^{c_{i}}
  \prod_{j=1}^{m}(Tr[(U^{-1})^{j}])^{d_{j}}+a_{c_{i},d_{j},l}\prod_{i}^{n}
  (Tr[(U^{-1})^{i}])^{c_{i}}\prod_{j=1}^{m}(Tr[U^{j}])^{d_{j}}\bigg]
\end{eqnarray}
where $\sum_{i=1}^{n}ic_{i}=\sum_{j=1}^{m}jd_{j}=l$ and the expression is symmetric in $U\to U^{-1}$. We choose $2\le l$ to differentiate this from case considered earlier . Using similar procedure for deriving the Schwinger Dyson equation earlier, we obtain in this case:
\begin{eqnarray}
  &&\sum_{c_{i},d_{j}}\bigg[\sum_{i=1}^{n}ic_{i}\tilde{a}_{c_{i},d_{j},l}
  \prod_{k=1,k\ne i}^{n}(Tr[U^{k}])^{c_{k}}(Tr[U^{i}])^{c_{i}-1}Tr[U^{i}
  (1-zU)^{-1}]\prod_{j=1}^{m}(Tr[(U^{-1})^{j}])^{d_{j}}\notag\\
  &&-\sum_{j=1}^{m}jd_{j}\tilde{a}_{c_{i},d_{j},l}\prod_{i}^{n}(Tr[U^{i}])^{c_{i}}
  \prod_{k=1,k\ne
    j}^{m}(Tr[(U^{-1})^{k}])^{d_{k}}(Tr[(U^{-1})^{j}])^{d_{j}-1}
  Tr[(U^{-1})^{j}(1-zU)^{-1}] \notag\\
  &&-\sum_{j=1}^{m}ic_{i}\tilde{a}_{c_{i},d_{j},l}\prod_{k=1,k\ne i}^{n}
  (Tr[(U^{-1})^{k}])^{c_{k}}(Tr[(U^{-1})^{i}])^{c_{i}-1}Tr[(U^{-1})^{i}(1-zU)^{-1}]
  \prod_{j=1}^{m}(Tr[U^{j}])^{d_{j}}\notag\\
  &&+\sum_{j}^{m}jd_{j}\tilde{a}_{c_{i},d_{j},l}\prod_{i}^{n}(Tr[(U^{-1})^{i}])^{c_{i}}
  \prod_{k=1,k\ne j}^{m}(Tr[U^{k}])^{d_{k}}(Tr[U^{j}])^{d_{j}-1}Tr[U^{j}(1-zU)^{-1}]\bigg]
\end{eqnarray}
Noting $R^{i}(0)=i{!}\langle Tr[U^{i}]\rangle=i{!}\langle
Tr[(U^{-1})^{i}]\rangle$, we have the following as the redefinition of
$\beta_{i}$ taking into account sum over $i$, $2\le i\le l$ (i.e. also
including generic zero coupling case discussed earlier).
\begin{eqnarray}
  &&\beta_{i}=\tilde{a}_{1,1,i}\frac{R^{i}(0)}{(i-1){!}}+\sum_{l=3}^{L}
  \sum_{c_{i},d_{j}}\bigg[ic_{i}\tilde{a}_{c_{i},d_{j},l}\prod_{k=1,k\ne
    i}^{n}\bigg(\frac{R^{k}(0)}{k{!}}\bigg)^{c_{k}}\bigg(\frac{R^{i}(0)}{i{!}}
  \bigg)^{c_{i}-1}\prod_{j=1}^{m}\bigg(\frac{R^{j}(0)}{j{!}}\bigg)^{d_{j}}
  \notag\\
  &&+id_{i}\tilde{a}_{c_{i},d_{j},l}\prod_{i=1}^{n}\bigg(\frac{R^{i}(0)}{i{!}}
  \bigg)^{c_{i}}\prod_{k=1,k\ne 
    j}^{m}\bigg(\frac{R^{k}(0)}{k{!}}\bigg)^{d_{k}}\bigg(\frac{R^{j}(0)}{j{!}}
  \bigg)^{d_{j}-1}\bigg] 
\end{eqnarray}
where as usual $\sum_{i}^{n}ic_{i}=\sum_{j}jd_{j}=l$. Under this
redefinition, the non-zero coupling equation for the Resolvent becomes
exactly same as zero coupling and so the spectral density in terms of
$\beta_{i}$ will be same. But now the relation between the $\tilde{a}$
parameters is extremely complicated. So to eliminate the variables
$\beta_{i}$ in favour of the coupling constants is a non trivial
task as there no longer remains a linear set of equations to solve. Nevertheless the spectral density takes exactly the same
functional form in these variables, which is indeed non trivial. In
fact any model which has an action, polynomial in traces of $n^{th}$
order, and invariant under the operation $U\to U^{-1}$ will have the
identical functional form of the eigenvalue distribution and the
Resolvent to the placquette model of $n^{th}$ order, with the parameters
$\beta_{i}$ mapped to the coupling constants of the placquette model. \\

Notably when we have the specific case where :
\begin{eqnarray}
Z=\int[dU]\exp\bigg(N^{2}\sum_{i=1}^{l}\bigg(\frac{1}{N^{2}}Tr[U]Tr[U^{\dagger}]\bigg)^{i}\bigg)
\end{eqnarray}
which is the generalization of the phenomenological $a,b$ model. This will be equivalent to the simple one placquette model under the redefinition:
\begin{eqnarray}
\beta=\sum_{i=1}^{l}i\tilde{a}_{i}(R'(0))^{2i-1}
\end{eqnarray}

\section{Different Phases of zero coupling model : approximation 2}\label{app:app1}

\subsection{ Solution class 1: No-cut solution}

For this class, the resolvent is analytic inside the unit
circle and also $R(0)=1$, which means the negative powers of $z$ must
cancel out. We note this is the case if $F(z)$ is a perfect square,
and from above we have
\begin{eqnarray}
-4\lb\beta_{1}z+\beta_{2}z^{2}\rb \lb \frac{\beta_{1}}{z}+\frac{\beta_{2}}{z^{2}}\rb
+4\beta_{1}R'(0)+4\beta_{2} \lb \frac{R'(0)}{z}+\frac{R''(0)}{2}+R'(0)z\rb=0 
\end{eqnarray}
From this we find the resolvent is given by (we choose $+$ sign in
eqn. (\ref{Rzform}) otherwise $R(z)$ would have isolated singular
points),
\begin{eqnarray}\label{nocut-a1a2}
R(z)=1+\beta_{1}z+\beta_{2}z^{2}.
\end{eqnarray}
We shall discuss the conditions on the parameters $\beta_1$ and
$\beta_2$ for which this class of solution is valid in the next
subsection.

The spectral density for this branch, is given by
\be\label{specden-a1a2-nocut}
2\pi \s (\theta) = 1+2\b_1\cos\theta+2\b_2\cos2\theta.
\ee
Defining $\cos\theta :=x$, this equation can be written as,
\be
2\pi \s(x) = 4\b_2 x^2 + 2\b_1 x+(1-2\b_2), \quad x\in [-1,+1]
\ee
and $\s(x)\geq 0$ in that range of $x$. $\s(x)$ defines a parabola in
$\s(x)-x$ plane it can have only one extremum.  Since $\s(x)>0$ there are
three possibilities. 
\begin{enumerate}[(a)]
\item
$\s(x)$ has a minimum between $-1$ and $+1$ and the value of the
function is greater than or equal to $zero$ at minimum.
\item
$\s(x)$ has no minimum for $x\in [-1,+1]$.
\item
$\s(x)$ has a maximum  $-1$ and $+1$ and the value of the
function is greater than or equal to $zero$ at maximum.
\end{enumerate}
For the first case $\b_2>0$ (since $2\pi\s''(x) = 8\b_2$) and the minimum
occurs at $x=-\frac{\b_1}{4\b_2}$. If the minimum lies between $-1$
and $+1$ we get,
\be
|\b_1| \leq 4 \b_2 \quad (\text{since} \ \b_2 >0).
\ee
Also demanding that the value of the function at this point is greater
than or equal to $zero$, we get,
\be
2\pi\s\lb -\frac{\b_1}{4\b_2}\rb = \frac{4\b_2(1-2\b_2)-\b_1^2}{4\b_2}
\geq 0 \quad \Rightarrow \quad \b_1^2 \leq 4\b_2(1-2\b_2).
\ee
\begin{figure}
\begin{subfigure}{.32\textwidth}
  \centering
\includegraphics[width=5cm,height=4cm]{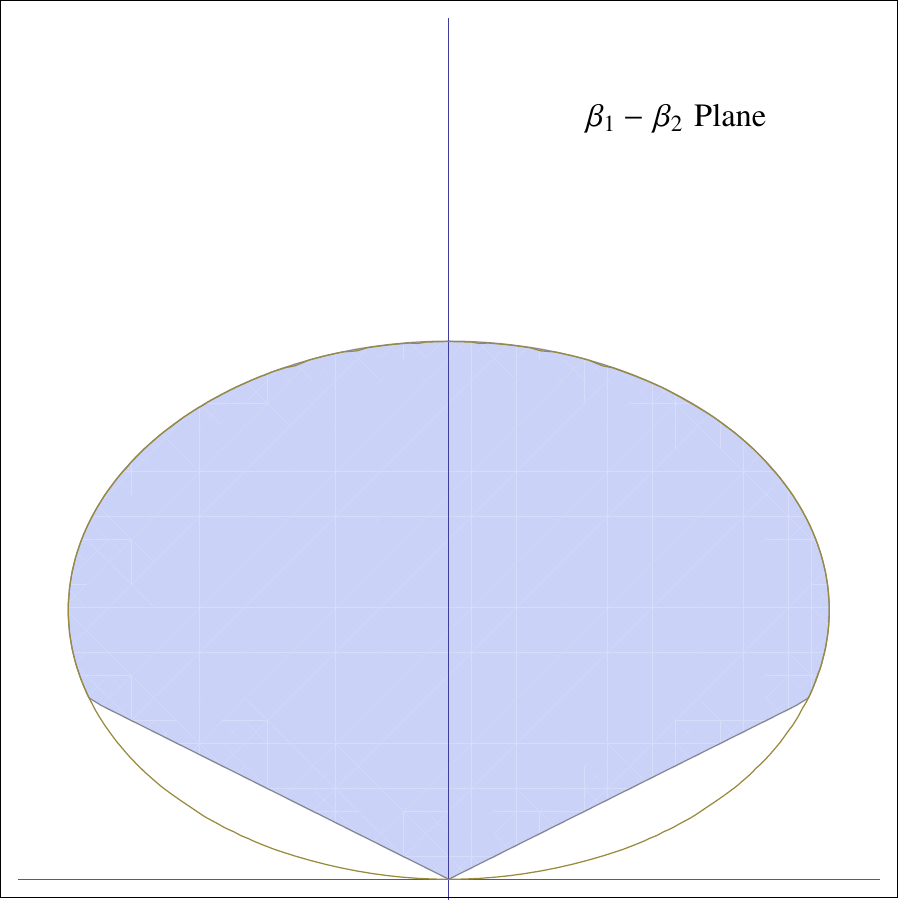}
\caption{}
\label{fig:a1a2-1}
\end{subfigure}%
\begin{subfigure}{.32\textwidth}
  \centering
\includegraphics[width=5cm,height=4cm]{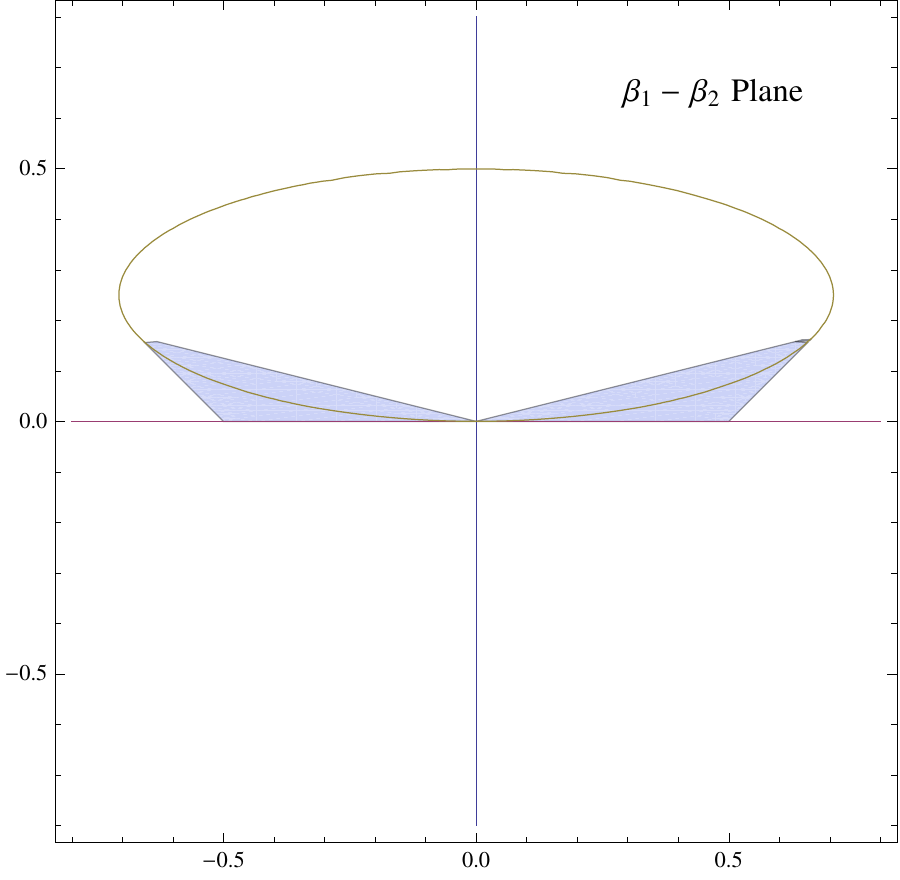}
\caption{}
\label{fig:a1a2-2}
\end{subfigure}%
\begin{subfigure}{.32\textwidth}
  \centering
\includegraphics[width=5cm,height=4cm]{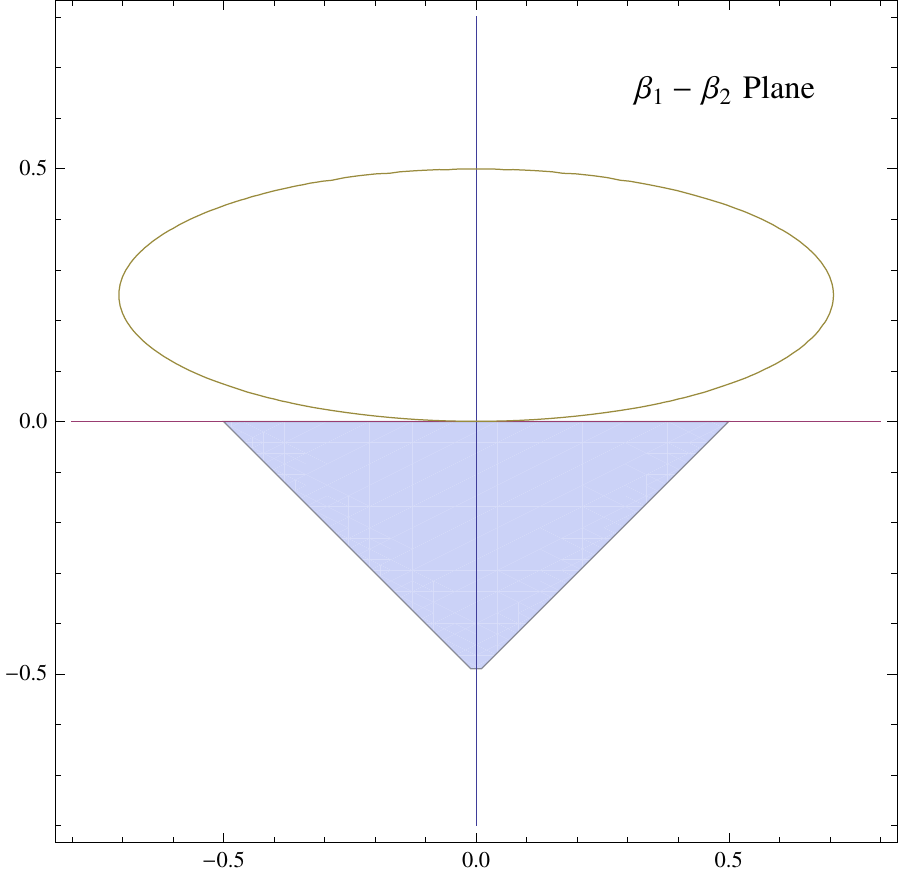}
\caption{}
\label{fig:a1a2-3}
\end{subfigure}%
\caption{}
\end{figure}
The blue area in figure (\ref{fig:a1a2-1}) denotes the corresponding
region in $(\b_1,\b_2)$ plane. The end points are given by $(\pm
\frac23,\frac16)$.

If there is no minimum for $\b_2>0$ case then the only condition
we have from the positivity of the function and these conditions are,
\be
1-2|\b_1|+2|\b_2| \geq 0.
\ee
along with
\be
|\b_1|\geq 4 \b_2.
\ee
Again the left most and right most end points are given by $(\pm
\frac23,\frac16)$. See figure (\ref{fig:a1a2-2}).

For $\beta_2<0$ the function can have maximum in $[-1,+1]$. The
minimum occurs at $x= \frac{\b_1}{4|\b_2|}$ and the value of the
function at this point is given by
\be
2\pi\s\lb \frac{\b_1}{4|\b_2|}\rb = \frac{4|\b_2|(1+2|\b_2|)+\b_1^2}{4\b_2}
\ee
which is always positive. Hence we do not get any further condition
from this. The only condition we have the function is positive in that
range {\it i.e.} $1-2|\b_1|-2|\b_2|>0$. This domain is depicted in
figure (\ref{fig:a1a2-3}).

\begin{figure}
\centering
\includegraphics[width=6cm,height=5cm]{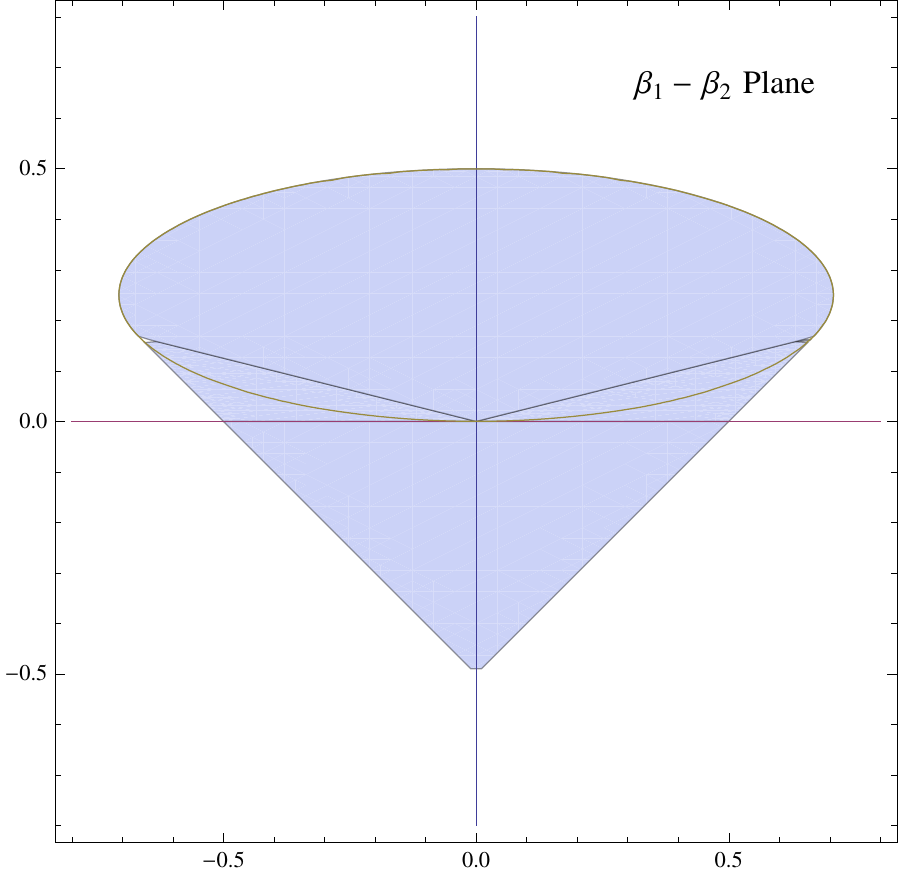}
\caption{}
\label{fig:no-cut-domain}
\end{figure}

Finally adding all these three domains we find the domain of validity
of no-cut solution depicted in (\ref{fig:no-cut-domain})

\bc
{\it Solution class 2: One-cut solution}
\ec

Assuming a one cut solution, the following can be taken as a
generalized ansatz which can be noted from the quadratic form of the
equation for the resolvent and the factorization of the
discriminant\footnote{Note that the function $F(z)$ in
  eqn. (\ref{eq:Fz-a1a2-twocut}) has a form like
  $\frac{P_8(z)}{z^4}$. Therefore, for one cut solution, this is the
  most generic way to choose the form of $F(z)$.}
\begin{equation}
  R(z)=\frac{1}{2}\lB 1-\beta_{1}\lb\frac{1}{z}-z\rb-\beta_{2}\lb
  \frac{1}{z^{2}}-z^{2}\rb 
  +\frac{\beta_{2}}{z^{2}}(z^{3}+a z^{2}+b z+1)\sqrt{z^{2}-2 z \,\cos\theta_1
    +1}\rB 
\end{equation}
Here we should note one important issue about the factorization. As we
can see all the coefficients of the polynomial $F(z)$ are
real. Therefore, when it is factorized into a quadratic polynomial
times a square of a cubic polynomial, with real coefficients, the
complex roots of the quadratic part have same modulus as the sum of
roots is real. More over their modulus must be $1$ as the product of
the roots is also $1$. Also one must note that the $z\to\frac{1}{z}$
symmetry of $F(z)$ demands the coefficients of the highest power of
$z$ and lowest power of $z$ in each of the factorized piece should be
same. Thus there can be a multiplicative constant, with any of the
factors, which can be re-absorbed in redefinition of $a$ and
$b$. Hence the above factorization is of the most generic form. 

From the condition $R(0)=1$ and using the following taylor series,
\begin{eqnarray}
 \sqrt{z^{2}-2\,\cos\theta_1z+1}=
  1+\frac{1}{2}(z^{2}-2\,\cos\theta_1z)-\frac{1}{8}(z^{2}-2\,
  \cos\theta_1z)^{2}+\frac{1}{16}(z^{2}-2\, \cos\theta_1z)^{3}+\cdots
\end{eqnarray}
we find,
\begin{eqnarray}
-\beta_{1}+b \beta_{2}-\beta_{2} \cos\theta_1 &=& 0\notag\\
a \beta_{2} -\beta_{2}b\, \cos \theta_1+ \beta_{2} \lb\frac{1}{2}
-\frac{1}{2} \cos^{2} \theta_1 \rb
&=& 1
\end{eqnarray}
Solving these two equations we find,
\begin{eqnarray}
b &=& \frac{\beta_{1}+\beta_{2} \cos \theta_1}{\beta_{2}}\notag\\
a &=&\frac{1}{\beta_{2}}+b\, \cos\theta_1-\frac{1}{2}(1- \cos^{2}
\theta_1 )
\end{eqnarray}
Since the measure is invariant under complex conjugation, and hence
the spectral density is invariant under $z\to1/z$ which implies
$a=b$. This in turn gives an equation for $\cos \theta_1$.
\be
3\beta_{2}\, \cos^{2}\theta_{1} +2(\beta_{1}-\beta_{2})
\cos\theta_{1}-2\beta_{1}-\beta_{2}+2=0 
\ee
%
As a consistency check we see that for $\beta_2=0$ we get back
condition (\ref{onecutcondition-a1}). Defining
$x=\sin^2\frac{\theta_1}{2}$ the above equation can be written as,
\be \label{x-eqn-a1a2-one-cut}
2x\lB \b_1+\b_2\lb 2-3 x\rb\rB =1.
\ee
On the other hand, the spectral density for this branch is given by,
\ben
2\pi\s(\theta) &=& 2 \Re[R(z)]-1 \nonumber\\
&=& 2 \cos\frac{\theta
  }{2}\sqrt{\sin ^2\frac{\theta _1}{2}-\sin
  ^2\frac{\theta }{2}} \left(\beta _1+\beta _2 \left(2 \cos \theta +\cos\theta
      _1-1\right)\right)
 \een
Therefore we need to find a domain in $(\b_1,\b_2)$ plane for which
there exists a real solution for $x$ between $0$ and $1$ in
eqn. (\ref{x-eqn-a1a2-one-cut}) and $\s(\theta)\geq 0$. The phase space
has been throughly discussed in \cite{Jurkiewicz:1982iz}. Interested
reader are referred to that paper. \\

\bc
{\it Solution class 3: Two-cut solution}
\ec

The most generic way to choose $F(z)$ for two-cut solution is
\be
F(z) = \left(\frac{\alpha}{z^2}(z^2+ b z+1)\sqrt{(z^2-2z
  \cos\theta_1+1)(z^2-2z\cos\theta_2+1)} \right)^2.
\ee
This form is invariant under $z\ra\frac1z$. $R(z)$ is given
by,
\be
R(z) = \frac12\lB 1-\beta_1\lb \frac1z-z\rb -\beta_2\lb
\frac1{z^2}-z^2\rb + \sqrt{F(z)} \rB.
\ee
From the regularity of $R(z)$ at $z=0$ we find that,
\ben
\alpha= \beta_2 \quad \text{and} \quad b= \cos\theta_1+\cos\theta_2
+\frac{\beta_1}{\beta_2}.  
\een
From the normalization of $R(z)$, {\it i.e.} $R(0)=1$ we find,
\be
-\beta_1(\cos\theta_1+\cos\theta_2) +
\beta_2\lb 2-(\cos\theta_1+\cos\theta_2)^2
-\frac12(\cos\theta_1-\cos\theta_2)^2\rb =2.
\ee
The spectral density for this distribution is given by,
\be 
2\pi \s(\theta) = 2 \Re[R(z=e^{i\theta})]-1 =
2\sqrt{(\cos\theta_1-\cos\theta)(\cos\theta_2 -\cos\theta_1)}\lb
\beta_1 + \beta_2 (2\cos\theta +\cos\theta_1+\cos\theta_2) \rb \ee
Since the spectral density has no support for
$\theta_1<\theta<\theta_2$ (assuming $\theta_2>\theta_1$) the
following integral must vanish,
\be
\int_{\theta_1}^{\theta_2} \s(\theta) d\theta =0.
\ee
This gives further condition of $\theta_1$ and $\theta_2$. 

The phase diagram can also be discussed in the similar way as in
\cite{Jurkiewicz:1982iz}. 

Thus we see that the complete phase diagram of ${\cal N}=4$ super
Yang-Mills theory can be captured studying the analytic properties of
the resolvent in a complex $z$ plane. 

\section{Derivation of properties of resolvent $H(h)$}\label{app:hprop}
Here we discuss the properties of the resolvent $H(h)$. Firstly lets take the case for $1-$cut solution for simplicity. The asymptotic expansion of $H(h)$, for $h\to\infty$, can simply be obtained as:
\begin{eqnarray}
H(h)&&=\int^{h_{U}}_{h_{L}}dh'\frac{u(h')}{h-h'}\notag\\
&&=\frac{1}{h}\int^{h_{U}}_{h_{L}}dh'\frac{u(h')}{1-\frac{h'}{h}}\notag\\
&&\sim\frac{1}{h}\int^{h_{U}}_{h_{L}}dh' u(h')\bigg(1+\frac{h'}{h}+\dots\bigg)\notag\\
&&=\frac{1}{h}\int^{h_{U}}_{h_{L}}dh' u(h')+\frac{1}{h^{2}}\int^{h_{U}}_{h_{L}}dh' u(h')h'+\dots\notag\\
\end{eqnarray}
Using definitions:
\begin{eqnarray}
\int^{h_{U}}_{h_{L}}dh' u(h')=1;\quad\quad\quad\int^{h_{U}}_{h_{L}}dh' u(h')h'=k'+\frac{1}{2}
\end{eqnarray}
we get the asymptotic expansion of $H(h)$.

Next we show the defining property of $H(h)$ :
\begin{eqnarray}
lim_{\epsilon\to0}H(h+i\epsilon)+H(h-i\epsilon)=2\ln\bigg[\frac{h}{\zeta}\bigg]
\end{eqnarray}
To show this we take the definition of $H(h)$ with discontinuous support :
\begin{eqnarray}
H(h)=\int_{p}^{q}dh'\frac{\tilde{u}_{1}(h')}{h-h'}+\int_{r}^{s}dh'\frac{\tilde{u}_{2}(h')}{h-h'}
\end{eqnarray}
Now if $h$ fall on the first branch $q>h>p$, then:
\begin{eqnarray}
H(h+i\epsilon)=\int_{p}^{q}dh'\frac{\tilde{u}_{1}(h')}{h+i\epsilon-h'}+\int_{r}^{s}dh'\frac{\tilde{u}_{2}(h')}{h+i\epsilon-h'}
\end{eqnarray}
The second term is automatically equal to :
\begin{eqnarray}
\Xint-_{r}^{s}dh'\frac{\tilde{u}_{2}(h')}{h-h'}
\end{eqnarray}
This is because $h$ does not fall between $[r,s]$, and so in the $\epsilon\to0$ limit this automatically gives the above. Next the first term can be expanded in a series in $\epsilon$. Keeping only up to order $\epsilon$ we have for the first term:
\begin{eqnarray}
\Xint-_{p}^{q}dh'\frac{\tilde{u}_{1}(h')}{h-h'}-i\epsilon\Xint-_{p}^{q}dh'\frac{\tilde{u}_{1}(h')}{(h-h')^{2}}+O(\epsilon^{2})
\end{eqnarray}
Thus adding both terms we have:
\begin{eqnarray}
H(h+i\epsilon)=\Xint-_{p}^{q}dh'\frac{\tilde{u}_{1}(h')}{h-h'}-i\epsilon\Xint-_{p}^{q}dh'\frac{\tilde{u}_{1}(h')}{(h-h')^{2}}+O(\epsilon^{2})+\Xint-_{r}^{s}dh'\frac{\tilde{u}_{2}(h')}{h-h'}
\end{eqnarray}
From similar consideration we obtain, for $q>h>p$ :
\begin{eqnarray}
H(h-i\epsilon)=\Xint-_{p}^{q}dh'\frac{\tilde{u}_{1}(h')}{h-h'}+i\epsilon\Xint-_{p}^{q}dh'\frac{\tilde{u}_{1}(h')}{(h-h')^{2}}+O(\epsilon^{2})+\Xint-_{r}^{s}dh'\frac{\tilde{u}_{2}(h')}{h-h'}
\end{eqnarray}
Adding the two equations above, we obtain:
\begin{eqnarray}
lim_{\epsilon\to 0}H(h+i\epsilon)+H(h-i\epsilon)=2\bigg[\Xint-_{p}^{q}dh'\frac{\tilde{u}_{1}(h')}{h-h'}+\Xint-_{r}^{s}dh'\frac{\tilde{u}_{2}(h')}{h-h'}\bigg]=2\ln\bigg[\frac{h}{\zeta}\bigg]
\end{eqnarray}
Where we have used the equation for the cut integral in the last substitution. Thus the asymptotic expansion and the the above property, gives a strong tool to describe the resolvent without actually computing the contour integral, which for general cases with discontinuous support is extremely cumbersome.

\section{Young Tableaux distribution for $(a_1,a_2)$ model}\label{app:yt-a1a2}

We discuss the derivation of the effective action and the equations of motion, if there are two terms in the action of the type, $Tr[U]Tr[U^{\dagger}]$ and $Tr[U^{2}]Tr[U^{\dagger 2}]$. The partition function takes the following form:  
\begin{eqnarray} Z=\sum_{k_{1},k_{2}}\frac{a_{1}^{k_{1}}a_{2}^{k_{2}}}{k_{1}{!}k_{2}{!}}\sum_{R}[\chi(C(k_{1},k_{2})]^{2} \end{eqnarray} The character in this case is explicitly written as: \begin{eqnarray} \sum_{q_{j}}\frac{k_{1}{!}k_{2}{!}}{\prod_{j}^{N}(l_{j}-2q_{j}){!}q_{j}{!}}\prod_{r<s}(l_{r}-l_{s}-2(q_{r}-q_{s}))\delta(\sum_{j}q_{j}-k_{2}) \end{eqnarray} where $l_{i}=n_{i}+N-1$, $n_{i}$ being the number of boxes in the $i^{th}$ row of the Young diagram. Using the fact that the sum over representations can be understood as sum over Young diagrams, we have the partition function equal to: \begin{eqnarray}
  &&Z=\notag\\
  &&\sum_{k_{1},k_{2},\{n_{i}\}=0}^{\infty}\delta(\sum_{i=1}^{N}\{n_{i}\}-k_{1}-2k_{2})\frac{a_{1}^{k_{1}}a_{2}^{k_{2}}}{k_{1}{!}k_{2}{!}}\bigg[\sum_{q_{j}}\frac{k_{1}{!}k_{2}{!}}{\prod_{j}^{N}(l_{j}-2q_{j}){!}q_{j}{!}}\prod_{r<s}(l_{r}-l_{s}-2(q_{r}-q_{s}))\delta(\sum_{j}q_{j}-k_{2})\bigg]^{2}\notag\\
\end{eqnarray} Rewriting the sum, we have: \begin{eqnarray}
&&\sum_{k_{1},k_{2},l_{i},q_{i},s_{i}}\frac{a_{1}^{k_{1}}a_{2}^{k_{2}}k_{1}{!}k_{2}{!}}{\prod_{j}^{N}(l_{j}-2q_{j}){!}q_{j}{!}(l_{j}-2s_{j}){!}s_{j}{!}}\prod_{r<p}(l_{r}-l_{p}-2(q_{r}-q_{p}))(l_{r}-l_{p}-2(s_{r}-s_{p}))\notag\\
&&\quad\quad\quad\quad\times\delta(K-k_{1}-2k_{2})\delta(\sum_{j}q_{j}-k_{2})\delta(\sum_{j}q_{j}-\sum_{j}s_{j})\notag\\
\end{eqnarray}
In the large $N$ limit then, the partition function takes the following form:
\begin{eqnarray}
Z=\int dk_{2}\,dl(x)\,dq(x)\,ds(x)\,d\lambda\,exp[-N^{2}S_{eff}(l(x),q(x),s(x),k_{2},\lambda)]
\end{eqnarray}
Where $\lambda$ is a Lagrange multiplier and $S_{eff}$ is given by:
\begin{eqnarray}
&&S_{eff}=-[(K'-2k'_{2}){\rm{ln}}a_{1}+k_{2}{\rm{ln}}a_{2}+(K'-2k'_{2}){\rm{ln}}(K'-2k'_{2})-(K'-2k'_{2})+k'_{2}{\rm{ln}}k'_{2}-k'_{2}\notag\\
&&-\int_{0}^{1} dx\,(l(x)-2q(x)){\rm{ln}}(l(x)-2q(x))-\int_{0}^{1} dx\,(l(x)-2s(x)){\rm{ln}}(l(x)-2s(x))\notag\\
&&+\int_{0}^{1} dx(l(x)-2q(x))+\int_{0}^{1} dx(l(x)-2s(x))\notag\\
&&-\int_{0}^{1}dx\,q(x){\rm{ln}}q(x)+\int_{0}^{1}dx\,q(x)-\int_{0}^{1}dx\,s(x){\rm{ln}}s(x)+\int_{0}^{1}dx\,s(x)\notag\\
&&+\frac{1}{2}\Xint-_{0}^{1}dx\Xint-_{0}^{1}dy\,{\rm{ln}}|l(x)-l(y)-2(q(x)-q(y))|+\frac{1}{2}\Xint-_{0}^{1}dx\Xint-_{0}^{1}dy\,{\rm{ln}}|l(x)-l(y)-2(s(x)-s(y))|\notag\\
&&+\lambda(\int_{0}^{1}(q(x)-s(x)))]
\end{eqnarray}
Where the variables have been defined in the large $N$ limit analogously to the case for one cycle. Also $K=N^{2}K'$ etc.. Note that we have performed the integration over $k_{1}$ in the path integral and thus replaced $k'_{1}$ by $K'-2k'_{2}$. More over one can use the following relation $\int dx\,l(x)=K'+\frac{1}{2}$ and $\int_{0}^{1}dx\,q(x)=k'_{2}=\int_{0}^{1}dx\,s(x)$. This gives:
\begin{eqnarray}
&&S_{eff}=-[(K'-2k'_{2}){\rm{ln}}a_{1}+k_{2}{\rm{ln}}a_{2}+(K'-2k'_{2}){\rm{ln}}(K'-2k'_{2})+k'_{2}{\rm{ln}}k'_{2}\notag\\
&&-\int_{0}^{1} dx\,(l(x)-2q(x)){\rm{ln}}(l(x)-2q(x))-\int_{0}^{1} dx\,(l(x)-2s(x)){\rm{ln}}(l(x)-2s(x))\notag\\
&&-\int_{0}^{1}dx\,q(x){\rm{ln}}q(x)-\int_{0}^{1}dx\,s(x){\rm{ln}}s(x)\notag\\
&&+\frac{1}{2}\Xint-_{0}^{1}dx\Xint-_{0}^{1}dy\,{\rm{ln}}|l(x)-l(y)-2(q(x)-q(y))|+\frac{1}{2}\Xint-_{0}^{1}dx\Xint-_{0}^{1}{\rm{ln}}dy\,|l(x)-l(y)-2(s(x)-s(y))|\notag\\
&&+\lambda(\int_{0}^{1}(q(x)-s(x)))+K'+1-k'_{2}]
\end{eqnarray}
\subsection{equations of motion}
Next we calculate the equations of motion for this action by taking variation of the above with respect to the variables $l(x),s(x),q(x)$ and $k'_{2}$. These are respectively as follows:
\begin{eqnarray}
&&-\frac{\delta S_{eff}}{\delta l(x)}={\rm{ln}}a_{1}+{\rm{ln}}(K'-2k'_{2})-{\rm{ln}}(l(x)-2q(x))-{\rm{ln}}(l(x)-2s(x))\notag\\
&&\quad\quad\quad\quad+\Xint-_{0}^{1}dy\frac{1}{l(x)-l(y)-2(q(x)-q(y))}+\Xint-_{0}^{1}dy\frac{1}{l(x)-l(y)-2(s(x)-s(y))}=0
\end{eqnarray}
\begin{eqnarray}
&&-\frac{\delta S_{eff}}{\delta q(x)}=-2{\rm{ln}}a_{1}+{\rm{ln}}a_{2}-2{\rm{ln}}(K'-2k'_{2})+{\rm{ln}}k'_{2}+2{\rm{ln}}(l(x)-2q(x))-{\rm{ln}}q(x)-1\notag\\
&&\quad\quad\quad\quad-2\Xint-_{0}^{1}dy\frac{1}{l(x)-l(y)-2(q(x)-q(y))}+\lambda=0
\end{eqnarray}
\begin{eqnarray}
&&-\frac{\delta S_{eff}}{\delta s(x)}=2{\rm{ln}}(l(x)-2s(x))-{\rm{ln}}s(x)+1-2\Xint-_{0}^{1}dy\frac{1}{l(x)-l(y)-2(s(x)-s(y))}-\lambda=0
\end{eqnarray}
\begin{eqnarray}
&&-\frac{\delta S_{eff}}{\delta k'_{2}}=-2{\rm{ln}}a_{1}+{\rm{ln}}a_{2}-2{\rm{ln}}(K'-2k'_{2})-2+{\rm{ln}}k'_{2}+1-1+\lambda=0
\end{eqnarray}
Using the equations of motion for $q(x)$ and $s(x)$ we get:
\begin{eqnarray}
&&{\rm{ln}}\bigg[\frac{a_{2}k'_{2}}{a_{1}^{2}(K'-2k'_{2})^{2}}\bigg]+2{\rm{ln}}(l(x)-2q(x))-{\rm{ln}}(q(x)s(x))-2\Xint-_{0}^{1}dy\frac{1}{l(x)-l(y)-2(q(x)-q(y))}\notag\\
&&+2{\rm{ln}}(l(x)-2s(x))-2\Xint-_{0}^{1}dy\frac{1}{l(x)-l(y)-2(s(x)-s(y))}=0\notag\\
\end{eqnarray}
Using equation of motion for $l(x)$ we get:
\begin{eqnarray}
&&{\rm{ln}}\bigg[\frac{a_{2}k'_{2}}{a_{1}^{2}(K'-2k'_{2})^{2}}\bigg]+2[{\rm{ln}}a_{1}+{\rm{ln}}(K'-2k'_{2})]-                                                                                                                                                                                                                                                                                             {\rm{ln}}(q(x)s(x))=0\notag\\
&&\Rightarrow q(x)s(x)=a_{2}k'_{2}
\end{eqnarray}
Again using equation of motion for $k'_{2}$ we can calculate the value of the Lagrange multiplier:
\begin{eqnarray}
{\rm{ln}}\bigg[\frac{a_{2}k'_{2}}{a_{1}^{2}(K'-2k'_{2})^{2}}\bigg]=2-\lambda
\end{eqnarray}
The above set of equations are extremely difficult to solve, primarily because there is no apriori information about the behaviour of $q(x)$ or $s(x)$. If we assume that either $q(x)$ or $s(x)$ is monotonically increasing then the other is of course monotonically decreasing. Then looking at the equation for $s(x)$, we observe if the support of $s(x)$ is continuous, (let us say $[\alpha,\beta]$) then evidently one can write $H(s)$ using similar ansatz as earlier:
\begin{eqnarray}
H(s)=2\ln\bigg[\frac{g(l-2s)-\sqrt{g^{2}(l-2s)-f^{2}(l-2s)}}{\kappa v(l-2s)}\bigg]-\ln\bigg[\frac{g'(s)-\sqrt{g^{\prime2}(s)-f^{\prime2}(s)}}{\kappa^{\prime} v'(s)}\bigg]
\end{eqnarray}
Where the functions $g,f,v,g',f',v'$ and the constants $\kappa,\kappa'$ are to be determined from the property of the resolvent :
\begin{eqnarray}
H(s+i\epsilon)+H(s-i\epsilon)=2\ln(l-2s)-\ln(s)+1-\lambda
\end{eqnarray}
and the asymptotic expansion using:
\begin{eqnarray}
\int_{s_{L}}^{s_{U}}ds\, \tilde{u}=1;\quad\quad\quad\int_{s_{L}}^{s_{U}}ds\, (l-2s)\tilde{u}=K'+\frac{1}{2}-2k'_{2}
\end{eqnarray}
where $\tilde{u}$ is the Young Tableaux density for $s$. It will be interesting to solve these set of equations, and obtain $u(l)$. This will allow one to check the phase space relationship more precisely and possibly generalize this to the whole class of models discussed in the eigenvalue sector. We leave this problem for later purpose.


\end{document}